\pdfoutput=1
\documentclass[11pt,reqno,preprint]{article}
\usepackage{jheppub,epsfig,amssymb,amsmath,mathrsfs,hyperref,graphicx}
\usepackage{multirow,bbm}
\usepackage[makeroom]{cancel}
\usepackage{caption}
\usepackage{subcaption}
\usepackage{ulem}
\usepackage[dvipsnames]{xcolor}
\usepackage{comment}
\usepackage{physics}
\usepackage{diagbox}
\def\be{\begin{equation}}
\def\ee{\end{equation}}
\def\ba{\begin{eqnarray}}
\def\ea{\end{eqnarray}}
\newcommand{\bea}{\begin{eqnarray}}
\newcommand{\eea}{\end{eqnarray}}

\def\Li{\textrm{Li}_2}

\def\e{\epsilon}
\def\a{{\textbf{a} }}
\def\b{{\textbf{b} }}
\def\u{{\textbf{u} }}
\def\v{{\textbf{v} }}

\def\emph#1{{\it #1}}

\def\KK{\mathrm{K}}
\def\NN{\mathrm{N}}

\def\Det{{\rm Det}}

\def\str{\textrm{Str}}
\def\Sph{\textrm{Sph}}
\def\AdS{\textrm{AdS}}

\newcommand{\qg}{\textrm{g}}
\newcommand{\qPhi}{\Phi}
\newcommand{\qpi}{\Pi}
\newcommand{\qK}{K}
\newcommand{\bqK}{\bar{K}}
\newcommand{\tqK}{\tilde{K}}

\newcommand{\qw}{\omega}

\newcommand{\qZ}{\mathbf{Z}}

\newcommand{\bbO}{\mathbb{O}}

\newcommand{\qU}{\mathscr{U}}
\newcommand{\GMN}{G}
\newcommand{\bGMN}{\bar{G}}
\def\sfchi{\begin{boldsymbol} \chi \end{boldsymbol}}
\def\bsfchi{\bar{\boldsymbol{\chi}}}

\newcommand{\zz}{q}

\newcommand{\cA}{\begin{cal}A\end{cal}}
\newcommand{\cB}{\begin{cal}B\end{cal}}
\newcommand{\cC}{\begin{cal}C\end{cal}}
\newcommand{\cD}{\begin{cal}D\end{cal}}
\newcommand{\cE}{\begin{cal}E\end{cal}}
\newcommand{\cF}{\begin{cal}F\end{cal}}

\newcommand{\cK}{\begin{cal}K\end{cal}}
\newcommand{\cL}{\begin{cal}L\end{cal}}
\newcommand{\cM}{\begin{cal}M\end{cal}}
\newcommand{\cN}{\begin{cal}N\end{cal}}

\newcommand{\cP}{\begin{cal}P\end{cal}}

\newcommand{\cR}{\begin{cal}R\end{cal}}

\newcommand{\cY}{\begin{cal}Y\end{cal}}
\newcommand{\cZ}{\begin{cal}Z\end{cal}}
\DeclareMathOperator*{\ordprod}{\prod\limits^{\vbox to -.5ex{\kern-0.5ex\hbox{$\leftharpoonup$}\vss}}}
\DeclareMathOperator*{\ordprodopp}{\prod\limits^{\vbox to -.5ex{\kern-0.5ex\hbox{$\rightharpoonup$}\vss}}}

\title{Classical correlation functions at strong coupling from hexagonalization}

\author{Benjamin Basso$^{1}$, Erkan Kalu\c c$^{1,2}$}
\author{and Didina~Serban$^{1,2}$}

\affiliation{$^{1}$ Laboratoire de Physique de l'\'Ecole Normale Sup\'erieure, ENS, Universit\'e PSL, CNRS, Sorbonne Universit\'e, Universit\'e Paris Cit\'e,
F-75005 Paris, France}

\affiliation{$^{2}$  Université Paris–Saclay, CNRS, CEA, Institut de Physique Théorique, 91191 Gif-sur-Yvette, France}

\abstract{We study correlation functions of half-BPS operators in planar $\mathcal{N}=4$ Super-Yang-Mills at strong coupling, in the classical limit where operator dimensions scale with the coupling. We focus on the two-dimensional kinematics corresponding in the dual description to strings propagating in $AdS_{3}\times S^{3}$. Using the hexagon formalism, we show that correlation functions exponentiate in this regime and are governed by the free energy of an associated set of Thermodynamic Bethe Ansatz (TBA) equations. These equations are structurally equivalent to the Gaiotto--Moore--Neitzke equations encoding BPS spectra in $\mathcal{N}=2$ supersymmetric field theories. Exploiting this correspondence, we apply wall-crossing techniques to extend the TBA framework and formulate a $\chi$-system applicable both to polygonal hexagon tilings and to closed geometries describing correlators of single-trace operators. In particular, for four-point functions, this construction generalizes the results of Caetano and Toledo for minimal surfaces in $AdS_{2}\times S^{1}$.}

\emailAdd{benjamin.basso@phys.ens.fr}
\emailAdd{osman-erkan.kaluc@ipht.fr}
\emailAdd{didina.serban@ipht.fr}

\begin{document}
\hypersetup{pageanchor=false}
\maketitle
\hypersetup{pageanchor=true}


\newcommand\scalemath[2]{\scalebox{#1}{\mbox{\ensuremath{\displaystyle #2}}}}

\section{Introduction and summary of results}
\label{sec:intro}

Integrability provides a powerful framework for computing correlation functions of local operators in planar $\cN=4$ Super-Yang-Mills (SYM) theory. In the case of two-point functions, it has led to the Quantum Spectral Curve~\cite{Gromov:2013pga,Gromov:2014caa}, which enables precise computations of the scaling dimensions of single-trace operators across the full range of the ’t Hooft coupling $g^2 = \lambda/(4\pi)^2$. Recently, it has been used to map out the low-lying spectrum of the theory from weak to strong coupling~\cite{Gromov:2023hzc,Marboe:2016yyn}, providing compelling evidence for the AdS/CFT correspondence.

In some cases, precise control over spectral data has also led to significant progress in the study of higher-point functions. Notable recent examples include correlation functions on the circular Wilson-Maldacena loop and four-point functions of stress-tensors~\cite{Cavaglia:2021bnz,Cavaglia:2022qpg,Caron-Huot:2022sdy,Caron-Huot:2024tzr,Cavaglia:2024dkk}, obtained using conformal bootstrap methods. At strong coupling, detailed knowledge of the string energy levels---combined with insights from the string worldsheet theory and analyticity constraints---have driven further advances, for instance in the computation of curvature corrections to Virasoro–Shapiro amplitudes~\cite{Alday:2022xwz,Alday:2023jdk,Alday:2023mvu,Fardelli:2023fyq,Wang:2025pjo}, which are directly related to four-point functions of gravitons and Kaluza–Klein modes.

While spectral data provide powerful constraints, a direct integrability-based treatment of general multi-point correlators must go beyond the spectral problem. The state-of-the-art framework for correlators is the hexagon form factor formalism for structure constants~\cite{Basso:2015zoa}, together with its higher-point extension, hexagonalization~\cite{Fleury:2016ykk,Eden:2016xvg,Fleury:2017eph,Bargheer:2017nne,Eden:2017ozn}. Geometrically, the con- struction parallels the triangulation of punctured surfaces, with the punctures representing the operator insertions and the edges of the triangulation their planar Wick contractions. Each hexagon has three mirror edges obtained by cutting along the corresponding propa- gators, while the remaining edges correspond to the operators themselves, represented as spin chains or strings. The full correlator is obtained by gluing hexagons, with the gluing implemented by summing over complete sets of mirror magnons propagating across the cuts.

This framework applies at arbitrary coupling and has led to numerous results, particularly for three- and four-point functions, where hexagonalization reproduces perturbative data to high loop orders~\cite{Basso:2015eqa,Eden:2015ija,Fleury:2016ykk,Eden:2016xvg,Fleury:2017eph,Basso:2017khq,Basso:2017muf,Coronado:2018ypq,Basso:2022nny}. In special limits, most notably the large-charge regime of the four-point function of half-BPS operators (the octagon), the hexagon series can be fully resummed, yielding an exact all-loop expression in terms of determinants of Bessel-type kernels~\cite{Coronado:2018cxj,Coronado:2018ypq,Bargheer:2019kxb,Kostov:2019stn,Belitsky:2020qrm,Belitsky:2019fan}. Studies of the octagon further reveal nontrivial structure at strong coupling~\cite{Bargheer:2019exp}, with connections to the string-theory description, and point more broadly to links between large-charge correlators and Coulomb-branch scattering amplitudes~\cite{Caron-Huot:2021usw,Caron-Huot:2023wdh,Bargheer:2025tcw}.

While inspiring, the methods developed for the octagon are difficult to extend to higher-point functions or to configurations involving sums and integrals over mirror states on multiple cuts. In practice, such computations are extremely challenging and typically limited to the lowest nontrivial orders at weak coupling~\cite{Fleury:2017eph,Bargheer:2018jvq,DeLeeuw:2019dak,Crisanti:2024vnd}. Progress may instead be achieved in the strong-coupling regime, where substantial simplifications arise and more direct resummation techniques become available~\cite{Jiang:2016ulr,Bargheer:2019exp}. 

The regime of interest corresponds to the classical limit, in which both the coupling constant and the operator dimensions scale to infinity simultaneously. In this regime, a classical string description becomes reliable, and correlation functions are expected to be exponentially suppressed and governed by minimal surfaces in $AdS_{5}\times S_5$. String theory suggests that a general integrable description should exist in this limit, rooted in the integrability of the worldsheet theory.

Before the development of the hexagonalization approach, an important step in that direction was already taken by Caetano and Toledo~\cite{Caetano:2012ac} who, working directly from the worldsheet theory, studied four-point functions of half-BPS operators in a one-dimensional kinematic setup. Their methods were based on earlier approaches for three-point functions of spinning strings~\cite{Janik_2011,Kazama_2011,Kazama_2014}. In that case, they obtained a general solution for arbitrary string configurations in an $AdS_2\times S^1$ subsector, in the form of a $\chi$-system---a set of functional equations closely related to the Thermodynamic Bethe Anstaz (TBA) equations introduced by Gaiotto, Moore, and Neitzke (GMN)~\cite{Gaiotto:2009hg} in the study of BPS spectra of $\mathcal{N}=2$ supersymmetric field theories. This construction parallels earlier results for gluon scattering amplitudes---or equivalently, null polygonal Wilson loops---described by open string worldsheets ending on polygonal contours at the AdS boundary~\cite{Alday:2007hr,Alday:2010vh,Alday:2009dv}. In that context, the minimal-surface problem can be reformulated in terms of a Hitchin system and analyzed via WKB methods~\cite{Gaiotto:2009hg}, mapping the string area to the free energy of a TBA system.

In this paper, we revisit the classical problem from the hexagonalization perspective and generalize the findings of~\cite{Caetano:2012ac} to higher-point functions of half-BPS operators in the large-charge regime $L_i\sim g \gg 1$. We restrict to two-dimensional kinematics~\cite{Fleury:2017eph}, corresponding to operator insertions in a plane and within an $O(4)$ scalar subsector, or equivalently to string configurations in an $AdS_3 \times S^3$ subspace.

Our main result is a set of TBA equations governing the classical strong-coupling limit of these correlation functions. Starting from hexagonalization, the correlators are expressed as partition functions $\qZ$ of magnon systems associated with  triangulations of the worldsheet. In the classical regime, this representation simplifies dramatically: the system reduces to weakly interacting magnons with purely diagonal interactions, in contrast to the finite-coupling case where the matrix structure of the form factors leads to substantial technical complications.

These simplifications allow the partition function to exponentiate, with a free energy governed by a Yang–Yang functional associated with a set of TBA equations, in close analogy with null polygonal Wilson loops~\cite{Alday:2010ku,Alday:2009dv,Alday:2010vh,Basso:2013vsa,BSV,Fioravanti_2015,Bonini_2019}. The final result takes the compact form
\be
\qZ \approx \exp{-2g \cA}\,,\qquad \cA = \sum_{\gamma} \cA_{\gamma}\,,
\ee
with
\be
\label{action_intro}
\cA_\gamma =  \int \frac{d\theta}{2\pi \cosh^2 \theta} \, \left[R(Y_{\gamma}^{\Sph}(\theta))+R(\bar{Y}^{\Sph}_{\gamma}(\theta))-R(Y^{\AdS}_{\gamma}(\theta))-R(\bar{Y}^{\AdS}_{\gamma}(\theta))\right]\, ,
\ee
where $\theta$ is the spectral parameter and $R(Y)$ denotes the Rogers dilogarithm,
\be
\label{eq:RogerDi_intro}
R(Y) \equiv \textrm{Li}_{2}(-Y) +\frac{1}{2} \log{(Y/\cY)}\log{(1+Y)}\, .
\ee
Here the $Y$-functions $Y^I_\gamma(\theta)$ come in four species, reflecting the $PSU(2|2)$ symmetry of the hexagon construction~\cite{Basso:2015zoa}. In the dual string description, they naturally split into AdS and sphere sectors, mirroring the decomposition of string degrees of freedom in $AdS_{3}\times S^{3}$. They satisfy TBA equations of the form
\be
\label{eq:TBA_intro}
\log{Y^{I}_{\gamma}(\theta)} = \log{{\cY}_{\gamma}^{I}(\theta)} + \sum_{\gamma'}\sum_{J}(-1)^{F_{J}+1}\langle \gamma, \gamma' \rangle \qK^{IJ}\star \log{(1+ Y^{J}_{\gamma'}(\theta))}\, ,
\ee
where $(-1)^{F_{J}}$ is a grading factor ($F_{\Sph} = 0, F_{\AdS} = 1$). The driving terms $\cY_{\gamma}^{I}(\theta)$ encode the dependence on kinematics and operator insertions---namely, cross ratios and R-charges---while the kernels $\qK^{IJ}(\theta,\theta')$ capture magnon interactions. Remarkably, these kernels are directly determined by the strong-coupling limit of the magnon $S$-matrix, essentially arising from its logarithm.
We present several equivalent formulations of these equations. In one, the kernel matrix becomes block-diagonal, effectively decoupling the AdS and sphere sectors at the cost of redefining the driving terms. In another, the kernels are recast in a relativistically invariant form.

A key aspect of the construction is the sum over magnon states labeled by $\gamma$. In simple cases, these labels correspond one-to-one with the cuts of the triangulation, with each cut supporting four $Y$-functions coupled to those on neighboring cuts. The structure of these interactions is encoded in skew-symmetric pairings $\langle \gamma,\gamma'\rangle$, which define the quiver of the triangulation.

This quiver structure is closely related to the convolution terms in the TBA equations. Since the kernels develop pole singularities when rapidities coincide, one must prescribe an ordering of the integration contours $\{\cC_\gamma\}$ associated with the $Y$-functions, consistent with kinematical limits. For linear quivers this ordering is straightforward, but it becomes nontrivial in the presence of cycles. Wall-crossing transformations~\cite{Gaiotto:2010okc,Alday:2010vh,Alday:2009dv,Toledo:PhDthesis} resolve this issue by allowing contour reordering at the cost of introducing composite $Y$-functions from pole contributions, which enter on the same footing as the elementary ones. With wall-crossing, the equations extend to arbitrary triangulations and reproduce those of GMN for $SU(2)$ Hitchin systems.

As usual, the integral equations admit an equivalent formulation in terms of functional relations of $Y$-system type. In our case, this system reflects the singularity structure described above and can be written as
\begin{align}
\label{eq:Ysys_intro}
Y_\gamma^{++}\,\bar{Y}_\gamma =\prod_{\gamma'\succ \gamma}(1+Y_{\gamma'}^{++})^{\langle \gamma',\gamma\rangle}\prod_{\gamma'\prec \gamma}\left(1+\bar{Y}_{\gamma'}\right)^{\langle \gamma,\gamma'\rangle}\, ,
\end{align}
together with the conjugate system obtained by $Y\leftrightarrow \bar Y$, where $Y^{++}\equiv Y(\theta+i\pi)$ and $\succ$ denotes the contour ordering. In this form, the equations are identical in the AdS and sphere sectors, which therefore appear completely decoupled.

Following~\cite{Gaiotto:2010okc}, the system can be further reformulated in terms of a $\chi$-system, providing an alternative set of functional equations with a universal geometric structure. We carry out this analysis in detail for polygonal hexagon tilings, which generalize the octagon case. We then discuss the extension to closed-string geometries corresponding to correlators of single-trace operators, focusing in particular on the four-point function. In the $AdS_{2}\times S^1$ subkinematics, we find perfect agreement with the functional equations and minimal-area results obtained by Caetano and Toledo~\cite{Caetano:2012ac}, after performing the final sum over the moduli (bridge lengths) entering the hexagonalization procedure.

The paper is organized as follows. Section~\ref{sec:review} reviews the basic elements of hexagonalization and introduces the notation related to triangulations, quivers, and cluster algebras. Section~\ref{sec:fromonecuttoNcutresum} presents the key ingredients of the strong-coupling analysis. We begin by reviewing the one-cut case, corresponding to the derivation of the octagon at strong coupling. We then turn to the two-cut case, where we explain the main simplifications in the hexagon sums that enable an exact computation at strong coupling. We subsequently extend the analysis to an arbitrary number of cuts and formulate an effective field theory description of the partition function. The derivation of the TBA equations from the saddle-point analysis of the effective action is carried out in Section~\ref{sec:tbafreeenergyforlinearquivers}. Two alternative formulations of the TBA equations and of the free energy are presented, together with their common universal $Y$-system reformulation. Strictly speaking, the analysis in this section applies to linear quivers arising from the simplest triangulations of a polygon; however, the notation is chosen to facilitate extension to more general quivers. In Section~\ref{sec:generaltriangulations}, we address the complications that arise for quivers with cycles, dual to triangulations containing internal triangles, and present its solution using wall crossing transformations for polygons. Motivated by the GMN approach, Section~\ref{sec:chi-system-lin-quivers} introduces the $\chi$-functions, the Kontsevich-Soibelman transformations, and the spectrum generating operator, culminating in the derivation of the $\chi$-system for polygons. The analysis is extended to closed-string correlators in Section~\ref{sec:closed}, with emphasis on the four-point function in the triangulation chosen by Caetano and Toledo~\cite{Caetano:2012ac}. To enable a precise comparison with the corresponding classical string results, we also analyze the sum over bridge lengths and its extremization. This leads to a fully symmetric form of the equations, in which the AdS and sphere degrees of freedom are treated on an equal footing. We conclude in Section~\ref{sec:conclusions}. Several appendices are devoted to more specialized topics. 

\textbf{Note added:} While completing this work, we learned of an upcoming paper by Till Bargheer, Thiago Fleury, and Davide Lai~\cite{DTT}, where they perform a strong-coupling analysis of higher-point functions via hexagonalization, which partially overlaps with our results.

\section{Hexagonalization and cluster algebras}\label{sec:review}

\subsection{Hexagonalization of correlation functions}
\label{sec:introtohexagonalization}

In this section we review the hexagonalization framework introduced in~\cite{Fleury:2016ykk,Fleury:2017eph}. In this work we apply it to $n$-point correlation functions of the simplest protected operators, namely the chiral primary operators built from the six real scalar fields $\Phi_{I}(x)$ of the theory, with $I=1, \ldots, 6$. In the integrability picture these operators correspond to vacuum states of the associated spin chains.
They take the form
\be\label{eq:1-2BPSoperators}
\mathcal{O}_{L_i}(x_i,y_i)=\Tr\, \left(y_i\cdot \Phi(x_i)\right)^{L_{i}} :=y_i^{I_1}\ldots y_i^{I_{L_{i}}}\Tr\, \Phi_{I_1}(x_i)\ldots \Phi_{I_{L_{i}}}(x_i) \;,
\ee
where $L_i$ is the R-charge of the operator (equivalently the spin-chain length). Since these operators are protected, their scaling dimension is $\Delta_i = L_i$ to all loops. The coordinates $x^\mu_i$ denote spacetime positions, while $y^I_i$ are null polarization vectors in six dimensions satisfying $y_i\cdot y_i:=\sum^6_{I=1} y_i^I y_i^I=0$.

Throughout this paper we restrict to two-dimensional kinematics, $x_{i}^{0} = x_{i}^{3} = 0$, so that the operator insertions lie in a Euclidean plane $\mathbb{R}^{2}$. We then parametrize each insertion point as $x_i = (\mathbf{x}_i, \bar{\mathbf{x}}_i)$ with complex coordinates
\be
\mathbf{x}_i=x_i^1+ix_i^2\,, \qquad \bar{\mathbf{x}}_i=x_i^1-ix_i^2\,.
\ee
We also restrict the polarization vectors to an $O(4)$ subsector by imposing $y^{5}_i = y^{6}_i = 0$. The resulting vectors $y_i=(y^a_i,0,0)$ define complex null rays in four dimensions and can be parametrized by points on another Riemann sphere,
\be
\mathbf{y}_i = y_{i}^{1}+iy_{i}^{2}\,, \qquad  \bar{\mathbf{y}}_i = y_{i}^{1}-iy_{i}^{2}\, .
\ee
In this way, the R-symmetry variables mirror the structure of the spacetime coordinates. On the dual gravity side, these kinematical restrictions correspond to focusing on strings propagating on $AdS_3\times S^3$.

\begin{figure}[h]
    \centering
    \includegraphics[width=0.6\linewidth]{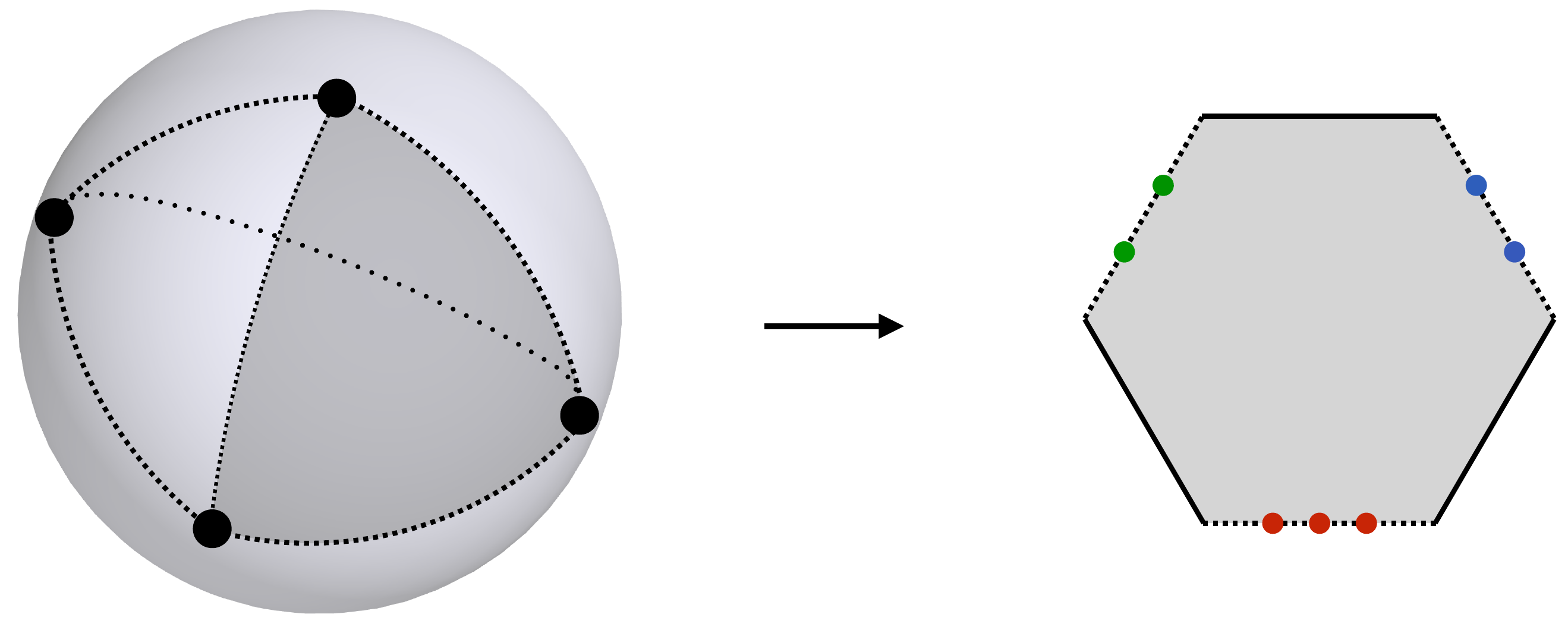}
    \caption{A triangulation of the four-punctured sphere. One triangle (shaded in gray) is represented in the right panel by a hexagon form factor. The edges of the triangle correspond to the mirror edges (dotted lines) of the hexagon, along which mirror magnons (colored dots) propagate. The physical edges (solid lines) of the hexagon originate from the vertices of the triangle.}
    \label{fig:hexagonalisationofthesphere}
\end{figure}

The hexagonalization conjecture expresses the $n-$point correlation functions at large $N$ as a sum over planar graphs dressed by dynamical factors,
\be
\label{eq:defccorrelator}
\langle \mathcal{O}_{L_1}(x_1, y_1)\ldots \mathcal{O}_{L_n}(x_n, y_n) \rangle_{c} =\frac{\sqrt{L_1\ldots L_n}}{N^{n-2}}\sum_{\text{graphs}}\left(\prod_{E}(d_{E})^{\ell_{E}}\right)\qZ[\{\qg_{E}\}]\, ,
\ee
where the formula applies to the connected part of the correlator and for properly normalized operators (see~\cite{Fleury:2016ykk}). In the weak coupling limit, $g^2\rightarrow 0$, the dynamical factors reduce to $\qZ = 1+\mathcal{O}(g^2)$, and the sum reproduces the standard Feynman graph decomposition of the tree-level correlator. Each graph corresponds to a pattern of Wick contractions, while each edge $E$ encodes a set of contractions between pairs of operators. More precisely, a Wick-contraction between scalar fields in $\mathcal{O}_{L_i}(x_i, y_i)$ and $\mathcal{O}_{L_j}(x_j, y_j)$ contributes a factor $d_{E}$ associated with the edge $E\equiv ij$,
\be
d_{E}:=\frac{y_{ij}^2}{x_{ij}^2}\;,\label{eq:superdistance}
\ee
where $x_{ij}^2=(x_i-x_j)^2$ is the spacetime distance squared and $y_{ij}^2 = (y_i-y_j)^2 = -2y_{i}\cdot y_{j}$ the invariant product of polarization vectors. The number of such contractions defines the bridge length $\ell_{E}$ of the edge $E$. 

The factor $\qZ[\{\qg_{E}\}]$ encodes the all-loop dynamical contributions associated with a given graph in~\eqref{eq:defccorrelator}. By construction, it is invariant under conformal and R-symmetry transformations and therefore depends only on spacetime and R-symmetry cross-ratios. These enter through a set of group elements $\{\qg_E\}$ associated with the edges, as explained below. According to the hexagon proposal, this dynamical factor can be computed at any value of the coupling by decomposing the graph into hexagons and gluing them along their mirror edges.

To obtain this decomposition, it is useful to view the planar graph as defining a triangulation $\mathcal{T}$ of a $n$-punctured sphere, where each puncture corresponds to an operator and each graph edge maps to an edge of the triangulation. This identification is straightforward when all bridge lengths are nonzero. When some bridge lengths vanish, however, the triangulation is no longer unique: the graph defines an equivalence class of triangulations related by flips of zero-bridge-length edges (see Figure~\ref{triangulationflip}). In such cases, one may choose any triangulation compatible with the graph, as the final result is independent of this choice.

An example of triangulation for the four-point function is shown in Figure~\ref{fig:hexagonalisationofthesphere}. The hexagon picture emerges by replacing each puncture with a small circle, as in the spin-chain or string description of the operator, so that each triangle becomes a hexagon representing a non-local form factor. Each hexagon thus has three physical edges, corresponding to arcs of these circles, and three mirror edges, corresponding to the edges of the triangulation. The latter are also referred to as mirror cuts, and in what follows we use the terms \textit{edge} and \textit{cut} interchangeably. 

The nontrivial step is to glue hexagons along their common mirror edges. This amounts to inserting a complete basis of mirror magnons $\psi_{E}$ on each edge. Schematically, one obtains
\be
\label{eq:Zdecomposition}
\qZ[\{\qg_E\}] = \sum_{\{\psi_E\}}\prod_{E\, \in\, \mathcal{T}}\mu_{\psi_E}\; \qg_{\psi_{E}}(\ell_E, z_E, \bar{z}_E, \alpha_E, \bar{\alpha}_E) \prod_{E_i,E_j,E_k \, \in\, \mathcal{T}}\mathfrak{h}(\psi_{E_i}|\psi_{E_j}|\psi_{E_k})\;,
\ee
where $\psi_{E_{i,j,k}}$ are the states associated with the three edges of a triangle in $\mathcal{T}$, $\mathfrak{h}(\psi_{E_i}|\psi_{E_j}|\psi_{E_k})$ is the corresponding hexagon form factor, and $\mu_{\psi_E}$ is the measure for the sum over mirror magnons.

Each state $\psi_{E}$ propagating along an edge $E$ acquires a weight $\qg_{\psi_{E}}$ associated with the relative displacement between adjacent hexagons. This factor encodes the geometric data of the correlator through the bridge length $\ell_E$ and the spacetime and R-symmetry cross-ratios $z_E,\,\bar z_E\, ,\alpha_E,\, \bar \alpha_E$. In the two-dimensional kinematics considered here, these parameters can be packaged into a group element $\qg_E=\qg(\ell_E, z_E, \bar{z}_E, \alpha_E, \bar{\alpha}_E)$ of the (centrally extended) $PSU(2|2)$ symmetry group of the magnons~\cite{Fleury:2016ykk}, such that the weight $\qg_{\psi_{E}} = \langle \psi_{E}|\qg_E|\psi_{E} \rangle$ in a basis where $\qg_E$ is diagonal.

\begin{figure}[h]
    \centering
    \includegraphics[width=0.35\linewidth]{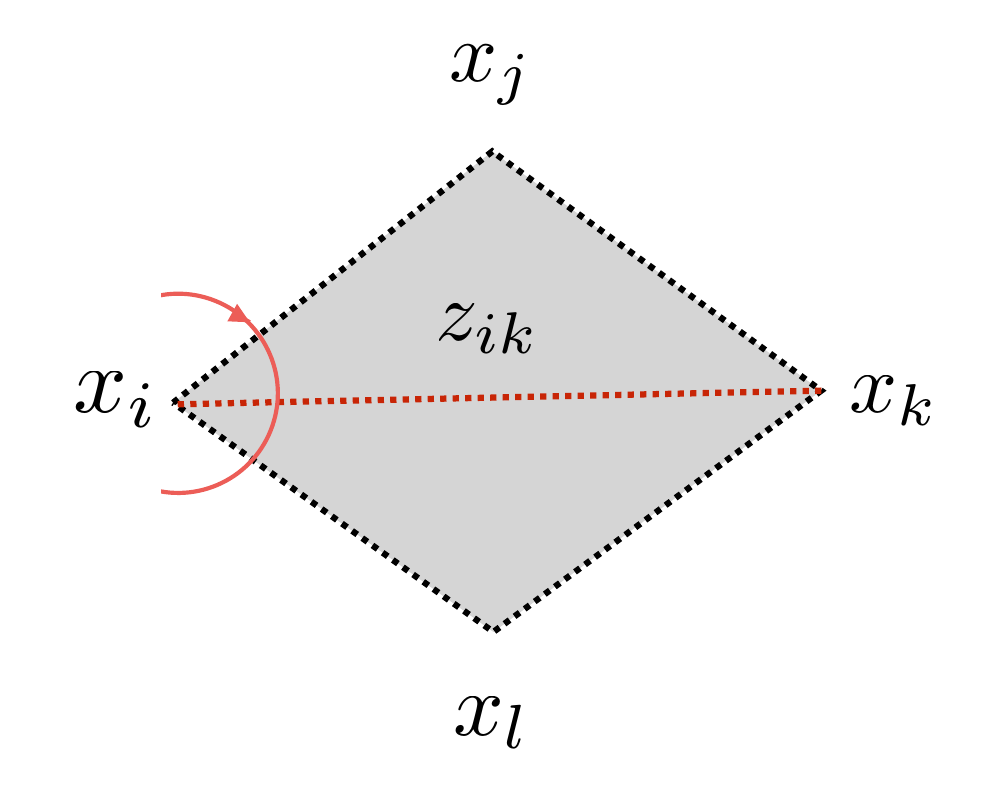}
    \caption{Illustration of the definition~\eqref{eq:crossratios} of the spacetime cross-ratio $z_{ik}$ associated with the diagonal of the quadrilateral $ijkl$~\cite{Fleury:2016ykk}. The orientation indicated by the orange arrow determines which distances appear in the numerator and which in the denominator in~\eqref{eq:crossratios}.}
    \label{fig:crossratio}
\end{figure}

The cross ratios associated to the edge $E=ik$ are defined locally in terms of the spacetime or R-symmetry coordinates of the four vertices of the quadrilateral surrounding that edge in the triangulation, as illustrated in Figure~\ref{fig:crossratio}.
An orientation must also be specified, indicated by the orange arrow in the figure, and can be chosen consistently across the entire triangulation. One then defines%
\footnote{Observe that our convention differs from \cite{Fleury:2016ykk,Fleury:2017eph} by $z\to-z$ and $\bar{z}\to-\bar{z}$, and similarly for $\alpha$ and $\bar{\alpha}$.} 
\be
\label{eq:crossratios}
z_{ik} = -\frac{(\mathbf{x}_i-\mathbf{x}_j)(\mathbf{x}_k-\mathbf{x}_l)}{(\mathbf{x}_i-\mathbf{x}_l)(\mathbf{x}_k-\mathbf{x}_j)}\, ,\qquad\alpha_{ik}=-\frac{(\mathbf{y}_i-\mathbf{y}_j)(\mathbf{y}_k-\mathbf{y}_l)}{(\mathbf{y}_i-\mathbf{y}_l)(\mathbf{y}_k-\mathbf{y}_j)}\, ,
\ee
and similarly for $\bar{z}_{ik}$ with $\mathbf{x}_{i,j,k,l} \rightarrow \bar{\mathbf{x}}_{i,j,k,l}$ and for $\bar{\alpha}_{ik}$ with $\mathbf{y}_{i,j,k,l} \rightarrow \bar{\mathbf{y}}_{i,j,k,l}$.

Although the expressions above are valid for arbitrary coupling $g$ and charge $L_i$ the sums are difficult to evaluate exactly. In this work we focus on the classical limit, where the coupling $g$ and charge $L_i$ are taken to infinity in a correlated way. For this reason, it is convenient to introduce
\be
\epsilon=\frac{1}{2g} = \frac{2\pi}{\sqrt{\lambda}} \ll 1 \, ,
\ee
and define the classical bridge lengths
\be
l_E=\ell_E\, \epsilon\, ,
\ee
which are held fixed.

Since the geometry of the punctured sphere introduces additional complications, we begin with an open geometry obtained by gluing hexagons into a polygon on the plane. This arises naturally as a limit of large-charge correlation functions in which some classical bridge lengths $l$ become infinite, as exemplified by the octagon introduced and studied in~\cite{Coronado:2018ypq, Coronado:2018cxj}.

\begin{figure}[h]
    \centering
    \includegraphics[width=1\linewidth]{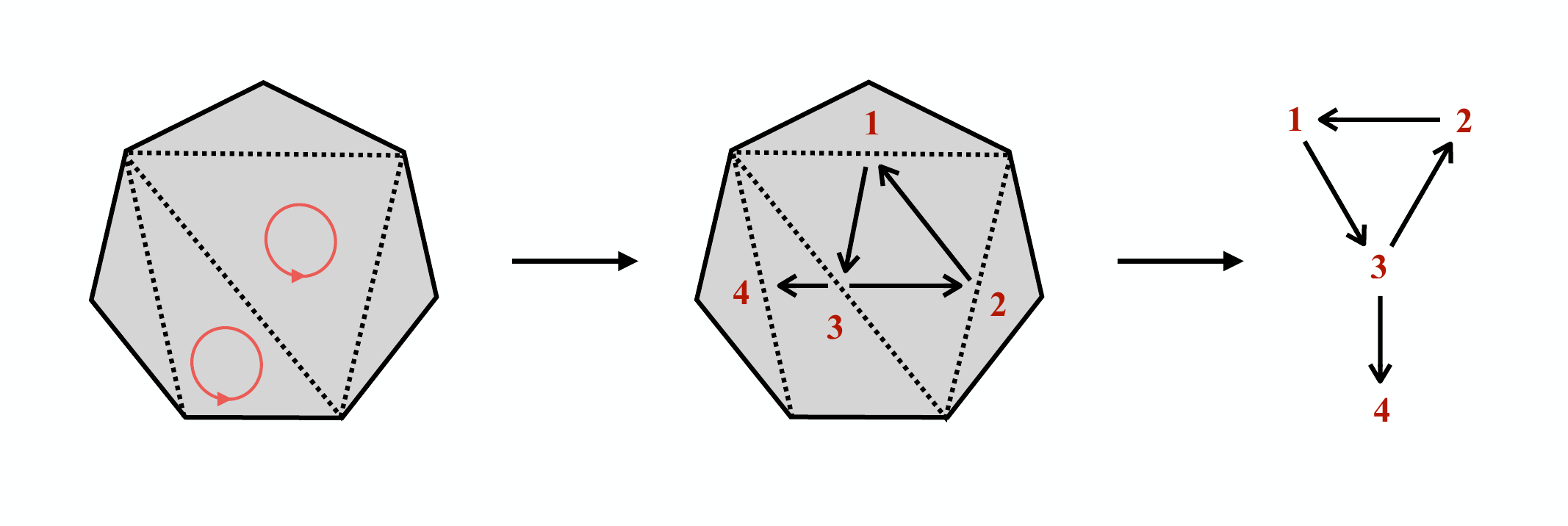}
    \caption{Open geometry with operator insertions at the vertices of a polygon. A representative triangulation is shown in the \textbf{left} panel. The triangles are assigned an orientation consistent with that in Figure~\ref{fig:crossratio}. The \textbf{middle} panel displays both the triangulation and the corresponding quiver determined by this orientation, while the \textbf{right} one shows only the resulting quiver.}
    \label{fig:genericopengeometry}
\end{figure}

Edges with infinite classical bridge length project onto the vacuum, effectively splitting the sphere into polygons with operators inserted at their vertices, as illustrated in Figure~\ref{fig:genericopengeometry}. The resulting polygonal observable has a structure similar to \eqref{eq:defccorrelator}, but with the external edges fixed to the vacuum and the sum running only over the internal edges (the diagonals of the polygon) with fixed bridge lengths.

\subsection{Preliminaries on quivers and cluster algebras\label{sec:triangulationsquiversclusteralgebras}}

To handle more efficiently the triangulation data underlying the hexagonalization, it is useful to introduce the related notions of quivers and cluster algebras. These structures have been extensively studied in the mathematical literature, particularly in the context of marked surfaces and the surface realization of cluster algebras (see, for instance, the foundational work~\cite{Fomin2008} and the pedagogical introduction~\cite{iwaki2014}). Here, we restrict attention to the structures relevant for our purposes and refer the reader to the literature for further details.

\paragraph{Quivers from triangulations.}

In our setup, we consider a marked surface $\mathcal{S}$, possibly with boundary, where the marked points correspond to operator insertions. These points may lie either on the boundary (vertices) or in the interior of the surface (punctures). We are interested in triangulations of the surface $\mathcal{S}$, and we denote by $\NN$ the number of edges in a triangulation and by $\mathcal{T}_{\mathcal{S}}$ the set of all triangulations of $\mathcal{S}$. As mentioned above, to define cross-ratios unambiguously, we fix an orientation for each triangulation, indicated by an orange arrow in Figure~\ref{fig:crossratio} and by orange circles in Figure~\ref{fig:genericopengeometry}.%
\footnote{To ensure the existence of a well-defined cluster algebra structure, certain degenerate cases must be excluded, such as a monogon with zero or one puncture. In the present work, we also disregard pathological configurations in which triangulations contain self-folded triangles. The latter necessitates tagged triangulations. Further details may be found in the references cited above.}

The combinatorial data of a triangulation, together with the choice of orientation, can be encoded in a quiver, i.e., an oriented graph. Given a triangulation $\mathcal{T} \in \mathcal{T}_{\mathcal{S}}$, the associated quiver is constructed as follows:
\begin{itemize}
    \item Each edge of the triangulation corresponds to a node of the quiver.
    \item For each triangle, the corresponding nodes are connected by arrows oriented according to the chosen orientation, as illustrated in Figure~\ref{fig:genericopengeometry}.
\end{itemize}
The quiver can then be described by a set of nodes $\{i \in I\}$, indexed by a subset $I\subset \mathbb{N}$ of cardinality $\NN$, and an antisymmetric matrix $B$, known as the signed adjacency (or exchange) matrix. Its entries $B_{ij} (=-B_{ji})$ count the arrows between nodes $i$ to $j$ with sign: an arrow $i\leftarrow j$ contributes $+1$ to $B_{ij}$, while an arrow $i\to j$ contributes $-1$. In this way, the matrix $B$ encodes both the connectivity of the triangulation and the chosen orientation.

Beyond labeling nodes by $i\in I$, it is convenient to introduce a charge lattice (\emph{cf.} \cite{Gaiotto:2009hg}). 
To each node \(i\), we associate a basis element \(\gamma_i\) in an abstract vector space (or, more precisely, a module). Henceforth, we label the nodes by their corresponding charges, denoted by \(\gamma_i\). The integer span of these vectors defines the charge lattice $\Lambda$. On $\Lambda$, we define a skew-symmetric bilinear form, or \textit{pairing}, by its values on the basis vectors,
\begin{align}
\label{eq:defsignedB}
    \langle \gamma_i,\gamma_j \rangle\equiv B_{ij}\,,
\end{align}
and extend it to all of $\Lambda$ by linearity. In this formulation, the adjacency matrix $B$ is completely encoded in the bilinear form.\footnote{More formally, in the spectral network and cluster algebra literature one speaks of a $\mathbb{Z}$-module $\Lambda$ freely generated by the nodes of the quiver; the data $(\Lambda, \langle\cdot,\cdot\rangle, (\gamma_i)_{i\in I})$ is referred to as a seed in this context~\cite{kineider2025spectralnetworksbridginghigherrank,fock2009clusterensemblesquantizationdilogarithm}.}

\begin{figure}[h]
    \centering
        \includegraphics[width=0.7\linewidth]{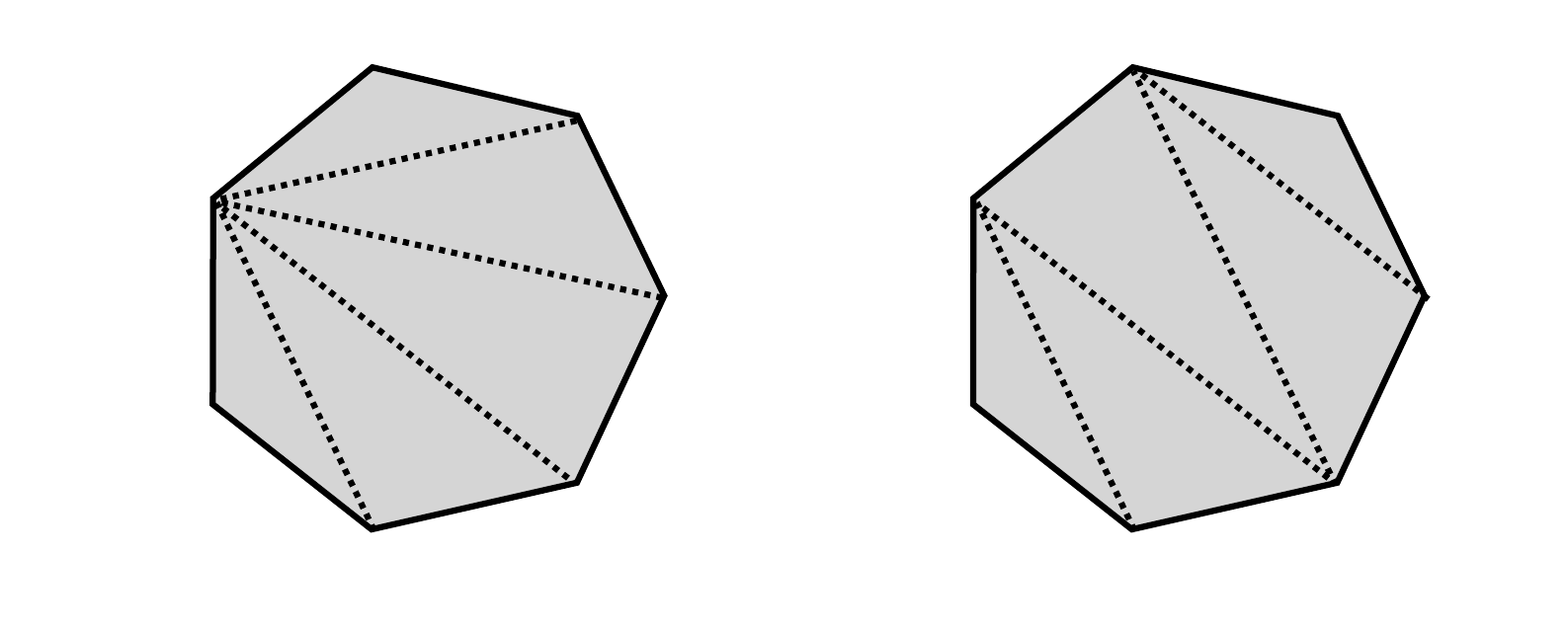}
    \caption{Two canonical examples of triangulations giving rise to linear quivers: the fan triangulation (\textbf{left}) and the zig-zag triangulation (\textbf{right}). The quivers associated with these triangulations are equivalent to the $A_{n-3}$ Dynkin diagram, up to arrow orientation.}
    \label{fig:fan}
\end{figure}

For the open geometry illustrated in Figure~\ref{fig:genericopengeometry}, we consider triangulations of an $n$-gon with operator insertions at its vertices. Each such triangulation consists of $n-2$ triangles and $\NN=n-3$ internal edges (diagonals). In the classification of cluster algebras, these triangulations correspond to type $A_{n-3}$, and we will occasionally refer to our systems by this type. They can be further distinguished according to whether the associated quiver contains oriented cycles. We refer to the acyclic case as \textit{linear quivers}, see Figure~\ref{fig:fan}. These satisfy the property that each triangle contains at most two internal edges, a feature that significantly simplifies the hexagon resummation. For this reason, linear quivers play a special role in Section \ref{sec:tbafreeenergyforlinearquivers}, while the general (cyclic) case is treated in Section~\ref{sec:generaltriangulations}.

\paragraph{Cluster algebras.}

Along with the charges, one may also introduce additional variables associated with the quiver through a labeled seed, i.e. a pair $(\{\chi_{\gamma_i}\}, B)$, where $\{\chi_{\gamma_i}\}$ is an $\NN$-tuple of variables and $B$ is the exchange matrix. The set $\{\chi_{\gamma_i}\}$ is called a cluster, and its elements $(\chi_{\gamma_1}, \ldots, \chi_{\gamma_{\NN}})$ are referred to as cluster $\chi$-variables~\cite{scott2003grassmanniansclusteralgebras,fock2009clusterensemblesquantizationdilogarithm}, or $\chi$-coordinates in the sense of Fock and Goncharov.  

\begin{figure}[h]
\centering
         \includegraphics[width=0.9\textwidth]{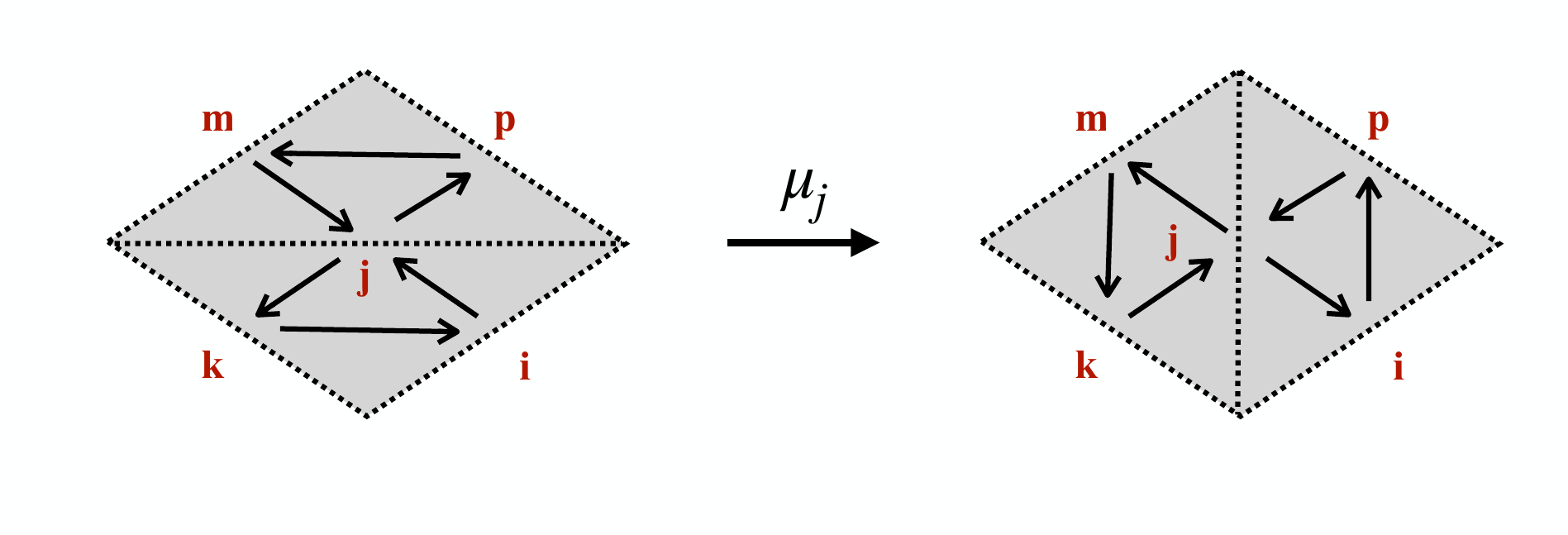}
        \caption{The change of a quiver under the flip of an edge. In the configuration shown in the figure the mutation $\mu_j$ takes places at the node $j$.}
         \label{triangulationflip}
\end{figure}

These variables transform by mutations, induced by flips of triangulations. An edge in a triangulation is generically shared by two adjacent triangles forming a quadrilateral. A flip replaces it with the other diagonal of the quadrilateral, as illustrated in Figure~\ref{triangulationflip}. This operation corresponds to a mutation of the quiver at the associated node $i$, under which the arrows transform according to the standard rules:
\begin{itemize}
    \item All arrows incident to \(i\) are reversed.
    \item For every oriented path \(j \to i \to k\), an arrow \(j \to k\) is added.
    \item Any resulting (oriented) two-cycles between nodes \(j\) and \(k\) are removed.
\end{itemize}
These transformations admit a concise reformulation as a change of basis of the charge lattice $\Lambda$. Namely, under a mutation $\mu_i$ at node \(i\), one has
\[
(\Lambda, \langle\cdot,\cdot\rangle, \{\gamma_j\}_{j\in I})\xrightarrow{\mu_i}(\Lambda, \langle\cdot,\cdot\rangle, \{\gamma'_j\}_{j\in I})\,,
\]
with the basis elements transforming as
\begin{align}\label{eq:mutationlatticebasis}
    \gamma_i &\xrightarrow{\mu_i} \gamma'_i = -\,\gamma_i\,,\\[0.3em]
    \gamma_j &\xrightarrow{\mu_i}\gamma'_j = 
    \gamma_j + \max\bigl(0,\langle \gamma_j, \gamma_i \rangle\bigr)\,\gamma_i\,,
    \qquad j \neq i\, . \nonumber
\end{align}
The pairing $\langle \cdot,\cdot\rangle$ itself is unchanged, but its matrix representation in the new basis, i.e.~the exchange matrix, transforms as $B_{ij}\to B'_{ij}=\langle \gamma'_i,\gamma'_j\rangle$.

The transformation of the cluster $\chi$-coordinates is more involved and takes the form
\begin{align}
\label{clusterchi}
\chi'_j=\begin{cases}
    \chi^{-1}_j, &j=i\,,\\
    \chi_j\,(1+\chi^{\,\textrm{sgn}\, B_{ij}}_i)^{B_{ij}},& j\neq i\,.
\end{cases}
\end{align}
In our setup, each edge of a triangulation is assigned two pairs of cross-ratios---holomorphic and anti-holomorphic---defined both in spacetime and in R-space, as in~\eqref{eq:crossratios}. These furnish the prototypical cluster $\chi$-variables and transform according to~\eqref{clusterchi} under a flip of the corresponding edge $i$.

The transformation of the lattice charges in~\eqref{eq:mutationlatticebasis} admits a deeper interpretation in relation to the cluster mutation rule~\eqref{clusterchi}. More precisely, it is naturally understood in terms of the Fock--Goncharov decomposition of a flip, reviewed in Appendix~\ref{app:comparison-GMN}.

\section{From one-cut to $\NN$-cut resummation}\label{sec:fromonecuttoNcutresum}

As reviewed above, hexagonalization reduces a planar correlator to the partition function $\qZ$ for a system of magnons living on the cuts of a triangulation. In this section, we introduce the ingredients needed to evaluate $\qZ$ at strong coupling.

\subsection{Octagon as a one-cut resummation}
\label{sec:Octonecut}

Before turning to more general cases, let us revisit the simplest example, in which two hexagons are glued into an octagon along a bridge of length $\ell$. This object first appeared in the study of four-point functions of large-charge BPS operators, where four large bridges effectively split the sphere into two octagons~\cite{Coronado:2018cxj}. At strong coupling, it takes the form of a TBA-like free energy for a four-flavor system~\cite{Bargheer:2019exp}. As this expression will serve as a basic building block for higher-point polygons, we briefly review its derivation and set up the notation.

Kinematically, this configuration can be viewed as a triangulation of a quadrilateral. It is characterized by the cross ratios $z,\, \bar z$ and $\alpha,\, \bar \alpha$, defined as in~\eqref{eq:crossratios}, which we parametrize as
\begin{align}
\label{eq:defzzbar}
    &z=e^{-\sigma^{\AdS}+i\phi^{\AdS}}\,,\quad \alpha=e^{-\sigma^{\Sph}+i\phi^{\Sph}}\, ,\\ \nonumber
     &\bar z=e^{-\sigma^{\AdS}-i\phi^{\AdS}}\,,\quad \bar\alpha=e^{-\sigma^{\Sph}-i\phi^{\Sph}}\,.
\end{align}
In this case, there is a single mirror cut and thus a single mirror sum is needed in~\eqref{eq:Zdecomposition}. The sum runs over a complete basis of mirror magnons and their bound states, characterized by their rapidities \(\textbf{u} = (u_{1}, \ldots, u_{N})\) and bound-state indices \(\mathbf{a} = (a_{1},\ldots,a_{N})\), with \(a_i = 1,2,\ldots\) and where $a_i = 1$ denotes the fundamental magnons. Since the magnons transform in non-trivial representations of the spin-chain symmetry group, an additional summation over their internal degrees of freedom (flavors) is required.

The result at arbitrary coupling reads
\begin{align}\label{eq:oct}
\mathbb{O}[\qg]
= \sum_{N \, =\, 0}^{\infty} \frac{1}{N!}
  \sum_{\mathbf{a}}\int d\u \,\mu_{\mathbf{a}}(\mathbf{u})
    \, T_{\mathbf{a}}(\mathbf{u}; \qg)\, ,
\end{align}
where \(N\) is the total number of magnons. The integration measure incorporates the contributions from the hexagon form factors and factorizes into one- and two-particle contributions,
\begin{align}\label{octintegrationmeas}
\mu_{\mathbf{a}}(\mathbf{u})
= \prod_{i\,=\,1}^{N} \mu_{a_i}(u_i)\;
  \prod_{1 \le i < j \le N} H_{a_i a_j}(u_i, u_j)\, ,\quad \textrm{and}\quad d\u  =  \prod_{j\, =\, 1}^N \frac{du_j}{2\pi\e} \,.
\end{align}
Both $\mu_{a_i}(u_i)$ and $H_{a_i a_j}(u_i, u_j)$ admit closed-form expressions in terms of the Zhukovsky variables 
\begin{align}\label{eq:zhukovsky}
x(u) = u + \sqrt{u^{2} - 1}\,, \qquad
x^{[\pm a]}(u) = x\!\left(u \pm \frac{i a \epsilon}{2}\right)\, ,
\end{align}
which can be found in~\cite{Basso:2015eqa,Jiang:2016ulr,Coronado:2018ypq}. For later convenience, we have rescaled the rapidities by the expansion parameter \(\epsilon = 1/2g\).

\subsubsection{Transfer matrices and dressings}

The sum over magnon flavors, together with the dependence on the cross ratios, is encoded in the average of the generalized transfer matrices $T^{\pm}$,
\be
\label{eq:Tmean}
T_{\textbf{a}}(\textbf{u} ;  \qg ) = \frac{1}{2}\left(\prod_{i\, =\, 1}^N T^+_{a_i}(u_i; \qg)+\prod_{i\, =\, 1}^N T^-_{a_i}(u_i;  \qg)\right)\, .
\ee
In the hexagon formalism, thanks to the diagonal symmetry of the hexagon form factors~\cite{Basso:2015zoa}, the flavor degrees of freedom can be organized into (short) representations $\mathscr{V}^\pm_{{a}}$ of the diagonal subalgebra $\frak{psu}(2|2)_{D}$ of the full spin-chain algebra. These modules are spanned by the states~\cite{Fleury:2016ykk}
\be\label{eq:dressedmodule}
|\psi_{\alpha_1} \ldots \psi_{\alpha_{a}} \rangle\, , | \psi_{\alpha_1} \ldots \psi_{\alpha_{a-2}} \phi_{1}\phi_{2}\rangle \,, |\psi_{\alpha_1} \ldots \psi_{\alpha_{a-1}}\cZ^{\pm \frac{1}{2}} \phi_{1}\rangle \,,  |\psi_{\alpha_1} \ldots \psi_{\alpha_{a-1}}\cZ^{\mp \frac{1}{2}} \phi_{2}\rangle \, ,
\ee
with symmetrization of $\alpha$-indices. Here $\psi_{\alpha}\, (\alpha = 1,2)$ and $\phi_{a}\, (a = 1,2)$ denote fundamental fermionic and bosonic components transforming as doublets under the spacetime and R-symmetry $\frak{su}(2)$ subalgebras of $\frak{psu}(2|2)_{D}$, respectively.

The superscripts $\pm$ correspond to the two possible dressings introduced in~\cite{Fleury:2016ykk, Fleury:2017eph}. Averaging over these dressings, as in~\eqref{eq:Tmean}, is required to reproduce perturbative gauge-theory data. In~\eqref{eq:dressedmodule}, the dressings amount to a redefinition of the fundamental bosonic components,
\be
\phi_{1} \rightarrow \cZ^{\pm \frac{1}{2}}\phi_{1}\,, \quad \phi_{2} \rightarrow \cZ^{\mp \frac{1}{2}}\phi_{2}\, ,
\ee
where $\cZ$ is known as the spin-chain $\cZ$-marker~\cite{Beisert:2006qh}. Its effect is to modify the contribution of each scalar flavor so that the resulting kinematical dependence matches gauge-theory expectations.

The transfer matrices $T^{\pm}_{a}(u; \qg)$ are defined as supertraces over $\mathscr{V}^\pm_{{a}}$. The argument $\qg\equiv\qg(u)$ denotes a group element of the (extended) symmetry group; in our conventions, it encodes both the dependence on the bridge length $\ell$ and on the cross ratios~\eqref{eq:defzzbar} which determine the relative orientation of the two hexagons in spacetime and R-symmetry space. Its explicit form in two-dimensional kinematics was derived in~\cite{Fleury:2016ykk},
 \begin{align}
\label{eq:gentrans}
T^{\pm}_{a}(u;\qg) &= \str_{\mathscr{V}^\pm_{a}}\, \qg(u)\\ \nonumber &=2(-1)^a\, e^{-\ell \cE_{a}(u)}e^{i\sigma^{\AdS}\cP_{a}(u)} 
\left(\cos \phi^{\AdS} -\cosh(\varphi\pm i\phi^{\Sph})\right)\,\frac{\sin a \phi^{\AdS}}{\sin  \phi^{\AdS}}\,,
\end{align}
where $\operatorname{Str} = \operatorname{Tr}\, (-1)^F$, with $F$ being the fermion number, and we introduced $\varphi=\sigma^{\AdS}-\sigma^{\Sph}$.

Equivalently, under the action of the group element $\qg$, each component in~\eqref{eq:dressedmodule} acquires a phase
\be
e^{\pm i\phi^{\AdS}} \,\, (\textrm{for $\psi$})\, , \qquad e^{\pm i\phi^{\Sph}} \,\, (\textrm{for $\phi$})\, ,
\ee
where the signs $+/-$ correspond to the components $\psi_{1}/\psi_{2}$ and $\phi_{1}/\phi_{2}$, respectively. The $\cZ$-marker further modifies the scalar contributions through
\be
\cZ^{\pm 1/2} \,\, \rightarrow\,\, e^{\pm \varphi}\, ,
\ee
with $\varphi$ defined above.

Finally, the flavor-independent prefactors in~\eqref{eq:gentrans} are governed by the central charges of the symmetry algebra, which measure the energy $\cE_{a}(u)$ and momentum $\cP_{a}(u)$ of the mirror magnon.
 
\subsubsection{Integration domain at strong coupling}

In the strong coupling limit $\epsilon\rightarrow 0$, the integrals above are dominated by contributions from the interval \(u\in (-1, 1)\), where the momentum scales as $\cP = O(1)$. The complementary region of the mirror kinematics ($|u|>1$) can be safely neglected. The latter corresponds to the giant-magnon regime, where $\cP = O(1/\epsilon)$, and the integrals exhibit rapid oscillations that average to zero.%
\footnote{This approximation is further supported by the scaling of the bridge length, $\ell = O(1/\epsilon)$, in the classical limit, which leads to an exponential suppression of contributions from giant magnons.} Equivalently, in terms of the Zhukovsky variable \(x\), the integration contour runs along the lower half of the unit circle.

\begin{figure}[h]
    \centering
    \includegraphics[width=1\linewidth]{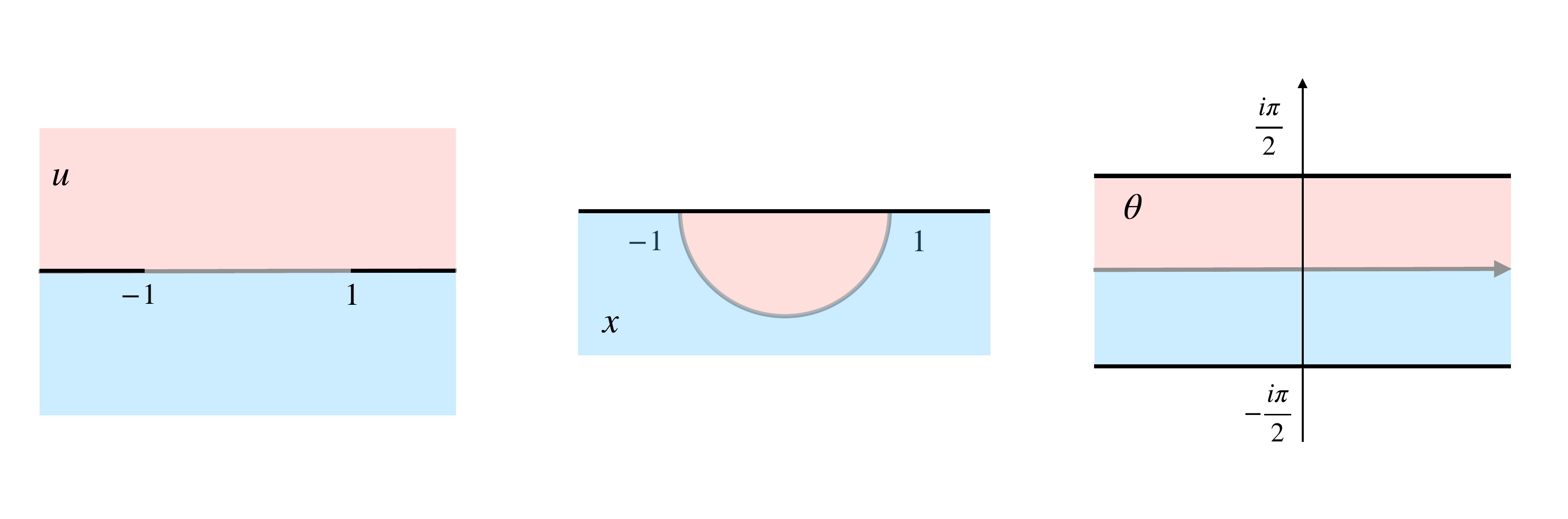}
    \caption{Maps between the variables $u\rightarrow x\rightarrow \theta$ in~\eqref{eq:x-to-theta} and the image of the fundamental strip in~$\theta$. The integration contour for the mirror integrals in the classical limit is shown in gray, while the boundaries of the fundamental domain are indicated by black lines.}
    \label{fig:domains}
\end{figure}

To better capture the physics of this regime, it is convenient to introduce the rapidity variable $\theta$, defined by
\be\label{eq:x-to-theta}
x = \textrm{coth}{\left(\frac{\theta}{2} + \frac{i\pi}{4}\right)} \qquad \rightarrow \qquad \quad u=\tanh \theta\,.
\ee 
In this parametrization, the integration contour maps to the entire real $\theta$-axis. The domain $\textrm{Im}\, \theta \in (-\pi/2, \pi/2)$ is commonly referred to as the \textit{fundamental strip}, and its images under the maps $u(\theta)$ and $x(\theta)$ are shown in Figure~\ref{fig:domains}. In what follows, we will freely switch between these rapidity variables whenever convenient.

As explained, for instance, in Section~5 of~\cite{Jiang:2016ulr}, a convenient way to access this regime is to start in the physical (spin-chain) kinematics, corresponding to rapidity $|u|>1$ just below the cut shown in the left panel of Figure~\ref{fig:domains}, and then analytically continue to the region $u\in (-1, 1)$ by passing through the blue region in the figure. Accordingly, the various ingredients in~\eqref{eq:oct} can be evaluated in the physical kinematics and subsequently continued to the relevant strong coupling domain.

In particular, for the dispersion relation of the $a$-th magnon bound state, one may use
\be\label{eq:dispersion-relation}
\cE_{a} 
 =  \log{\left(\frac{x^{[-a]}}{x^{[+a]}}\right)}\
=-i\cP_{a}^{\textrm{ph}}\, , \qquad \cP_{a} 
= -i a + \frac{1}{\epsilon} \left(\frac{1}{x^{[+a]}}-\frac{1}{x^{[-a]}}\right)
=-i\cE_{a}^{\textrm{ph}}\, , 
\ee
where $x^{[\pm a]}$ are defined in~\eqref{eq:zhukovsky}. Expanding at strong coupling,
\be
x^{[\pm a]} = x \pm \frac{i\epsilon a x^2}{x^2-1} +\mathcal{O}(\epsilon^2)\, ,
\ee
and expressing the result in terms of $\theta$ using~\eqref{eq:x-to-theta}, the dispersion relations simplify drastically. They reduce to those of relativistic particles with mass $a$,
\be
\cE_{a} \approx -\frac{2ix a \epsilon}{x^2-1}=a \epsilon \cosh \theta\, , \qquad \cP_{a} \approx -ia \frac{x^2+1}{x^2-1} =a \sinh \theta\, ,
\ee
up to an overall rescaling of the energy. This shows that, in the classical limit, a bound state behaves as a stack of $a$ fundamental particles, with vanishing binding energy. 
Thus, the abelian factors of the transfer matrix becomes
\be
e^{-\ell \cE_{a}}e^{i\sigma^{\AdS}\cP_{a}} = e^{-a \,l \cosh{\theta}}e^{i a \sigma^{\AdS} \sinh{\theta}}\,,
\ee
where we introduced the rescaled length $l = \epsilon \ell$.

The transfer matrices simplify accordingly and can be expressed in terms of the basic building blocks of the fundamental transfer matrix. Introducing
\be\label{ef:defcalY}
\qg
= \operatorname{diag}\left(\,\mathcal{Y}^{\AdS},\,
\bar{\mathcal{Y}}^{\AdS},\,
\mathcal{Y}_{\omega}^{\Sph},\,
\bar{\mathcal{Y}}_{\omega}^{\Sph}\,\right)\,,
\ee
in the basis of the fundamental representations $\mathscr{V}_1^{\pm}$ defined above, one finds
\begin{align}
\label{lYAdS}
\log{\cY^{\AdS}(\theta)} &= i\phi^{\AdS} +i \sigma^{\AdS} \sinh{\theta} - l\cosh{\theta}\, ,\\ \nonumber
\log{\bar{\cY}^{\AdS}(\theta)} &= -i\phi^{\AdS} +i \sigma^{\AdS} \sinh{\theta} - l\cosh{\theta}\,,
\end{align}
for the eigenvalues corresponding to the \(\AdS\) components (fermions). The expressions for the sphere components (scalars) read similarly but their form depends on the dressing ($\qw=\pm$)%
\footnote{We note that for both choices of dressing, the superdeterminant of the transfer matrix is equal to unity. In terms of the eigenvalues introduced above, this condition reads $\cY_{\qw}^{\Sph}(\theta)\,\bar{\cY}_{\qw}^{\Sph}(\theta) = \cY^{\AdS}(\theta)\,\bar{\cY}^{\AdS}(\theta)$.}
\be
\label{lYSph}
\begin{aligned}
\log{\cY_\qw^{\Sph}(\theta)} &= i\phi^{\Sph} +\qw \log{U} +i \sigma^{\AdS} \sinh{\theta} - l\cosh{\theta}\, , \\
\log{\bar{\cY}_\qw^{\Sph}(\theta)} &= -i\phi^{\Sph}-\qw \log{U} +i \sigma^{\AdS} \sinh{\theta} - l\cosh{\theta}\, ,
\end{aligned}
\ee
where we defined for later convenience,
\begin{align}
\label{eq:defUphi}
    \log U=\varphi=\sigma^{\AdS}-\sigma^{\Sph}\, .
\end{align}
Notice that the cross ratios can be read off by evaluating the $\cY$–functions at $\theta = i\pi/2$. In particular,
\begin{align}
\label{eq:AdS-special}
\cY^{\AdS}(i\pi/2) = z\, ,& \qquad \bar{\cY}^{\AdS}(i\pi/2) = \bar{z}\, ,\\
\label{eq:Sph-special}
\cY_{\qw}^{\Sph}(i\pi/2) = \alpha\, U^{-(1-\qw)}\, ,& \qquad \bar{\cY}_{\qw}^{\Sph}(i\pi/2) = \bar{\alpha} \, U^{-(1+\qw)}\, .
\end{align}
In these terms, the classical transfer matrices can be obtained from the generating function
\begin{align}
\frac{(1+t\,\cY_{\qw}^{\Sph})(1+t\,\bar \cY_{\qw}^{\Sph})}{(1+t\,\cY^{\AdS})(1+t\,\bar{\cY}^{\AdS})}=1+\sum_{a\geq 1} t^a\,T^{\qw}_a\,,
\end{align}
such that for fundamental magnons $T_1^\omega=\str_{\mathscr{V}_1^\omega}\qg =\cY_{\qw}^{\Sph}+\bar \cY_{\qw}^{\Sph}-\cY^{\AdS}-\bar \cY^{\AdS}$.

\subsubsection{Exponentiation and clustering}\label{Clustering}

The measure factors~\eqref{octintegrationmeas} also simplify enormously at strong coupling, approximating to
\begin{align}\label{eq:mu2}
H_{a_i,a_j}(u_i, u_j)\approx 
  \frac{(u_i - u_j)^{2} + \frac{(a_i - a_j)^{2}}{4}\epsilon^{2}}
       {(u_i - u_j)^{2} + \frac{(a_i + a_j)^{2}}{4}\epsilon^{2}}\, ,\qquad
\mu_{a}(u) \approx \frac{1}{a}\, ,
\end{align}
up to smooth terms of order \(\mathcal{O}(\epsilon^{2})\), which will be irrelevant here. This shows that interactions between magnons die off rapidly for $u_i-u_j = \mathcal{O}(1)$, so that the multi-particle sum in~\eqref{eq:oct} exponentiates.

However, there remains a subtle but important effect coming from regions where two rapidities approach each other within separations of order $u-v \sim \epsilon$ in~\eqref{eq:mu2}. In these regions, the rational factors in the two-particle measure develop poles that pinch the integration contours. These singularities modify the combinatorics of the bound-state expansion, giving rise to a phenomenon known as \emph{clustering}. This phenomenon was analyzed from both the combinatorial and functional perspectives in~\cite{Jiang:2016ulr}, and applied to the octagon computation in~\cite{Bargheer:2019exp}.%
\footnote{This phenomenon is common in integrable systems, appearing for instance in the Nekrasov partition function \cite{Nekrasov:2009rc,Bourgine_2014}, stochastic processes \cite{borodin2013macdonaldprocesses} and in resummations of the Pentagon OPE for null Wilson loops~\cite{Fioravanti_2015}.} In the functional approach, clustering amounts to combining the two-particle interaction terms into a Cauchy determinant~\cite{Bettelheim_2014}, which allows the full sum to exponentiate and be rewritten as a Cauchy–Fredholm determinant. A brief review is provided in Appendix~\ref{app:cauchy}.

The net effect of clustering at strong coupling is to modify the one-particle measure according to
\begin{align}\label{eq:renormeasure}
\mu_{a}(u)\, T^{\omega}_{a}(u)
\;\approx\; \str_{\mathscr{V}_a^\omega }\,\frac{\qg}{a}
\quad \buildrel\textrm{clustering}\over\longrightarrow\quad -\,\str_{\mathscr{V}_1^\omega}\,\frac{(-\qg)^{a}}{a^{2}}
=-\,\str\,\frac{(-\mathcal{Y}_{\omega})^{a}}{a^{2}}\, .
\end{align}
The interpretation of~\eqref{eq:renormeasure} becomes natural once bound states are viewed as unbound stacks of magnons. More precisely, in the classical limit the constituents of a bound state effectively become unbound, so that the sum over bound states and their associated transfer matrices can be reorganized as a sum over stacks built from the four fundamental excitations.

Here, the symbol \(\str\), when used without an explicit specification of the representation, denotes the graded sum over the four fundamental magnon flavors, with positive signs assigned to the sphere (Sph) excitations and negative signs to the AdS ones. More generally, \(\str\) is defined for an arbitrary function \(F[\mathcal{Y}]\) as
\begin{align}
\str\, F[\mathcal{Y}]
\equiv \sum_{I} (-1)^{F_I} F[\mathcal{Y}^{I}]
= F[\mathcal{Y}^{\Sph}] + F[\bar{\mathcal{Y}}^{\Sph}]
  - F[\mathcal{Y}^{\AdS}] - F[\bar{\mathcal{Y}}^{\AdS}]\, .
\end{align}

After clustering, the strong-coupling resummation yields a remarkably compact expression for the octagon free energy,
\be\label{eq:oct-sc}
\bbO \approx \frac{1}{2}\sum_{\qw\, =\, \pm}
\exp\!\left[ -\frac{1}{\epsilon}\cA_\omega\right]\,,\qquad 
  \cA_\omega =\int\frac{d\theta}{2\pi\cosh^{2}\theta}\,
    \str\,\Li\left(-\cY_\qw(\theta)\right)
\ ,
\ee
where the integration over the mirror rapidity $u$ has been replaced by the relativistic rapidity $\theta$ according to 
\be
du = \frac{d\theta}{\cosh^{2}\theta}\, .
\ee 
The appearance of the dilogarithm in the effective action is a direct consequence of the renormalization of the measure in~\eqref{eq:renormeasure} by clustering.

In this form, the result appears to differ slightly from that obtained in eq.~(2.21) of ref.~\cite{Bargheer:2019exp}. In fact, one can show that the two results coincide after performing an integration by parts in eq.~\eqref{eq:oct-sc} and going to the subkinematics $\alpha = \bar{\alpha} = -1$ considered in that reference. In this restricted kinematics, the two dressings are equivalent and it is possible to combine the various $Y$–functions into a single effective one, using
\be
1+Y_{\textrm{\cite{Bargheer:2019exp}}}\Big(\theta-\frac{i\pi}{2}\Big) \equiv \frac{(1+\cY^{\Sph}(\theta))\,(1+\bar{\cY}^{\Sph}(\theta))}
{(1+\cY^{\AdS}(\theta))\,(1+\bar{\cY}^{\AdS}(\theta))}\, .
\ee
To recover eq.~(2.23) from \cite{Bargheer:2019exp}, we substitute $\phi^{\AdS}\to \phi+\pi$ and $\sigma^{\AdS}\to \varphi$, as well as $\phi^{\Sph}\to \pi$ and $\sigma^{\Sph}\to 0$, and shift the rapidity variable as indicated above. 

As a final remark, let us note that the dressed transfer matrices~\eqref{eq:gentrans} exhibit a holomorphicity property with respect to the R-symmetry cross ratios, namely
\be
\partial_{\bar\alpha}\, T^+=\partial_\alpha\, T^-=0\, .
\ee
This can be seen, for instance, from the expressions for $\cY_{\qw}^{\Sph}$ and $\bar{\cY}_{\qw}^{\Sph}$ in~\eqref{eq:Sph-special}, taking into account that the factor $U = \sqrt{\alpha\bar{\alpha}/z \bar{z}}$ cancels the dependence on $\alpha$ or $\bar{\alpha}$ for $\qw=+1$ or $\qw = -1$, respectively. This property immediately implies an analogous relation for the free energy~\eqref{eq:oct-sc} in the two dressings,
\be\label{eq:holomorphicity-area}
\partial_{\bar\alpha}\, \cA_+=\partial_\alpha\, \cA_-=0\, .
\ee
Remarkably, this structure is quite general and persists for higher-point functions. It shows that the sum over dressings can be interpreted as a decomposition of the correlator into holomorphic and anti-holomorphic components with respect to the R-symmetry variables. 

From a string-theory perspective, this feature may seem surprising, as it suggests that each component is associated with its own area $\cA_{\qw}$ and hence with its own classical string solution. It is also somewhat unexpected that the spacetime and R-space dependence are treated asymmetrically, since no relation analogous to~\eqref{eq:holomorphicity-area} holds for the spacetime cross ratios, which enter in the same way in both dressings (see, e.g.,~\eqref{lYAdS}). We will return to this apparent puzzle and its resolution in Section~\ref{sec:sum-over-moduli}, where a precise comparison of the hexagon formalism with string-theory results will be carried out.

\subsection{Two-cut partition function}

\begin{figure}[h]
\centering
\includegraphics[width=10cm]{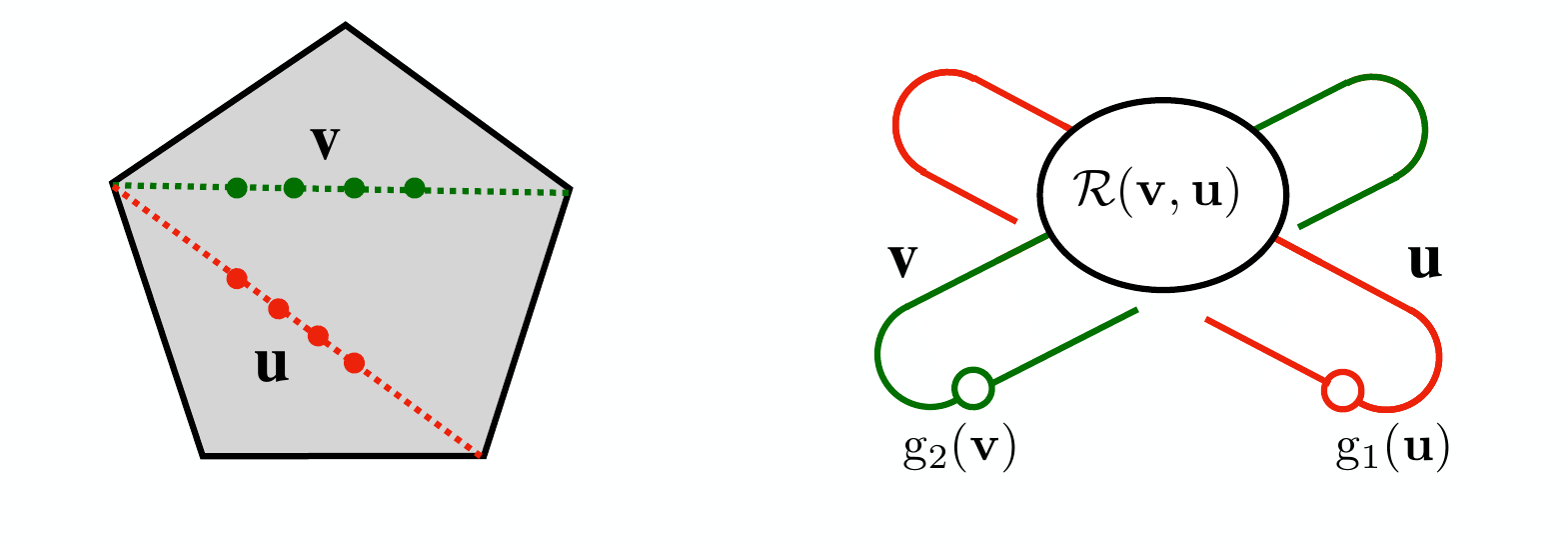}
\caption{\textbf{Left:} Two-cut configuration for the five-point function with magnon insertions of rapidities $\textbf{u}$ and $\textbf{v}$. 
\textbf{Right:} Schematic representation of the matrix part: the red line denotes a multi-magnon state on the lower (red) cut, while the green line denotes a multi-magnon state on the upper (green) cut. The circles indicate insertions of the corresponding group elements. 
}
\label{fig:twocut}
\end{figure}

Let us now turn to the next nontrivial case, namely a configuration involving two cuts. This setup corresponds to a pentagon, with five operators inserted at its vertices, as illustrated in Figure~\ref{fig:twocut}. 
Each cut is characterized by its own bridge length and associated set of cross ratios, parametrized as in~\eqref{eq:defzzbar}. The sums over magnons for this geometry give rise to the two-cut partition function~\cite{Fleury:2017eph}
\be\label{eq:two-cuts}
\qZ[\qg_{1}, \qg_{2}] = \sum_{N, M \, =\, 0}^{\infty} \frac{1}{N! M!} \sum_{\textbf{a}, \textbf{b}}  \int d\u\,d\v\,\mu_{\textbf{a}}(\textbf{u})\,\mu_{\textbf{b}}(\textbf{v}) \frac{\mathcal{M}_{\textbf{ab}}(\textbf{u}, \textbf{v};\qg_1,\qg_2)
}{h_{\textbf{a}\textbf{b}}(\textbf{u}, \textbf{v})}\, ,
\ee
where the integration measures on each cut are the same as in the one-cut case.

The new ingredients in this expression are the dynamical factors associated with the hexagons connecting the two cuts, 
\be
h_{\textbf{ab}}(\textbf{u}, \textbf{v}) = \prod_{i\, =\, 1}^{N}\prod_{j\, =\, 1}^{M}h_{a_i b_j}(u_i, v_j)\, ,
\ee
which appear in the denominator, as well as the corresponding matrix part,
\be
\mathcal{M}_{\textbf{ab}}(\textbf{u}, \textbf{v};\qg_1,\qg_2)
= \,\str_{\, \mathscr{V}_{\textbf{a}}\otimes \mathscr{V}_{\textbf{b}}} \, \left[\qg_{1}(\textbf{u})\, \qg_{2}(\textbf{v})\, \cR_{\textbf{ba}} (\textbf{v}, \textbf{u}) \right]\, .
\ee
Here, the supertrace is taken over the multiplets of excitations,
\be
\mathscr{V}_{\textbf{a}} = \bigotimes_{i\, =\, 1}^{N}\mathscr{V}_{a_{i}}\, \label{eq:thefullmodule}\,,
\ee
and similarly for $\mathscr{V}_{\textbf{b}}$. The operator $\cR_{\textbf{b}\textbf{a}}(\textbf{v}, \textbf{u})$ denotes the product of magnon $R$-matrices, normalized to unity on the highest-weight state. It is related to the more familiar $S$-matrix by a graded permutation, introduced to remove fermionic sign factors.

As in the octagon case~\eqref{ef:defcalY}, the group elements $\qg_1(\u)$ and $\qg_2(\v)$ encode the cross ratios and bridge lengths associated with the corresponding cuts. One must also include the $\mathcal{Z}$-marker dressing, which, as explained in~\cite{Fleury:2017eph}, modifies not only the scalar weights---as in the one-cut case---but also the matrix part of the hexagon form factors, through rapidity-dependent terms dressing the $R$-matrix (see Table~\ref{table:diagonal} in Appendix~\ref{app:r-matrix}). An average over the two dressings, $\qw=\pm$, is understood in~\eqref{eq:two-cuts-diagonal}. To avoid overloading the notation, we will keep the dependence on $\qw$ and the associated averaging procedure implicit throughout the remainder of this section.

\subsubsection{Poles and contours}\label{sec:decouplingpoleandorder}

As it stands, the expression for the hexagon sum contains several poles in the rapidity plane. To understand their structure, let us start by examining the zeros and poles originating from the dynamical factor $h_{ab}$. Its basic building block is the hexagon form factor for fundamental magnons~\cite{Basso:2015zoa}
\be\label{eq:h11}
h_{11}(u, v) = \frac{u-v}{u-v-i\epsilon} \frac{\left(1-1/x^{-}y^{+}\right)^{2}}{(1-1/x^{+}y^{+})(1-1/x^{-}y^{-})} \sigma(u, v)^{-1}\, ,
\ee
where $\sigma(u, v)$ is the Beisert-Eden-Staudacher dressing phase~\cite{Beisert:2006ez}. The bound-state expression is obtained by the usual fusion procedure of the fundamental form factor,
\be\label{eq:hab}
h_{ab}(u, v) = \prod_{j\, =\, 0}^{a-1}\prod_{k\, =\, 0}^{b-1} h_{11} \left(u^{[1-a+2j]}, v^{[1-b+2k]}\right)\, ,
\ee
where $u^{[\pm n]} = u\pm i\epsilon n/2$, and similarly for $v^{[\pm n]}$. Substituting the rational factor in $(u-v)$ from~\eqref{eq:h11} into~\eqref{eq:hab}, one finds that the rational part of $h_{ab}(u, v)$ is given by
\be
\label{eq:dynrat}
h_{ab}(u, v)\big|_{\textrm{rational part}} = \frac{\left(u-v+\frac{|a-b|}{2}\,i\e\right)\ldots\left(u-v+\frac{|a+b|-2}{2}\,i\e\right)}{\left(u-v-\frac{|a-b|+2}{2}\,i\e\right)\ldots\left(u-v-\frac{|a+b|}{2}\,i\e\right)}\,.
\ee
This introduces a finite set of poles and zeros in the integrand~\eqref{eq:two-cuts}, located at
\be
\label{eq:dynpoleszeros}
\begin{cases}
    {\textrm{ poles at }}\quad u-v = -i\frac{|a-b|\epsilon}{2} -ik\epsilon\, , \\{\textrm{ zeros at }}\quad u-v = i\frac{|a-b| \epsilon}{2}+i(k+1) \epsilon\,,
\end{cases}
\ee
with $k = 0, \ldots, \textrm{min} \{a, b\}-1$.

The matrix part $\cR_{ba}(v,u)$ likewise develops poles that couple the rapidities \(u\) and \(v\). These poles are more difficult to exhibit explicitly, as closed-form expressions for the $R$-matrix are rather involved~\cite{Fleury:2017eph,DeLeeuw:2019dak,DeLeeuw:2020ifb,Arutyunov:2008zt,Arutyunov:2009mi}. Nevertheless, one may verify---by direct inspection of its explicit components---that all poles of the $R$-matrix occur precisely at the locations where the dynamical part has zeros, and are therefore canceled by them.%
\footnote{We thank Thiago Fleury for insightful discussions on this point and for his assistance in verifying these features of the mirror bound-state $S$-matrix.} The simplest example is the fundamental $R$-matrix~\cite{Beisert:2006qh} with $a=b=1$, which exhibits poles only at $u-v = i\epsilon$.

The remaining factors in $\cR$ and $h$, including the dressing phase in~\eqref{eq:h11}, are smooth functions of the Zhukovsky variables within the fundamental domain and therefore do not generate singularities. We thus conclude that the only singularities in the integrand are the \(u-v\) poles listed in~\eqref{eq:dynpoleszeros}. Crucially, they all lie on the same side of the integration contours for both $u$ and $v$. This ensures that poles associated with different cuts cannot pinch any integration contour in the limit $\epsilon \to 0$.

The only potentially problematic case is the pole at $u=v$ when $a=b$. This pole governs the decoupling limit, in which two magnons with same rapidity and flavor pair up, and decouple from the remaining excitations on the hexagon~\cite{Basso:2015zoa}. Because it lies directly on the integration contours, it requires a careful prescription.

This prescription amounts to an ordering of the contours along the imaginary direction. Concretely, one introduces a small displacement such that
\be\label{eq:can-order}
\textrm{Im}\, \u > \textrm{Im}\, \v\, ,
\ee
i.e.~the $u$-contours lie slightly above the $v$-contours.

\begin{figure}[h]
    \centering
    \includegraphics[width=12cm]{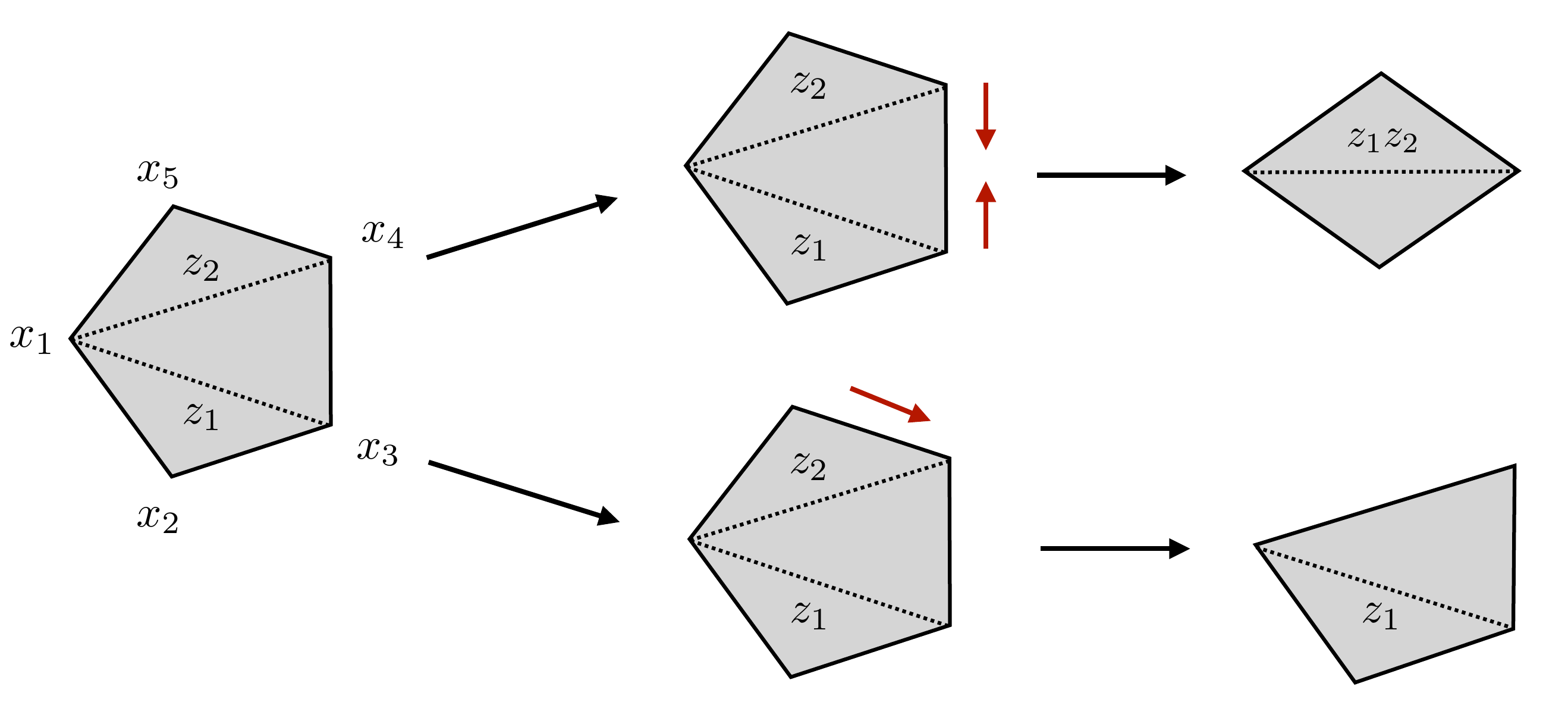}
    \caption{Two OPE limits of a pentagon. In the upper panel we send $x_4\to x_3$ by sending $z_1\to -\infty$ and $z_2\to 0$ while keeping $z_1 z_2$ fixed, while in the lower one we send $x_5\to x_4$ by sending $z_2\to \infty$.}
    \label{fig:decouplinglimits}
\end{figure}

The motivation for this choice comes from behavior of the integrals in the OPE limit~\cite{Fleury:2017eph,Bargheer:2018jvq} (see also~\cite{Basso:2013vsa} for similar considerations in the context of null polygonal Wilson loops). To illustrate it, fix three vertices of the pentagon at $x_1=(0,0)$, $x_2=(1,1)$ and $x_3=(\infty, \infty)$. The remaining vertices are parametrized by cross-ratios as
\be
x_4 = (-z_1, -\bar{z}_{1}) \,, \qquad x_5 = \left(-\frac{z_1z_2}{z_2+1}, -\frac{\bar{z}_1 \bar{z}_2}{\bar{z}_2+1}\right) \,.
\ee
Consider two OPE limits in which the pentagon degenerates into a quadrilateral: $x_{5}\rightarrow x_{4}$ and $x_{4}\rightarrow x_{3}$, shown in the lower and upper panels of Figure~\ref{fig:decouplinglimits}, respectively. Although similar in nature, these limits are realized differently in the chosen triangulation.

The first limit corresponds to $\sigma^{\AdS}_{2}\to-\infty$. In this regime, the upper cut approaches the outer edge of the pentagon and the second mirror channel is effectively projected onto the vacuum. Since the $v$-integrals become highly oscillatory when $\sigma^{\AdS}_{2}\to-\infty$, this behavior is correctly captured only if the contours can be shifted into the lower half-plane, where the integrands are exponentially suppressed, as ensured by~\eqref{eq:can-order}.%
\footnote{More precisely, one finds $e^{i\cP_{b}(v) \sigma^{\AdS}_{2}} \sim e^{b\sigma^{\AdS}_{2}} \rightarrow 0$ when $v\rightarrow -i\infty$.}%

By contrast, the limit $x_{4}\rightarrow x_{3}$, with $x_{5}\neq x_{4}$, requires $\sigma^{\AdS}_{2}\to \infty$ and $\sigma^{\AdS}_{1} \to -\infty$, with $\sigma^{\AdS}_{1}+\sigma^{\AdS}_{2}$ fixed. In this case, the cuts merge and the dominant contribution arises from pairs of magnons with identical rapidities and quantum numbers. The $v$-contours must then be deformed in the opposite direction, and the integral is governed by the decoupling poles when they collide with the $u$-rapidities, in agreement with~\eqref{eq:can-order}.%
\footnote{To be more precise, in both limits one must also take the corresponding limits in the R-symmetry coordinates, so as to eliminate any residual scalar contributions. In other words, in the OPE limits discussed here not only the operator positions but also their polarization vectors must coincide.}

These considerations hold at any coupling and provide a physical justification for the contour ordering. The same prescription extends to higher polygons, applying to any pair of cuts with the same orientation, and ensures the correct behavior in all OPE limits. 

\subsubsection{Diagonal approximation}\label{sec:Diagonal}

The absence of pinching singularities in \(u-v\) implies that all clustering and combinatorial structures arise exclusively from the integration measures associated with each cut individually, which remain unchanged. Interactions between different cuts, by contrast, can be evaluated directly in the limit $\epsilon \rightarrow 0$ with $u, v$ held fixed. In this regime, the hexagon form factors simplify drastically. In particular, one finds
\be\label{eq:h-to-K}
h_{ab}(u, v) \approx \sqrt{S_{ab}(u,v)} = 1 + \epsilon\, a b\, \qK(u, v) + \mathcal{O}(\epsilon^{2})\, ,
\ee
where $S_{ab}$ denotes the dynamical part of the magnon $S$-matrix in the $SL(2)$ sector. Here, $K$ is given by
\be
\label{kernelxy}
\qK(u, v) 
= \frac{2 i x y (x y - 1)}{(x^{2} - 1)(x - y)(y^{2} - 1)}\, ,
\ee
where $x = x(u)$ and $y = x(v)$. A similar simplification occurs for $R$-matrix, which asymptotes to the identity in this limit, leading to
\be\label{eq:classical-S-matrix}
\frac{\cR_{ba}(v,u)}{h_{ab}(u,v)}\approx  1-\e\,\mathbb{K}_{ab}(u,v) \, ,
\ee
where $\mathbb{K}$ relates to the classical $R$-matrix studied in refs.~\cite{deLeeuw:2008dp,Beisert:2007ty,Moriyama:2007jt,Torrielli:2007mc}. Here, we will refer to it as the kernel, since it enters as the integral kernel in the TBA equations.

According to~\eqref{eq:classical-S-matrix}, at zeroth order in $\epsilon$, interactions between cuts are trivial, and the hexagon sums factorize into a product of two decoupled octagons. At the next order, magnons on different cuts interact through the matrix kernels $\mathbb{K}_{ab}(u,v)$. However, since these interactions are weighted by two magnon measures, each scaling as $1/\epsilon$, they cannot be neglected.

\begin{figure}[h]
\centering
\includegraphics[width=9cm]{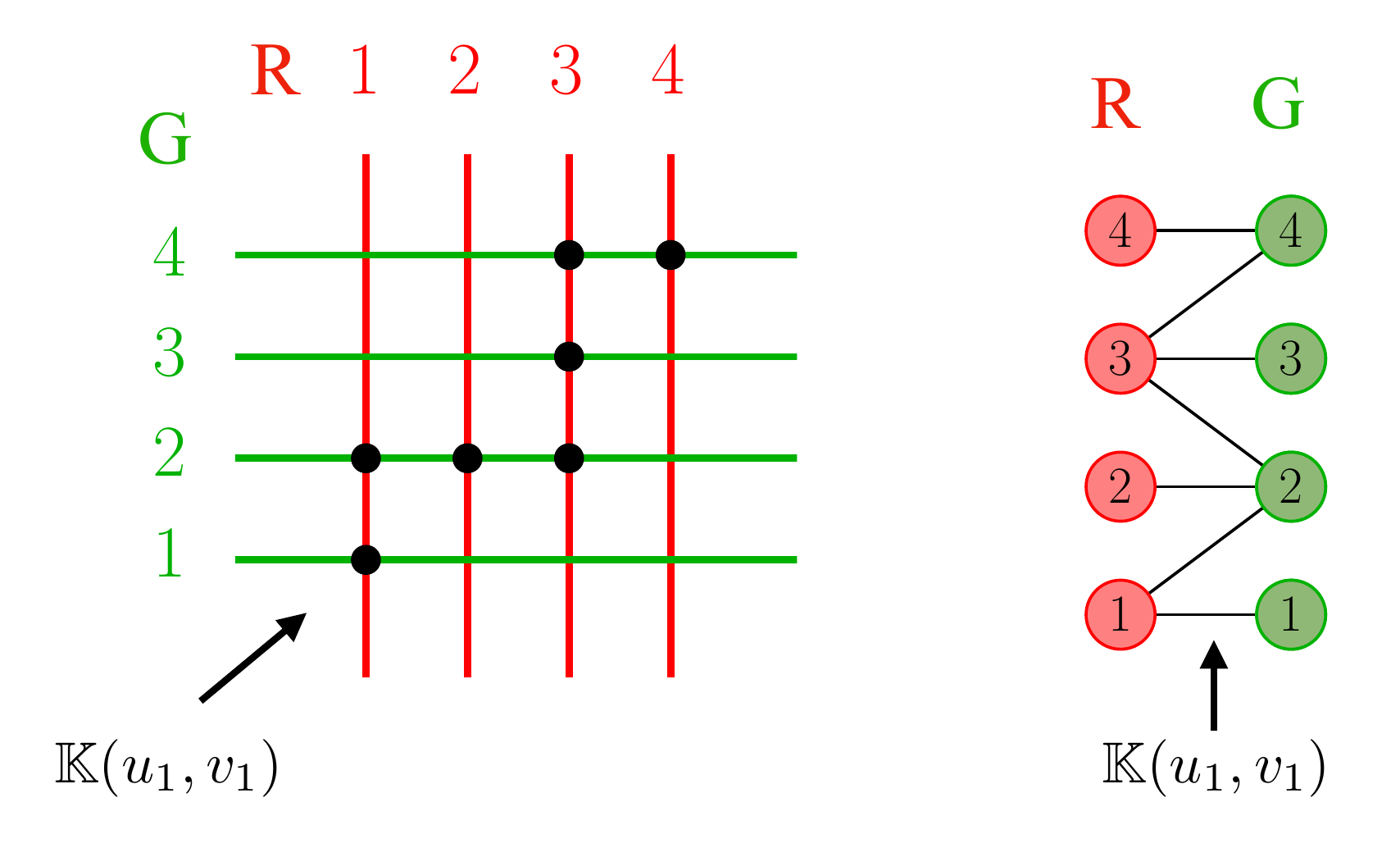}
\caption{\textbf{Left:} Sketch of a connected configuration contributing at leading order to the vertex-model representation of the matrix part in the two-cut case, with $N=M=4$ magnons on each cut. Elements belonging to the two cuts $(1,2)$  are represented by red ($R$) and green ($G$) colors, respectively.  Each of the $N+M-1$ black dots denotes the insertion of an element of the kernel $\mathbb{K}(u,v)$; the absence of a dot corresponds to inserting the identity. \textbf{Right:} The corresponding tree configuration, where magnons are represented by colored \emph{vertices} and kernels by \emph{branches}. A vertex is labeled by its color followed by its index. The vertices $R2,R4$ and $G1,G3$, which are attached to a single branch, are referred to as \emph{leaves}. For clarity, we have omitted the group-element insertions shown in Figure~\ref{fig:twocut}, as well as the orientation of the lines, since they are not essential to the argument.}
\label{fig:tree_example}
\end{figure}

To systematically isolate the contributions that survive in the strong-coupling limit of the correlator, one may employ a diagrammatic expansion similar to that developed in~\cite{Kostov:2017vwz}. In this setup, each magnon is represented as a vertex and each kernel as a branch connecting vertices, as depicted in the right panel of Figure~\ref{fig:tree_example}. When computing the free energy, i.e.~the logarithm of the correlator, we can restrict attention to the connected diagrams. The leading contributions then correspond simply to connected tree diagrams.

This can be seen through a simple power counting in $\e$. Each integral contributes a factor $\e^{-1}$, while each kernel contributes a factor $\e^{+1}$. Therefore, for a connected diagram arising from the expansion of  \(\cR_{\textbf{ba}}(\mathbf{v},\mathbf{u})\), we have
\begin{align}
\label{eq:tree_count}
\textrm{power of}\ \e\ \ \ {\buildrel \over =}\ \ &\# \ \textrm{of kernels (dots)}\,-\,\# \ \textrm{of integrals (lines)}\quad (\textrm{l.h.s. of  Figure } \ref{fig:tree_example})\\\nonumber 
{\buildrel \over =}\ \ &\# \ \textrm{of branches}\,-\,\# \ \textrm{of vertices}\geq -1 \quad \quad \quad \ \ \  (\textrm{r.h.s. of  Figure } \ref{fig:tree_example})\, .
\end{align}
The leading connected contribution is thus of order \(\epsilon^{-1} = 2g\) and is encoded by tree diagrams, such as one depicted in Figure~\ref{fig:tree_example}.

This structure leads to an important simplification: only the diagonal elements of \(\mathbb{K}_{ab}(u,v)\) contribute to the tree diagrams. To see this, consider the example in the right panel of Figure~\ref{fig:tree_example}, where \(N = M = 4\). The vertices of different colors are connected by \(N + M - 1 = 7\) kernels, represented as black lines. To compute the contribution of this configuration to \(\cM_{\boldsymbol{\a} \boldsymbol{\b}}(\mathbf{u},\mathbf{v}; \qg_1,\qg_2)\),
one inserts the group elements \(\qg_1(u)\) and \(\qg_2(v)\) in the appropriate spaces and then evaluates the supertrace as suggested in the right-hand side of Figure~\ref{fig:twocut}. Importanty, in two-dimensional kinematics, the group elements \(\qg_1(u)\) and \(\qg_2(v)\) commute. We may therefore compute all supertraces in a common eigenbasis of the tensor-product representation
\begin{align}
\mathscr{V}_{\text{tot}} = \mathscr{V}_{G1}\otimes \mathscr{V}_{G2}\otimes \cdots \otimes 
\mathscr{V}_{R1}\otimes \cdots .
\end{align}

The supertrace can be evaluated recursively by starting from the \emph{leaves} of the tree, i.e.~vertices attached to the rest of the diagram by a single \emph{branch}. Consider, for example, the vertex \(G1\) in Figure~\ref{fig:tree_example}, which is connected to (and thus scatters with) \(R1\). Taking the supertrace over the degrees of freedom of \(G1\) imposes flavor conservation
in $\mathscr{V}_{G1}$, which in turn implies flavor conservation in $\mathscr{V}_{R1}$. Consequently, only the diagonal part of the  scattering matrix between \(G1\) and \(R1\) contributes at this stage. After the trace over $\mathscr{V}_{G1}$ is performed, the leaf \(G1\) can be removed, and \(R1\) becomes a new leaf of the reduced tree.

This ``leaf stripping" procedure can be iterated. At each stage, tracing over a leaf enforces flavor conservation along the corresponding branch. Iterating until all leaves are removed shows that only the diagonal matrix elements of \(\mathbb{K}_{ab}(u,v)\) survive throughout the tree. 
 
\subsubsection{Clustering and simplified partition function}
\label{sec:simplify2cut}

According to the previous discussion, in the classical limit we can restrict ourselves to the diagonal part of the scattering matrix~\eqref{eq:classical-S-matrix}. For fundamental magnons $\in \mathscr{V}_1\otimes \mathscr{V}_1$, this amounts to retaining only
\be
K^{IJ}(u,v)\equiv \mathbb{K}_{IJ}^{IJ}(u,v)\, ,
\ee
where \(I,J \in \{\psi_{1}, \psi_{2} \,|\, \phi_{1}, \phi_{2}\}\). Equivalently, one may replace~\eqref{eq:classical-S-matrix} by a diagonal matrix $\mathbb{D}(u,v)$, defined by $\mathbb{D}_{IJ}^{KL} = \delta_{I}^{K}\delta_{J}^{L} \mathbb{D}^{IJ}$, with components
\begin{align}
\mathbb{D}^{IJ}(u,v) \equiv \exp\!\left[-\,\epsilon\, K^{IJ}(u,v)\right] \approx 1 - \epsilon\, K^{IJ}(u,v)\, ,
\end{align}
to leading order in $\epsilon$.

The same reasoning applies to bound states $\in \mathscr{V}_a\otimes \mathscr{V}_b$. At strong coupling, these decompose into stacks of fundamental magnons, so the diagonal components of their scattering matrix factorize into products of fundamental ones and can be expressed in terms of $\mathbb{D}$. This structure becomes most transparent after performing the clustering on each cut. The algebra is the same as for the octagon and allows one to reorganize the sums over bound states directly in terms of stacks of fundamental magnons with a given flavor. The net result is that the weights take the form~\eqref{eq:renormeasure}, while the interaction between a stack of $a$ magnons of flavor $I$ and a stack of $b$ magnons of flavor $J$ is simply given by
\begin{align}
\label{eq:Dpower}
\mathbb{D}^{IJ}_{ab}(u,v)=
\left(\mathbb{D}^{IJ}(u,v)\right)^{ab}\, ,
\end{align}
or equivalently, $\mathbb{D}_{ab}(u,v) = \mathbb{D}(u,v)^{ab}$, at the matrix level.

Introducing the shorthand notations,
\be
\frac{(-\qg_1)^\a}{\a^2} = \prod_{i\, =\, 1}^{N} \frac{(-\qg_1(u_i))^{a_i}}{a_i^2}\,, \qquad \frac{(-\qg_2)^\b}{\b^2} = \prod_{j\, =\, 1}^{M} \frac{(-\qg_2(v_j))^{b_j}}{b_j^2}\,,
\ee
and,
\be\label{eq:exponentialsimplified}
\mathbb{D}_{\a\b}(\u,\v) = \prod_{i,j} \mathbb{D}(u_i,v_j)^{a_i b_j}\,,
\ee
we conclude that, in the strong-coupling limit, the two-cut partition function~\eqref{eq:two-cuts} can be replaced by
\begin{align}\label{eq:two-cuts-diagonal}
\qZ[\qg_{1}, \qg_{2}] \approx \sum_{N, M \, =\, 0}^{\infty} \frac{(-1)^{N+M}}{N!\, M!} \sum_{\textbf{a}, \textbf{b}}
\int d\u\, d\v\,
\str_{\mathscr{V}_{\textrm{tot}}}
\!\left[
\frac{(-\qg_{1})^\a}{\a^2}\,\frac{(-\qg_{2})^\b}{\b^2}\,
\mathbb{D}_{\a\b}(\mathbf{u},\mathbf{v})
\right]\,,
\end{align}
where the supertrace is taken over the total space of fundamental magnons,
\be
\mathscr{V}_{\textrm{tot}}\equiv\mathscr{V}_1^{\otimes N}\otimes\mathscr{V}_1^{\otimes M}\,.
\ee
In this form, the partition function is expressed entirely in terms of the weights and scattering data of fundamental magnons. In particular, one can parametrize $\qg^{I}_{1,2}$ in terms of the functions $\cY^{I}_{1,2}$ as before. Explicit expressions for \(K^{IJ}\) are provided in Appendix~\ref{app:r-matrix}, including the effects of the dressing by the \(\mathcal{Z}\)-markers.

\subsection{Generalization to the $\NN$-cut partition function}
\label{sec:Ncut}

The arguments leading to the simplified expression in~\eqref{eq:two-cuts-diagonal} extend naturally to multi-cut systems obtained through more elaborate gluings of hexagons. The precise form of the integrand will not be needed here; what matters is that the rule governing interactions between magnons on the various cuts has a local and factorized structure~\cite{Basso:2015zoa}.

\begin{figure}[h]
\centering
\includegraphics[width=14cm]{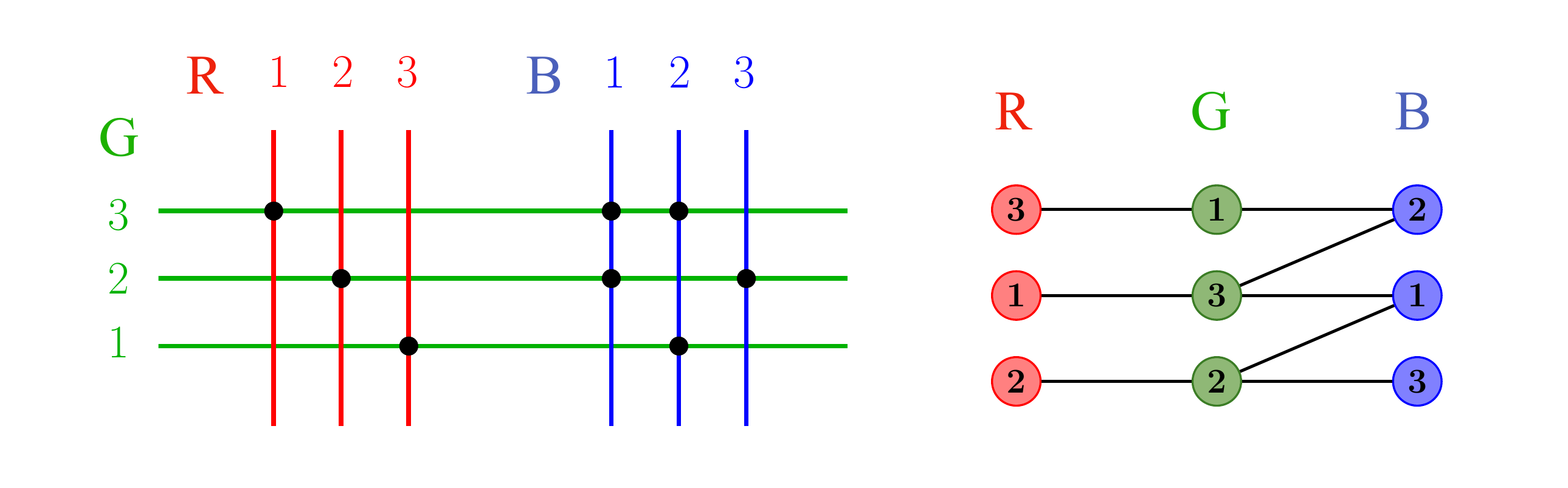}
\caption{\textbf{Left:} Connected configuration associated with a linear quiver with three cuts ($R,G,B$), labeled ``red'', ``green'', ``blue'', and $N=M=P=3$ magnons on each cut. \textbf{Right:} The corresponding tree configuration, with $N+M+P$ vertices and $N+M+P-1$ branches. The vertices $R1,R2,R3$ and $B3$ are leaves. A graph containing a cycle, where ``red'' and ``blue'' magnons also scatter, can be treated similarly; the only difference is that the tree will contain branches connecting ``red'' and ``blue'' vertices.}
\label{fig:tree_3}
\end{figure}

The basic observation is that any pair of magnons located on two cuts of a hexagon acquires an $R$-matrix contribution from that hexagon, exactly as in the case discussed above. The resulting matrix part for the $\NN$-cut system is then obtained schematically by assembling all such pairwise interactions arising from the individual hexagons. A concrete example is shown in Figure~\ref{fig:tree_3} for a three-cut configuration associated with a linear quiver (fan or zigzag).

The proof that the matrix part can be organized as a sum over trees relies solely on power counting, and not on the detailed pattern of $R$-matrices. As a result, the diagonal approximation applies to each $R$-matrix individually, regardless of the global gluing or the number of cuts involved. For the three-cut case, the relevant tree diagrams are illustrated in Figure~\ref{fig:tree_3}, where the same recursive leaf-stripping argument applies without modification.

It is nevertheless important to keep track of the connectivity of the mirror cuts in the triangulation, as it determines the resulting pattern of diagonal interactions. In addition, since the basic interaction is antisymmetric under the exchange of flavors and rapidities, $K^{IJ}(u, v) = -K^{JI}(v,u)$, one must also account for the relative orientation of the cuts.

To formalize this picture, we use the language introduced in Section~\ref{sec:triangulationsquiversclusteralgebras} and associate a quiver with \(\NN\) nodes to the triangulation \(\mathcal{T}\). Recall that each node of the quiver corresponds to an edge \(E \in \mathcal{T}\) and a charge \(\gamma_E\in \Lambda\). In this terminology, a mirror magnon on $E$ can be identified with a state of charge $\gamma_{E}$, and its couplings to other states are encoded in the signed adjacency matrix~\eqref{eq:defsignedB}, or equivalently in the pairings between edges, $\langle \gamma_{E}, \gamma_{E'}\rangle$.

The resulting diagonal interaction between magnons on a pair of edges $(\gamma_{E}, \gamma_{E'})$ can then be written compactly as
\be
\label{eq:twobody}
\mathbb{D}(u,u')^{\langle \gamma_{E},\gamma_{E'} \rangle}\,,
\ee
where the directionality of the interaction is encoded in the sign of the exponent. This expression trivially reduces to unity when the edges are not adjacent in the triangulation ($\langle \gamma_{E},\gamma_{E'} \rangle = 0$), as expected.

It is also convenient to introduce notation for the full set of states associated with the triangulation,
\be
\mathcal{B} \equiv \{\gamma_E \mid  E \in \mathcal{T} \}\,.
\ee
In principle, one should further equip $\mathcal{B}$ with a partial order encoding the ordering of the integration contours, and generalizing~\eqref{eq:can-order} to the multi-cut case. However, determining this poset structure is subtle, and we defer a detailed discussion to Sections~\ref{sec:tbafreeenergyforlinearquivers} and~\ref{sec:generaltriangulations}. For now, we ignore singularities arising from decoupling poles and the associated contour-ordering issues, and instead focus on the formal structure of the system.

One may then collect all factors and write down the general partition function. Dropping the edge index $E$ for notational simplicity, the group elements $\qg_{\gamma}$, bound-state indices $\a_\gamma$, and rapidity sets \(\mathbf{u}_\gamma\) associated with each cut all carry the state label \(\gamma\in \cB\). As in the two-cut case, we first perform clustering on each edge, reorganizing bound states into stacks, which now live in the full tensor product of fundamental modules,
\be
\mathscr{V}_{\mathrm{tot}} \equiv \bigotimes_{\gamma\, \in\, \mathcal{B}} \mathscr{V}_{1}^{\otimes N_\gamma}\, .
\ee
Using these notations together with~\eqref{eq:exponentialsimplified}, the two-cut expression~\eqref{eq:two-cuts-diagonal} for the partition function at strong coupling generalizes to the \(\NN\)-cut case, which can be written as
\begin{align}\label{eq:simplifiedNcutpartitionfunction}
\qZ[\{\qg_\gamma\}]
\approx   
\prod_{\gamma\, \in\, \mathcal{B}}\sum_{N_{\gamma}\, =\, 0}^{\infty}\frac{(-1)^{N_\gamma}}{N_\gamma!} \sum_{\a_\gamma}\int d\u_\gamma\,
\,\str_{\mathscr{V}_{\mathrm{tot}}}
\left[
\prod_{\gamma\, \in\, \mathcal{B}} \frac{(-\qg_\gamma)^{\a_\gamma}}{\a_\gamma^2}
\prod_{\substack{(\gamma , \gamma')}}
\mathbb{D}_{\a_\gamma\,\a_{\gamma'}}(\mathbf{u}_\gamma,\mathbf{u}_{\gamma'})^{\langle\gamma,\gamma'\rangle}
\right] \,.
\end{align}
Up to the aforementioned subtleties concerning the choice of integration contours, this expression applies to arbitrary configurations of $\NN$ cuts and is built from the same basic ingredients as in the one- and two-cut cases.

\subsection{Path integral description}
\label{sec:EFD}

Having simplified the partition function in the general $\NN$-cut case, we now proceed to its resummation. In the strict strong-coupling limit, \(\mathbb{D}\rightarrow 1\) in~\eqref{eq:simplifiedNcutpartitionfunction} and \(\qZ\) appears to factorize into independent one-cut contributions of the octagon type, \(\qZ \sim \prod_{\gamma}\mathbb{O}[\qg_\gamma]\). To recover the full free energy at strong coupling, however, one must go beyond this approximation and incorporate trees of $K$-interactions arising from the $\mathcal{O}(\epsilon)$-corrections to the individual \(\mathbb{D}\) factors, as discussed in Section~\ref{sec:Diagonal}. A description of this tree-based resummation is given in Appendix~\ref{app:TBAtree}. Here, we adopt an alternative approach based on an effective field description, which applies to diagonal pairwise interactions and allows us to recast $\qZ$ in the form of a path integral.

The starting point is to extract the \(\mathbb{D}\)-terms from the integrals and sums in~\eqref{eq:simplifiedNcutpartitionfunction}. This is achieved by representing them through the action of a suitable operator. To this end, we introduce auxiliary fields $\Phi^I_\gamma(u)$, labeled by a fundamental flavor index $I$ and a state $\gamma$, together with their conjugate functional derivatives satisfying
\be
\left[\frac{\delta}{\delta \qPhi^I_{\gamma}(u)}\,, \qPhi^J_{\gamma'}(u')\right] = \delta^{IJ} \delta_{\gamma \gamma'}\, \delta(u-u')\,,
\ee
where $\delta_{\gamma \gamma'}$ and $\delta^{IJ}$ denote Kronecker deltas. We then define the operator
\be\label{eq:def-cD}
\cD = \prod_{\gamma, I} \prod_{\gamma',J} \exp{-\frac{\epsilon}{2}\! \int du\, du' \,\qK_{\gamma \gamma'}^{IJ}(u, u') \frac{\delta^2}{\delta \qPhi^I_{\gamma}(u) \delta\qPhi^J_{\gamma'}(u')}}\, ,
\ee
where the kernel encodes both the interactions and the adjacency relations between cuts,
\be
\qK_{\gamma \gamma'}^{IJ}(u, u') = \langle \gamma, \gamma' \rangle \qK^{IJ}(u, u')\, ,
\ee
and satisfies the antisymmetry properties
\be\label{eq:propertiesofkernels}
\qK_{\gamma\gamma'}^{IJ}(u, u') = -\qK_{\gamma'\gamma}^{IJ}(u, u') = -\qK_{\gamma\gamma'}^{JI}(u', u)\, .
\ee
This operator is constructed so as to reproduce the $\mathbb{D}$-terms in~\eqref{eq:simplifiedNcutpartitionfunction} when acting on exponentials of the auxiliary fields. More precisely, one first applies $\cD$ and then sets the auxiliary fields to zero. Introducing the convenient notation
\be
\langle F[\{\Phi_\gamma\}]\rangle = \lim_{\{\qPhi_\gamma\} \rightarrow 0}\, \cD\, F[\{\Phi_\gamma\}]\, ,
\ee
for any functional $F[\{\Phi_\gamma\}]$, one finds---using~\eqref{eq:def-cD} and the Baker–Campbell–Hausdorff formula---that the $\mathbb{D}$-terms can be written compactly as
\be\label{eq:diff-op}
\prod_{(\gamma,\gamma')}\mathbb{D}(\u_\gamma,\u_{\gamma'})^{\langle \gamma,\gamma' \rangle} = \langle \exp\sum_{\gamma,i, I}\qPhi^I_{\gamma}(u_{\gamma,i})\rangle\, .
\ee
Substituting this representation into~\eqref{eq:simplifiedNcutpartitionfunction}, the resummation becomes straightforward, since all contributions are now fully factorized, as in the case of decoupled octagons. In particular, the dependence on the auxiliary fields in~\eqref{eq:diff-op} can be absorbed into a redefinition of the group elements,
\be\label{eq:group-shift}
\qg_\gamma\to \qg_\gamma e^{\qPhi_\gamma}\, ,
\ee
or equivalently, $\cY^I_{\gamma}\rightarrow \cY^I_{\gamma}\, e^{\qPhi_\gamma^I}$ for the eigenvalues, so that the $\NN$-cut partition function~\eqref{eq:two-cuts} takes the compact form
\be\label{eq:Z-to-octagons}
\qZ[\{\qg_\gamma\}] = \langle \bbO[\qg_{\gamma_1} e^{\qPhi_{\gamma_1}}] \,\bbO[\qg_{\gamma_2} e^{\qPhi_{\gamma_2}}]\ldots \bbO[\qg_{\gamma_{\NN}} e^{\qPhi_{\gamma_{\NN}}}] \rangle \,.
\ee 
Here, $\bbO[\qg_{\gamma} e^{\qPhi_{\gamma}}]$ denotes the octagon with the modified argument~\eqref{eq:group-shift}, namely
\be\label{eq:modified-oct}
\bbO[\qg_\gamma e^{\qPhi_\gamma}] \approx \exp{-\frac{1}{\epsilon}\int \frac{du}{2\pi}\, \str\, \textrm{Li}_{2}\left(-\cY_{\gamma} e^{\qPhi_{\gamma}(u)}\right)}\, .
\ee
To evaluate~\eqref{eq:Z-to-octagons}, it is then convenient to pass to Fourier space by introducing momenta $\qpi^I_{\gamma}$ conjugate to the fields $\qPhi^I_{\gamma}$. Denoting the functional measures by $D\qPhi$ and $D\qpi$, any functional can be expressed via its Fourier transform as
\begin{align}
&F[\{\qPhi_{\gamma}\}] \\ \nonumber
&\qquad =\int\! D\qpi\,  \exp\Big[\sum_{\gamma, I}\int \frac{du}{2\pi \epsilon}\, \qpi^I_{\gamma}(u)\, \qPhi^I_{\gamma}(u)\Big] \!\!\int\! D\qPhi' \, \exp\Big[- \sum_{\gamma, I} \int \frac{du}{2\pi \epsilon} \,\qpi^I_{\gamma}(u)\, {\qPhi'}^I_{\gamma}(u)\Big] F[\{\qPhi'_\gamma\}]\, ,
\end{align}
up to an overall normalization and with suitable integration contours. In this representation, the action of $\lim_{\Phi\rightarrow 0}\cD$ is straightforward. Namely, it amounts to replacing the exponent in the first exponential by
\be\label{QuadraticPi}
-\frac{1}{2}\sum_{\gamma, I}\sum_{\gamma',J}\int \frac{du du'}{(2\pi)^2\epsilon  }\, \qpi^{I}_{\gamma}(u) \,K_{\gamma\gamma'}^{IJ}(u, u')\, \qpi^{J}_{\gamma'}(u')\, .
\ee
Applying this to the product of octagons in~\eqref{eq:Z-to-octagons}, and using~\eqref{eq:modified-oct}, one finally obtains the path-integral representation
\be\label{eq:path-integrap-representation}
\qZ[\{\qg_\gamma\}] \approx \int D\qPhi\, D\qpi\, \exp{-\frac{1}{\epsilon} S[\qPhi, \qpi]}\, ,
\ee
with the action
\begin{align}\label{eq:general-action}
S &=\! \sum_{\gamma, I}\int \frac{du}{2\pi} \qpi_{\gamma}^{I}(u)\, \qPhi^{I}_{\gamma}(u) + \frac{1}{2} \sum_{\gamma, I} \sum_{\gamma', J}\int \frac{du du'}{(2\pi)^2}\,\qpi_{\gamma}^{I}(u)\, \qK^{IJ}_{\gamma \gamma'}(u, u')\, \qpi^{J}_{\gamma'}(u') \nonumber\\
&\qquad\qquad \qquad \qquad\qquad\qquad \qquad \qquad \qquad\qquad+  \sum_{\gamma}\int \frac{du}{2\pi}\, \str\,\textrm{Li}_{2}\left(-\cY_{\gamma} e^{\qPhi_{\gamma}(u)}\right)\, .
\end{align}
Here, the indices $\gamma, \gamma'$ run over the full set $\mathcal{B}$ and $I,J$ over the four fundamental flavors. At this stage, one could integrate out the momenta $\qpi^{I}_{\gamma}$ by formally inverting the kernels, thereby obtaining a path-integral representation involving the fields $\Phi^{I}_{\gamma}$ alone. This representation would make contact with the resummation method used in the context of null polygonal Wilson loops~\cite{BSV,Fioravanti_2015, Bonini_2019}. However, this step is unnecessary for the classical limit $\epsilon\rightarrow 0$, which can be obtained directly via a saddle-point analysis, as discussed in the next section.

\section{TBA equations and free energy for linear quivers}\label{sec:tbafreeenergyforlinearquivers} 

In what follows, starting from the action~(\ref{eq:general-action}), we derive the saddle-point equations for the fields $\Phi^I_\gamma$, as well as the value of the partition function $\qZ[\{\qg_\gamma\}]$ at the saddle point, and cast them into the form of a system of TBA equations. To facilitate comparison with the equations obtained in the string-theory context, we switch to the variable $\theta$ and rewrite the action~\eqref{eq:general-action} accordingly, keeping the rest of the notation unchanged.

Throughout this section, we restrict our attention to polygon triangulations corresponding to linear (acyclic) quivers, with $\NN = n-3$, such as those shown in Figure~\ref{fig:fan}. This restriction is necessary because the formulation of the TBA equations requires a definite prescription for the integration contours. As discussed in Section~\ref{sec:decouplingpoleandorder}, this prescription is fixed locally for each pair of cuts by the decoupling pole and is implemented by shifting the contours along the imaginary $\theta$-direction, thereby inducing a partial ordering on the states. Denoting by $\cC_{\gamma}$ the integration contour associated with a state $\gamma$, and writing $\gamma\succ \gamma'$ for $\textrm{Im}\, \cC_{\gamma} > \textrm{Im}\, \cC_{\gamma'}$, the decoupling pole condition then requires
\begin{align}
\label{partialorderdef}
\langle \gamma, \gamma' \rangle >0 \quad\Rightarrow\quad
\gamma \succ \gamma'\,.
\end{align}
This ordering is only partial: if two nodes in the quiver are not connected ($\langle \gamma, \gamma' \rangle = 0$), no ordering is imposed on the corresponding contours.

The main challenge is to implement this constraint consistently for all $\gamma\in \cB$. For linear quivers, this is immediate, as their structure induces a compatible ordering. In this case, the set $\mathcal{B} = \{\gamma_E \mid  E \in \mathcal{T} \}$ naturally acquires a poset structure determined by the orientation of arrows in the quiver, which we refer to as the \textit{canonical poset}.

The simplest example is provided by the fan triangulation shown in the left panel of Figure~\ref{fig:fan}. Labeling its diagonals as $\gamma_{i}$, with $i = 1, \ldots, n-3$ from bottom to top, one has $\langle \gamma_i, \gamma_{i+1}\rangle = 1$ for all $i<n-3$, leading to the ordering
\begin{align}
\label{eq:fanordering}
\gamma_1 \succ \gamma_2 \succ \cdots \succ \gamma_{n-3}.
\end{align}
More general linear quivers, such as those arising from zig-zag triangulations, are obtained by flipping some arrows in the fan quiver, which modifies the relative ordering while preserving consistency.

In summary, for acyclic quivers, the contour ordering is naturally encoded in the quiver itself, enabling a direct formulation of the TBA equations. For quivers with cycles, however, the pairings $\langle \gamma, \gamma'\rangle$ no longer define a consistent partial order, and a more elaborate construction is required, as discussed in Section~\ref{sec:generaltriangulations}. 

\subsection{First form of the equations}\label{tbalinearfromhexagonalization}
 
The path-integral representation~\eqref{eq:path-integrap-representation} provides a convenient starting point for analyzing the strong-coupling limit using a saddle-point approximation. Varying the action~\eqref{eq:general-action} with respect to the conjugate fields $\Pi$ and $\Phi$ leads directly to the equations of motion, 
\be
\label{eq:TBAlinquiv}
\begin{aligned}
&\qPhi^I_{\gamma}(\theta) = -\sum_{\gamma'\, \in\,  \cB}\sum_J\int_{\cC_{\gamma'}} \frac{d\theta'}{2\pi\cosh^2 \theta'} \,\qK^{IJ}_{\gamma\gamma'}(\theta, \theta') \, \qpi^J_{\gamma'}(\theta')\,, \\
&\qpi^I_{\gamma}(\theta) =  (-1)^{F_I}\, \log{\left(1+\cY^I_\gamma e^{\qPhi^I_{\gamma}(\theta)}\right)}\,.
\end{aligned}
\ee
Defining
\be
Y^I_\gamma\equiv \cY^I_\gamma \exp{\Phi^I_\gamma}\,,
\ee
we find that the equations can be recast into the form of TBA equations
\be\label{eq:tbafirstform}
\log{Y^{I}_{\gamma}}(\theta) = \log{{\cY}_{\gamma}^{I}(\theta)} + \sum_{\gamma'\, \in\, \mathcal{B}}\sum_{J}(-1)^{F_J+1}\qK_{\gamma\gamma'}^{IJ}\star L^{J}_{\gamma'}(\theta)\, .
\ee
Here the convolution $\star$ is given by
\be
\label{eq:defconvone}
\qK_{\gamma\gamma'}\star L^{J}_{\gamma'}(\theta) :=  \int_{\cC_{\gamma'}} \frac{d\theta'}{2\pi \cosh^2{\theta'}} \,\qK_{\gamma\gamma'}(\theta, \theta')\, L_{\gamma'}^{J}(\theta')\, ,
\ee
with $L^{I}_{\gamma}(\theta) = \log{(1+Y^{I}_{\gamma}(\theta))}$. The integration contours $\{\cC_{\gamma}, \gamma\in \cB\}$ run parallel to the real $\theta$-axis. As discussed earlier, they are slightly shifted in the imaginary direction and ordered appropriately to account for the singularities in the kernels (see~\eqref{eq:pole-tildeKIJ} below).

Substituting the solution~\eqref{eq:TBAlinquiv} back into the action, we obtain the free energy $\cA$ describing the strong-coupling limit of the multi-cut hexagon sum,
\be
\qZ[\{\qg_\gamma\}] \approx \exp{-\frac{1}{\epsilon} \cA}\, .
\ee
It decomposes into contributions from each cut, all of which are controlled by the same expression,
\be\label{eq:actionasrogers}
\cA = \sum_{\gamma\, \in \, \mathcal{B}} \cA_{\gamma}\, ,\qquad \textrm{with} \qquad
\cA_\gamma =  \sum_I (-1)^{F_{I}} \int \frac{d\theta}{2\pi \cosh^2 \theta} \, R(Y_{\gamma}^{I})\, .
\ee
The function
\be
\label{eq:RogerDi}
R(Y) \equiv \textrm{Li}_{2}(-Y) +\frac{1}{2} \log{(Y/\cY)}\log{(1+Y)}\, ,
\ee
is, up to a contribution depending on the driving term $\cY$, the expression of the Rogers dilogarithm.%
\footnote{More precisely, the function $R$ above coincides—up to an overall minus sign—with what is commonly known as the modified Rogers dilogarithm in the mathematical literature. Conventions differ among authors (see, e.g.,~\cite{Nakanishi:2024sum}). In what follows, we shall simply refer to $R$ as the Rogers dilogarithm.}
It naturally extends the formula found for the octagon~\eqref{eq:oct-sc}, where we have $Y=\cY$.

Inspecting the kernels in Appendix~\ref{app:r-matrix}, it is natural to separate their contributions into the two sectors, AdS and Sph, corresponding to the indices $I = \psi_{1}, \psi_{2}$ and $I=\phi_{1}, \phi_{2}$, respectively. The motivation for this decomposition is that the dominant interactions occur among magnons within the same sector and are governed by identical kernels.

Namely, writing
\be\label{eq:tildeK-deltaK}
\qK^{IJ}_{\gamma \gamma'}=(-1)^{1+F_{J}}(\tqK^{IJ}_{\gamma\gamma'}+\delta\tqK^{IJ}_{\gamma \gamma'})\, ,
\ee
with $\tilde{\qK}_{\gamma \gamma'}^{IJ}(\theta, \theta') = \langle \gamma,\gamma' \rangle\tilde{\qK}^{IJ}(\theta, \theta')$, etc., these interactions are encoded in the block-diagonal matrix
\be
\label{eq:Kerblock}
\tqK^{IJ} = \left(\begin{array}{cccc} \qK & \bqK & 0 & 0\\ \bqK & \qK & 0 & 0 \\ 0 & 0 & \qK & \bqK \\ 0 & 0 & \bqK & \qK \end{array}\right)\, , \qquad I, J\in \{\psi_{1}, \psi_{2}\vert \phi_{1}, \phi_{2}\}\, .
\ee
Here, $\qK = \tilde{\qK}^{\psi_{1}\psi_{1}}$ and $\bqK = \tilde{\qK}^{\psi_{1}\psi_{2}}$ define the $\AdS$ kernels,
\be\label{eq:kernelstheta}
\!\!\!\!\!\qK(\theta, \theta') = \frac{i}{2}\cosh{\theta} \cosh{\theta'}\, \textrm{coth}{\, \left(\frac{\theta-\theta'}{2}\right)}\, , \quad \bqK(\theta, \theta')=-\qK(\theta+i\pi, \theta')\, .
\ee
In particular, $\qK$ arises directly from the $SL(2)$ scattering phase in~\eqref{kernelxy}. It has a pole at coincident rapidities, $\theta = \theta'$,
\be\label{eq:pole-tildeKIJ}
\qK \sim \frac{i\cosh^2{\theta}}{\theta-\theta'} \, ,
\ee
while $\bqK$ has poles at shifted arguments, e.g.~$\theta\pm i\pi = \theta'$. Thus, the poles at $\theta=\theta'$ appear only on the diagonal in~\eqref{eq:Kerblock}, as for the decoupling poles of the hexagon form factors.

The remaining terms in~\eqref{eq:tildeK-deltaK} factorize into separate functions of $\theta$ and $\theta'$, and can be absorbed into a redefinition of the TBA driving terms,
\be\label{eq:shifted-driving-terms}
\log \cY^I_{\gamma}\to \log \tilde\cY^I_{\gamma} =\log \cY^I_{\gamma}+ \delta\log\cY^I_{\gamma}\, .
\ee
Adopting the notation used earlier for the flavor indices, 
\be
F^{\psi_{1}}=F^\AdS,\quad  F^{\psi_{2}}=\bar F^\AdS,\quad F^{\phi_{1}}=F^\Sph,\quad  F^{\phi_{2}}=\bar F^\Sph\,,
\ee
we can then rewrite the equations~\eqref{eq:tbafirstform} in the same compact form in each sector. Suppressing the sector label for simplicity, we obtain 
\be
\begin{aligned}
\label{eq:tbahexagon}
\log{Y_{\gamma}(\theta)} &= \log{\tilde{\cY}_{\gamma}(\theta)} + \sum_{\gamma'\, \in\, \mathcal{B}}\left[K_{\gamma \gamma'}\star L_{\gamma'}(\theta)+\bar K_{\gamma \gamma'}\star \bar L_{\gamma'}(\theta)\right]\, , \\
\log{\bar Y_{\gamma}(\theta)} &= \log{\tilde{\bar{\cY}}_{\gamma}(\theta)} + \sum_{\gamma'\, \in\, \mathcal{B}}\left[\bar K_{\gamma \gamma'}\star  L_{\gamma'}(\theta)+ K_{\gamma \gamma'}\star  \bar L_{\gamma'}(\theta)\right]\, ,
\end{aligned}
\ee
in both sector, where $\qK_{\gamma\gamma'} = \langle \gamma, \gamma'\rangle \qK$ and $\bqK_{\gamma\gamma'}= \langle \gamma, \gamma'\rangle \bqK$. 

The modified driving terms~\eqref{eq:shifted-driving-terms} retain their original dependence on $\theta$, up to shifts in the parameters. They take different forms in each sector:
\be\label{eq:shifts}
\delta\log \cY^{I}_{\gamma}(\theta) =
\begin{cases}
- \textcolor{red}{\delta l_{\gamma}}\cosh\theta\, , & I=\psi_1,\psi_2 \;(\mathrm{AdS}), \\[4pt]
- \textcolor{red}{\delta l_{\gamma}}\cosh\theta + \textcolor{blue}{\delta \log{U_{\gamma}}}\, (\pm \qw +i \sinh{\theta})\,  , & I=\phi_1,\phi_2 \;(\mathrm{Sph}),
\end{cases}
\ee
where the $+$ and $-$ signs corresponds to $I=\phi_{1}$ and $I=\phi_{2}$, respectively, and $\qw = \pm 1$ encodes the dressing. Finally, the parameter shifts are given by
\begin{align}
\textcolor{red}{\delta l_{\gamma}} = \sum_{\gamma'\, \in\, \mathcal{B}}&\langle \gamma, \gamma' \rangle \int \frac{d\theta'}{4\pi i \cosh^{2}{\theta'}}\left[\left(\sinh{\theta'}-i\qw \right)L^{\Sph}_{\gamma'}(\theta')+\left(\sinh{\theta'}+i\qw \right)\bar L^{\Sph}_{\gamma'}(\theta')\right] \, ,\label{eq:delta-l}
\end{align}
and
\begin{align}
&\textcolor{blue}{\delta \log{U_{\gamma}}} = - \frac{1}{2}\sum_{\gamma'\, \in\, \mathcal{B}} \langle\gamma, \gamma'\rangle \,\kappa_{\gamma'}\, ,\label{eq:delta-logU}
\end{align}
with
\be
\label{eq:kappa}
\kappa_{\gamma} =  \int \frac{d\theta}{2\pi \cosh{\theta}}\,\str\, L_{\gamma}(\theta)\, .
\ee

\subsection{Second form of the equations}
\label{sec:secondform}

The TBA equations~\eqref{eq:tbafirstform} derived in the previous section take a natural form dictated by the hexagonalization construction. In this formulation, the driving terms $\cY$ have a clear interpretation, directly encoding the weights of the magnons in terms of the bridge lengths and cross-ratios.

The main complication lies in the kernels~\eqref{eq:kernelstheta}, which are not of relativistic difference type. A similar situation arises in the study of null Wilson loops, where the kernels obtained from the Pentagon-OPE approach also exhibit a complicated structure~\cite{Basso:2013vsa,Bonini:2015lfr}, despite the fact that the TBA description derived from the minimal surface analysis admits a universal relativistic form~\cite{Alday:2009dv,Alday:2010vh}. This mismatch is only superficial and is resolved by simple redefinitions~\cite{Alday:2010ku,Gaiotto:2014bza}.

An analogous reformulation exists in the present case. The TBA equations can be recast in a ``relativistic frame'' where the kernels depend only on rapidity differences. This requires a further redefinition of the driving terms and leads to the equations
\be
\begin{aligned}
\label{eq:TBA-GMN}
\log{Y_{\gamma}(\theta)} &= \log{\sfchi_{\gamma}(\theta)} + \sum_{\gamma'\, \in\, \mathcal{B}}\left[G_{\gamma \gamma'}* L_{\gamma'}(\theta)+\bar G_{\gamma \gamma'}* \bar L_{\gamma'}(\theta)\right]\, , \\
\log{\bar Y_{\gamma}(\theta)} &= \log{\bar\sfchi_{\gamma}(\theta)} + \sum_{\gamma'\, \in\, \mathcal{B}}\left[\bar G_{\gamma \gamma'}*  L_{\gamma'}(\theta)+ G_{\gamma \gamma'}*  \bar L_{\gamma'}(\theta)\right]\, ,
\end{aligned}
\ee
with the same functions $L_\gamma(\theta)$ and $\bar L_\gamma(\theta)$ as before. Here $G_{\gamma \gamma'}=\langle \gamma,\gamma' \rangle G$ and $\bar G_{\gamma \gamma'}=\langle \gamma,\gamma' \rangle \bar G$, and the relativistic kernels are
\be\label{eq:def-relativistickernels}
\GMN(\theta) = \frac{i}{2}\, \textrm{coth}{\left(\frac{\theta}{2}\right)}\, , \qquad \bGMN(\theta) = -\GMN(\theta+i\pi) = -\frac{i}{2}\, \textrm{tanh}{\left(\frac{\theta}{2}\right)}\, .
\ee
Moreover, the new convolution $*$ now involves translationally invariant measure, e.g.
\be
G_{\gamma \gamma'}*L_{\gamma'}(\theta):=\int_{\cC_{\gamma'}} \frac{d\theta}{2\pi} \,G_{\gamma \gamma'}(\theta-\theta')\,L_{\gamma'}(\theta')\, ,
\ee
and similarly for $\bar G_{\gamma \gamma'}(\theta)$ and $\bar L_\gamma(\theta)$.
The kernel $G(\theta-\theta')$ has a simple pole at $\theta=\theta'$, with unit residue,
\be
G(\theta-\theta')\sim \frac{i}{\theta-\theta'}\, .
\ee
As before, the integration contours are ordered according to the poset structure of $\mathcal{B}$.

Again, the two sectors decouple and the equations depend on the sector only through the driving terms, that are  given by
\be\label{eq:GMNzeromodes}
\log{\sfchi^I_{\gamma}(\theta)} = C^I_{\gamma} + Z_{\gamma} e^{\theta} + \bar{Z}_{\gamma} e^{-\theta}\, .
\ee
The so-called central charges $Z_\gamma$ and $\bar{Z}_\gamma$ depend on the geometric data (cross-ratios and bridge lengths) and, remarkably, are the same for all flavors, both in the sphere and AdS sectors. They satisfy
\be
\label{eq:ZplusZbar}
Z_\gamma + \bar{Z}_{\gamma} = - l_{\gamma} - \textcolor{red}{\delta l_\gamma}\, ,
\ee
and
\be
\label{eq:ZmoinsZbar}
Z_\gamma - \bar{Z}_{\gamma} = i\sigma^{\AdS}_{\gamma} +i  \sum_{\gamma'\, \in\, \mathcal{B}}\langle\gamma, \gamma'\rangle\int \frac{d\theta'}{4\pi \cosh{\theta'}} \left[L_{\gamma'}^{\AdS}(\theta') + \bar{L}_{\gamma'}^{\AdS}(\theta') \right]\, .
\ee
The coefficients $C^I_{\gamma}$ are less universal and depend on the type of excitations. In the AdS and sphere sectors, respectively, we find
\be\label{eq:constant-C}
\begin{aligned}
C_{\gamma}^{\AdS} &= -\bar{C}^{\AdS}_{\gamma} = i\phi^{\AdS}_{\gamma} +i \sum_{\gamma'\, \in\, \mathcal{B}} \langle\gamma, \gamma'\rangle\int \frac{d\theta'}{4\pi} \textrm{tanh}{\,\theta'} \left[L_{\gamma'}^{\AdS}(\theta') - \bar{L}_{\gamma'}^{\AdS}(\theta') \right]\, , \\
C_{\gamma}^{\Sph} &= -\bar{C}^{\Sph}_{\gamma}\, = i\phi^{\Sph}_{\gamma}\, +i \sum_{\gamma'\, \in\, \mathcal{B}} \langle\gamma, \gamma'\rangle\int \frac{d\theta'}{4\pi} \textrm{tanh}{\,\theta'} \left[L_{\gamma'}^{\Sph}(\theta') - \bar{L}_{\gamma'}^{\Sph}(\theta') \right]+ \qw \log{\qU_{\gamma}}\, ,
\end{aligned}
\ee
where $\qw$ is the dressing parameter and we introduce
\be\label{eq:logU-TBA}
\log{\qU}_{\gamma} := \log U_\gamma +\textcolor{blue}{\delta \log U_\gamma}\,. 
\ee 

Finally, the free energy can be rewritten in the more canonical form
\be\label{nicer-action}
\cA =  \sum_{\gamma\, \in \, \mathcal{B}} \int \frac{d\theta}{2\pi} (Z_{\gamma} e^{\theta}+\bar{Z}_{\gamma} e^{-\theta})\, \str\, L_{\gamma}(\theta) +   \sum_{\gamma\, \in\,  \mathcal{B}}l_{\gamma} \,\kappa_{\gamma}\, ,
\ee
which is standard for this class of TBA equations, up to the presence of the $\kappa$-terms. This expression follows from~\eqref{eq:actionasrogers} after integrating by parts the dilogarithmic terms and using the TBA equations to simplify the result. As the intermediate steps are lengthy and not particularly illuminating, we defer the full derivation to Appendix~\ref{app:simpler-action}.

\subsection{$Y$-system}\label{sec:Y-sys}

As emphasized at the beginning of this section, the construction of the TBA system comes equipped with a canonical set of integration contours, such that each function $Y^{I}_{\gamma}$ appearing in~\eqref{eq:tbafirstform} is naturally defined on its corresponding contour $\cC_{\gamma}$. As is standard in the TBA framework, however, these functions admit analytic continuation beyond their original domains, allowing one to explore their behavior in other regions of the rapidity plane.

In particular, by considering the continuation under the shift $\theta\rightarrow \theta+i\pi$, one finds that the $Y$-functions functions satisfy a set of functional relations of $Y$-system type, similar to those arising in conventional TBA equations~\cite{Zamolodchikov:1991et, Kuniba:1993cn}. The derivation of the $Y$-system does not depend on which form of the TBA we adopt, so here we use equations~\eqref{eq:tbahexagon}. Moreover, since the result is identical in the AdS and sphere sectors, we suppress the sector index in what follows.

The key observation is that the kernels and driving terms transform simply under the shift $\theta\rightarrow \theta+i\pi$. This shift implements relativistic crossing symmetry, flipping the sign of the energy and momentum of the magnons and relating different $R$-matrix elements (see eq.~\eqref{eq:kernel-crossing-sym}). In terms of the TBA data~\eqref{eq:shifted-driving-terms}, one has
\begin{align}
   \log \cY_\gamma(\theta + i\pi)=- \log \bar{\cY}_\gamma (\theta)\,, \qquad \log \bar{\cY}_\gamma(\theta + i\pi)=- \log \cY_\gamma (\theta)\, ,
\end{align}
and similarly for $\delta \log{\cY}$'s in~\eqref{eq:shifts}, as well as
\be\label{eq:kernel-shift}
\qK(\theta+i\pi, \theta') = -\bqK(\theta, \theta') \qquad \bqK(\theta+i\pi, \theta') = -\qK(\theta, \theta')\, .
\ee
Naively, this suggests that
\be
Y^{++}_{\gamma} \equiv Y_{\gamma}(\theta+i\pi)
\ee
maps to $1/\bar{Y}_{\gamma}$, and vice versa. However, this ignores that the kernel~\eqref{eq:kernelstheta} has a pole at coincident rapidities,
\begin{align}\label{eq:K-pole}
   \qK_{\gamma\gamma'}(\theta,\theta')\equiv \langle\gamma,\gamma'\rangle\,\qK(\theta,\theta')\ \ {\buildrel \theta\to \theta' \over \simeq}\ \ \frac{i\, {\langle\gamma,\gamma'\rangle}}{\theta-\theta'}\,\cosh^2 \theta
   \,.
\end{align}
As a result, the right-hand side of the TBA equation for $\log{Y_{\gamma}(\theta)}$ jumps whenever $\theta$ crosses a contour $\cC_{\gamma'}$ with $\gamma'\succ \gamma$. Taking the residue at $\theta = \theta'$ yields
\be\label{eq:disc-logY}
\left(\qK_{\gamma\gamma'}\star L_{\gamma'}\right)(\theta-i0) = \left(\qK_{\gamma\gamma'}\star L_{\gamma'}\right)(\theta+i0) - \langle \gamma, \gamma'\rangle \log{(1+Y_{\gamma'})}\, ,
\ee
for $\theta \in \cC_{\gamma'}$. Similar poles arise from $\bar{K}_{\gamma \gamma'}\star \bar{L}_{\gamma'}$ at the end of the continuation, that is when the rapidity reaches $\theta+i\pi$, as seen in~\eqref{eq:kernel-shift}. These cross all contours with $\gamma' \prec \gamma$ and contribute with the opposite sign.

After combining all these effects, we obtain that the $Y$-functions obey the relations
\begin{align}
\label{YsysAdSSph}
Y_\gamma^{++}\,\bar{Y}_\gamma =\prod_{\gamma'\succ \gamma}(1+Y_{\gamma'}^{++})^{\langle \gamma',\gamma\rangle}\prod_{\gamma'\prec \gamma}\left(1+\bar{Y}_{\gamma'}\right)^{\langle \gamma,\gamma'\rangle}\,,
\end{align}
together with a similar equation obtained by exchanging $Y\leftrightarrow \bar{Y}$. Since the driving terms cancelled out in the derivation, the AdS and sphere sectors are completely decoupled in this formulation. In turn, the $Y$-system encodes all the information about the TBA equations, except for the driving terms: these are zero-mode solutions of the homogeneous difference equations, $\cY_{\gamma}^{++}\bar{\cY}_{\gamma} = 1$, which must be fixed by additional conditions.

We note that the $Y$-system~\eqref{YsysAdSSph} has a somewhat unusual structure since its right-hand side contains both shifted and unshifted $Y$-functions. This feature reflects the presence of decoupling poles near the initial contour of integration in the hexagon construction. It would be interesting to investigate whether this system can be brought to a more canonical form by redefining some $Y$-functions through shifts of their arguments. 

Similar structures have been observed in the context of null polygonal Wilson loops~\cite{Bonini:2015lfr}.
As in that case, the $Y$-system is extremely efficient for solving the TBA equations in particular regimes~\cite{Alday:2010vh,Alday:2009dv}, such as the zero bridge-lengths limit, where the $Y$-functions become constant ($Y^{++} = Y = \text{const}$). We refer the reader to Appendix~\ref{app:constant-Y} for a construction of this solution.

\section{Quiver cycles and wall crossing}\label{sec:generaltriangulations}

Up to this point we have restricted our attention to triangulations associated with linear quivers. This allowed us to work with the contour ordering induced from the quiver~\eqref{partialorderdef}. However, when the quiver contains oriented cycles, the integration contours can no longer be consistently ordered in this way.

\begin{figure}[h]
\centering
\includegraphics[width=8cm]{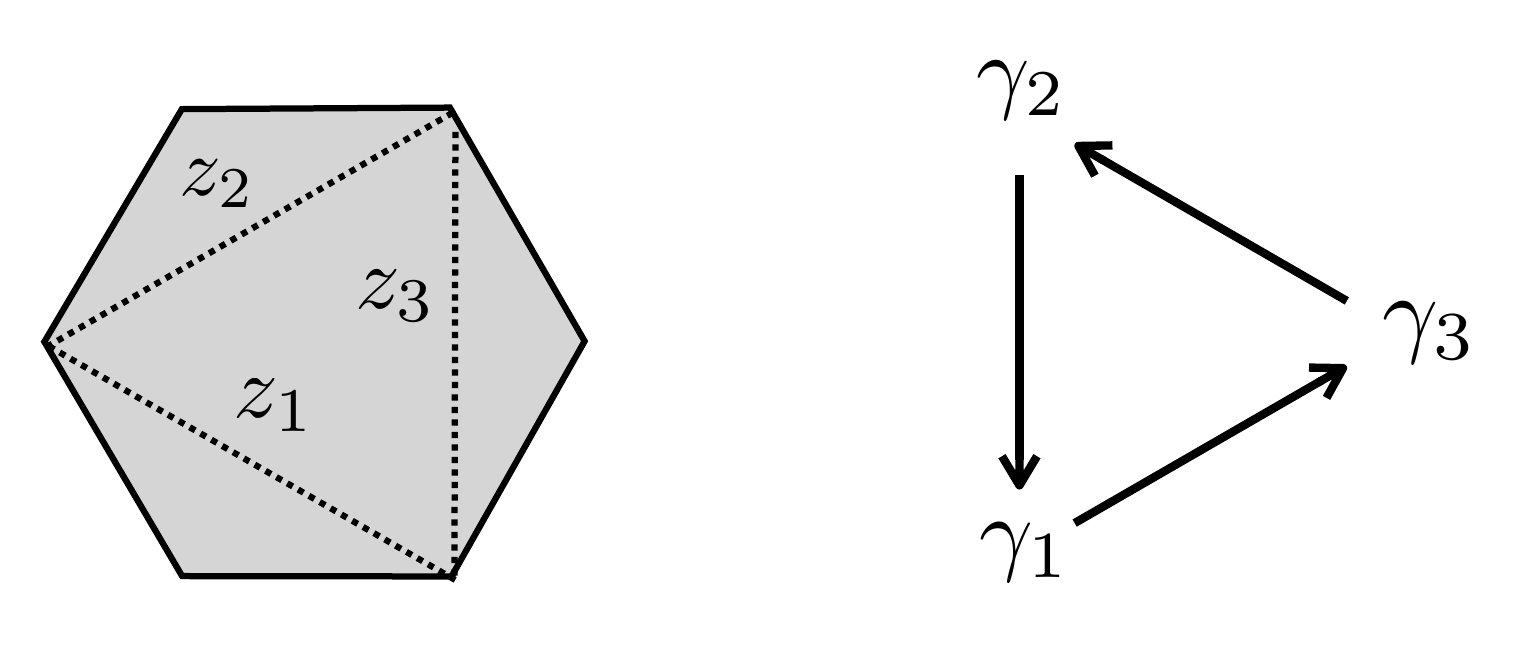}
\caption{\textbf{Left:} Symmetric triangulation of a hexagon, with implicit orientation as in Figure~\ref{fig:genericopengeometry}. \textbf{Right:} The corresponding cyclic quiver.  }
\label{fig:A3cycle}
\end{figure}

The simplest example is the 3-cycle shown in Figure~\ref{fig:A3cycle}, corresponding to a symmetric triangulation of a hexagon. In this case,
\be
\langle \gamma_{i}, \gamma_{i+1}\rangle = 1\, , \qquad i=1,2,3\, ,
\ee
with $\gamma_{i+3} = \gamma_{i}$, which makes the prescription~\eqref{partialorderdef} incompatible with any global contour ordering.

A natural attempt to resolve this issue is to adopt a Feynman-like prescription, integrating around each pole separately by introducing infinitesimal imaginary shifts determined by the pairings,
\be\label{eq:Feynman-presc}
\qquad K_{\gamma \gamma'}(\theta, \theta') \rightarrow K_{\gamma \gamma'}(\theta+i\langle \gamma, \gamma' \rangle 0, \theta')\, .
\ee
In this formulation, no global contour ordering is required, and the TBA equations can be written for arbitrary quivers, with all contours taken along the real $\theta$-axis. The drawback is that this prescription is not fully compatible with the hexagon construction and, in fact, violates the decoupling condition.

\begin{figure}[h]
\centering
\includegraphics[width=12cm]{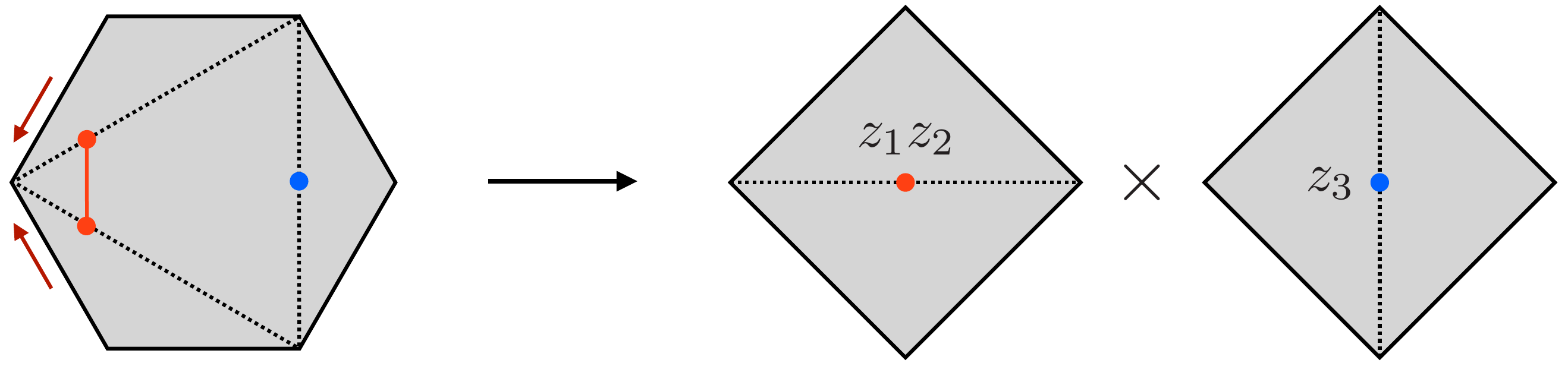}
\caption{In the double OPE limit where the three left vertices approach each other, the six-point function factorizes into a product of two four-point functions. This limit is controlled by two types of excitations: pairs of magnons (in red), analogous to those appearing in the OPE limit of the five-point function, and isolated magnons on the right cut (in blue). Ultra-local interactions~\eqref{eq:threemagcontact} involving magnons on three cuts simultaneously violate this factorization and are therefore incompatible with the decoupling limit.}
\label{fig:hexagonOPElimit}
\end{figure}

The origin of this violation can be traced to contact terms arising in products of kernels. The first instance appears in the three-magnon configuration, with one magnon on each cut of the inner triangle. The hexagon decoupling property requires that the \textit{connected} contribution to this process be smooth at coincident rapidities. Instead, expanding the free energy to order $\propto \cY_{\gamma_{1}}\cY_{\gamma_{2}}\cY_{\gamma_{3}}$ yields the following kernel combination%
\footnote{The regular part in the product of distributions is $\frac{1}{4}\prod_{i=1,2,3}\cosh{\theta_{\gamma_i}}(\cosh{\theta_{\gamma_1}} + \cosh{\theta_{\gamma_2}} + \cosh{\theta_{\gamma_3}})$.}
\be\label{eq:threemagcontact}
K_{\gamma_{1}\gamma_{2}} K_{\gamma_{2}\gamma_{3}} + \textrm{cyclic} \, =\,  (2\pi)^2 \cosh^4{\theta_{\gamma_{1}}}\delta (\theta_{\gamma_{1}}-\theta_{\gamma_{2}}) \delta (\theta_{\gamma_{2}}-\theta_{\gamma_{3}}) +\textrm{regular} \, ,
\ee
which exhibits unphysical $\delta$-function singularities when combined with~\eqref{eq:Feynman-presc}.%
\footnote{This follows by inserting the pole~\eqref{eq:K-pole} of $K(\theta, \theta')$ into the left-hand side of the equation. Summing over cyclic images cancels all poles, up to delta-function terms.}
Equivalently, these singularities signal a breakdown of the OPE factorization depicted in Figure~\ref{fig:hexagonOPElimit}.

Since these contributions are ultra-local, they can in principle be cancelled by introducing additional degrees of freedom in the form of new $Y$-functions. In the present case the singular term can be removed by a $Y$-function describing a three-magnon state localized on the cycle. However, while this approach is workable in simple cases, it quickly becomes impractical for triangulations with multiple cycles, due to the uncontrolled proliferation of composite states. We refer the interested reader to Appendix~\ref{app:TBAcycles} for illustrations of this construction.

A more elegant resolution is obtained by inverting the pole prescription for a pair of states, e.g., $i\langle \gamma_1,\gamma_2\rangle 0 \rightarrow -i\langle \gamma_1,\gamma_2\rangle 0$. This removes the contact terms in~\eqref{eq:threemagcontact} and restores a consistent contour ordering. Such a prescription was used in  ref.~\cite{Ferrando:2025qkr} to treat the cubic contribution to the three-point function in the BPS sector of $\mathbb{Z}_k$ orbifold $\mathcal{N}=2$ SYM theories. Operationally, it corresponds to exchanging the integration contours.

In the remainder of this section, we explore the consequences of such contour exchanges, which relate to \emph{wall-crossing} transformations of the TBA equations~\cite{Gaiotto:2010okc,Alday:2010vh,Alday:2009dv,Toledo:PhDthesis}. This approach provides a systematic resolution of quiver cycles and yields TBA equations and free energy for arbitrary triangulations.

\subsection{Wall-crossing transformation}\label{sec:wallcrossingtransformation}

As noted above, handling cycles requires consistently swapping integration contours in the TBA equations, introducing a partial ordering for the set of contours. This prevents contact terms like~\eqref{eq:threemagcontact} but introduces composite excitations, analogous to bound-state formation in wall-crossing of 4D $\mathcal{N}=2$ theories~\cite{Gaiotto:2010okc}. We briefly review how this arises in TBA, following~\cite{Alday:2010vh,Toledo:PhDthesis}.

\subsubsection{Revisiting $A_2$ quiver}
\label{sec:A2WC}

The wall-crossing mechanism is most transparently illustrated in the TBA equations for the $A_{2}$ system, and extends naturally to more general quivers. Since the mechanism works the same for each flavor independently, we suppress flavor indices in the following.

Consider then a two-cut configuration with labels $\gamma_1$ and $\gamma_2$ such that $\langle\gamma_1,\gamma_2\rangle=1$. This fixes the canonical ordering of contours, $\textrm{Im}\, \cC_{\gamma_{1}} > \textrm{Im}\, \cC_{\gamma_{2}}$, or, in shorthand notation, $\gamma_1 \succ \gamma_2$. Permuting the contours requires crossing the decoupling pole of the kernel~\eqref{eq:K-pole}, whose residue produces an additional local term, as in~\eqref{eq:disc-logY} with $\{\gamma, \gamma'\}\leftrightarrow \{\gamma_{1},\gamma_{2}\}$.

\begin{figure}[h]
    \centering
    \includegraphics[width=0.8\linewidth]{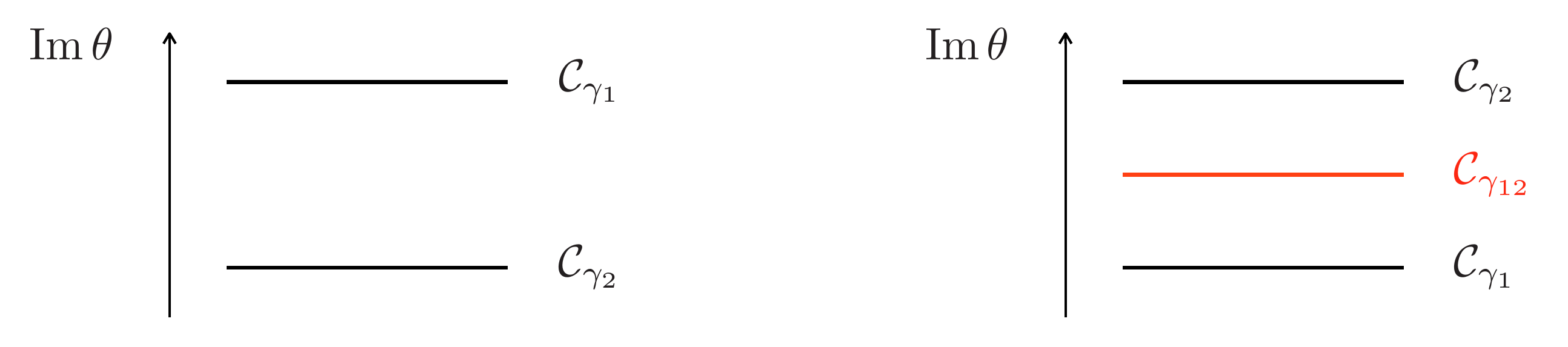}
    \caption{\textbf{Left:} Canonical contour ordering for $\langle \gamma_{1}, \gamma_{2} \rangle = 1$. \textbf{Right:} Swapping the contours generates the state $\gamma_{12}$, with $\cC_{\gamma_{12}}$ located between $\cC_{\gamma_{1}}$ and $\cC_{\gamma_{2}}$ along the imaginary $\theta$-direction.}
    \label{fig:crossing-contours}
\end{figure}

The main point is that this local term can be absorbed into a redefinition of the $Y$-functions while keeping the structure of the TBA equations intact, at the cost of introducing a new function $Y'_{\gamma_{12}}$. Defining
\be\label{eq:newcompositeYfunction}
Y'_{\gamma_1} = \frac{Y_{\gamma_1}}{1+Y_{\gamma_2}}\, , \qquad Y'_{\gamma_2} = \frac{Y_{\gamma_{2}}}{1+Y_{\gamma_{1}}}\, , \qquad Y'_{\gamma_{12}} = \frac{Y_{\gamma_{1}}Y_{\gamma_{2}}}{1+Y_{\gamma_{1}}+Y_{\gamma_{2}}}\,,
\ee
one can verify that the $Y'$'s satisfy the TBA equations~\eqref{eq:tbafirstform} for the enlarged set $\cB' = \{\gamma_{1}, \gamma_{2}, \gamma_{12}\}$ with $\gamma_{12} = \gamma_{1}+\gamma_{2}$. The redefinitions of $Y_{\gamma_1}$ and $Y_{\gamma_2}$ in~\eqref{eq:newcompositeYfunction} account for the local term while the composite function $Y'_{\gamma_{12}}$ ensures that the $L$-terms in~\eqref{eq:disc-logY} decompose correctly,
\be
\label{YtoYprime}
(1+Y_{\gamma_{1}}) = (1+Y'_{\gamma_{1}})(1+Y'_{\gamma_{12}})\, , \qquad (1+Y_{\gamma_{2}}) = (1+Y'_{\gamma_{2}})(1+Y'_{\gamma_{12}})\, .
\ee
The equation for $Y'_{\gamma_{12}}$ follows from applying the same reasoning to $\log{(Y_{\gamma_{1}}Y_{\gamma_{2}})}$, with driving term
\be
\cY_{\gamma_{12}}= \cY_{\gamma_1}\, \cY_{\gamma_2}\, .
\ee
Equivalently, the state $\gamma_{12}$ carries weights $l_{\gamma_{12}} = l_{\gamma_{1}}+l_{\gamma_{2}}$ and $z_{\gamma_{12}} = z_{\gamma_{1}}z_{\gamma_{2}}$, and similarly for $\bar{z}, \alpha$ and $\bar{\alpha}$.
Consistency further requires its contour to lie between the original ones,
\be\label{eq:wall-crossed-order}
\gamma_{2} \succ \gamma_{12} \succ \gamma_{1}\, ,
\ee
a characteristic feature of wall-crossing~\cite{Gaiotto:2010okc} (see Figure~\ref{fig:crossing-contours}).

Physically, this describes the formation of a bound state $\gamma_{12} = \gamma_{1}+\gamma_{2}$ from two elementary magnons $\gamma_{1}$ and $\gamma_{2}$ with identical rapidity and flavor, in line with the discussion in Section~\ref{sec:decouplingpoleandorder}.%
\footnote{Relatedly, the wall-crossed system provides a better starting point for the study of the OPE limit discussed in Section~\ref{sec:decouplingpoleandorder}. In particular, the limit displayed in the top panel of Figure~\ref{fig:decouplinglimits} is equivalent to the limit where $Y'_{\gamma_1}, Y'_{\gamma_2} \rightarrow 0$ with $Y'_{\gamma_{12}}$ held fixed. In this regime, one easily verifies that the TBA equations reduce to those of the octagon, with weights $\cY_{\gamma_{12}} = \cY_{\gamma_{1}}\cY_{\gamma_{2}}$, etc.} Once created, this state contributes on the same footing as the others and interacts through the usual couplings $\sim \langle\gamma,\gamma' \rangle K^{IJ}$.

The same holds for the free energy~\eqref{eq:actionasrogers}, which retains its form up to an additional contribution from the composite state,
\be
\cA = \sum_{\gamma\, \in \, \cB'} \cA'_{\gamma} =  \cA'_{\gamma_1}+\cA'_{\gamma_2}+\cA'_{\gamma_{12}}\, .
\ee
This follows directly from Abel’s identity for Rogers dilogarithms~\eqref{eq:RogerDi},
\be
\label{eq:RogerPentagon}
R(Y_{\gamma_1})+R(Y_{\gamma_2}) = R(Y'_{\gamma_1})+R(Y'_{\gamma_2})+R(Y'_{\gamma_{12}})\, ,
\ee
using the definitions~\eqref{eq:newcompositeYfunction}.

Finally, if the two states do not interact, i.e.~$\langle \gamma_{i},\gamma_{j}\rangle = 0$, no pole is encountered, the contours may be freely permuted, and no bound state is generated.

\subsubsection{Cyclic $A_3$ quiver}
\label{sec:A3cycle}

We are now ready to address the ordering of states in the presence of cycles. The basic building block is the cyclic $A_{3}$ quiver introduced above, which we now analyze in detail. As noted earlier, in this case the pairings do not define a consistent ordering of the states.

\begin{figure}[h]
\centering
\includegraphics[width=8cm]{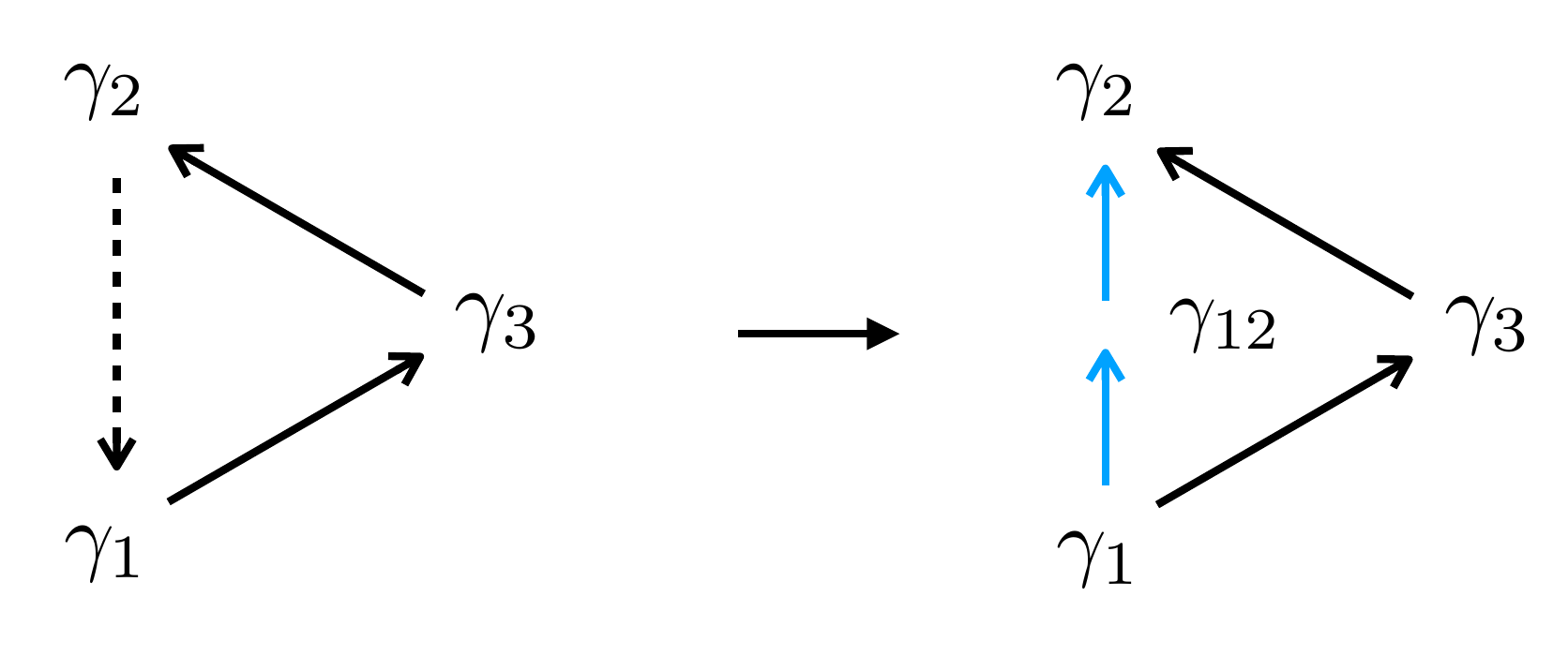}
\caption{Illustration of the wall-crossing method for the cyclic $A_3$ quiver. \textbf{Left:} The arrow $\gamma_1\leftarrow \gamma_2$ is selected to perform the wall-crossing transformation, producing the bound state $\gamma_{12} = \gamma_{1}+\gamma_{2}$. \textbf{Right:} Blue arrows indicate the ordering after wall-crossing, which in our conventions is $\gamma_{2}\succ \gamma_{12}\succ \gamma_{1}$. The full poset is obtained by placing $\gamma_{3}$ in the sequence according to its pairings with the other states. There is no arrow between between $\gamma_{3}$ and $\gamma_{12}$, since $\langle \gamma_{3}, \gamma_{12}\rangle = 0$.}
\label{fig:A3WallCrossing}
\end{figure}

To resolve this issue, we begin by performing a wall-crossing transformation on an $A_{2} \subset A_{3}$ subquiver, as illustrated in Figure~\ref{fig:A3WallCrossing}. Concretely, this amounts to selecting an arrow, say $\gamma_{1}\leftarrow \gamma_{2}$, and imposing an ordering of states opposite to the orientation encoded in the quiver. As discussed above, this procedure is consistent provided one introduces the composite excitation $\gamma_{12} = \gamma_{1}+\gamma_{2}$, with the ordering~\eqref{eq:wall-crossed-order}.

We then incorporate the remaining state $\gamma_{3}$. Its couplings to $\gamma_{1}$ and $\gamma_{2}$ follow directly from the quiver, $\langle \gamma_{3}, \gamma_{1} \rangle = -\langle \gamma_{3}, \gamma_{2} \rangle = 1$. For $\gamma_{12}$, we use its decomposition in terms of $\gamma_{1}$ and $\gamma_{2}$, leading to the coupling
\be\label{eq:decoupling-A3}
\langle \gamma_{12}, \gamma_{3}\rangle = \langle \gamma_{1}+\gamma_{2}, \gamma_{3}\rangle = 0\, .
\ee
Thus, $\gamma_{3}$ does not couple to $\gamma_{12}$. This contrasts with the ultra-local interactions arising in the Feynman-like prescription~\eqref{eq:threemagcontact}; here no such issue appears, and the decoupling~\eqref{eq:decoupling-A3} is consistent with the OPE limit shown in Figure~\ref{fig:hexagonOPElimit}.

The local ordering constraints must still be satisfied for the contour of $\gamma_{3}$ relative to those of $\gamma_{1}$ and $\gamma_{2}$. This can be achieved without further wall-crossing by placing the contour $\cC_{\gamma_3}$ between $\cC_{\gamma_{1}}$ and $\cC_{\gamma_{2}}$. The resulting configuration is then consistent with the decoupling limits in all OPE channels.

The corresponding TBA equations, $Y$-system, and free energy take the usual forms for the partially ordered set
\be\label{eq:order-cyclic-A3}
\mathcal{B}
=\{\,
\gamma_{2} \succ \gamma_{3},\;
\gamma_{12} \succ \gamma_{1}
\,\}\, .
\ee
Since $\langle \gamma_{12},\gamma_3 \rangle=0$, the functions $Y_{\gamma_{12}}$ and $Y_{\gamma_{3}}$ do not enter each other’s TBA equations, and the relative ordering of the contours $\cC_{\gamma_{12}}$ and $\cC_{\gamma_{3}}$ is unconstrained.

An important consequence of this analysis is that a triangulation containing a 3-cycle cannot support a consistent system built solely from the three elementary states $\gamma_{1}, \gamma_{2}, \gamma_{3}$. The wall-crossing analysis instead shows that at least four states are required, with one possible partially ordered set given in~\eqref{eq:order-cyclic-A3}. This observation is further supported by considering the flip from the \(A_3\) fan quiver to the \(A_3\) cyclic quiver, as discussed in Appendix~\ref{app:comparison-GMN}.

Other poset structures can be obtained by further reorganizing the contours through additional wall-crossing transformations. This makes it possible to verify that the full system is, in fact, cyclically symmetric, as suggested by the symmetry of the triangulation. A detailed discussion is deferred to Section~\ref{sec:chi-system-lin-quivers}, where this property is shown to follow from the Kontsevich–Soibelman wall-crossing formula.

\subsection{General solution to state ordering}
\label{sec:generalsolution}

\begin{figure}[h]
\centering
\includegraphics[width=6cm]{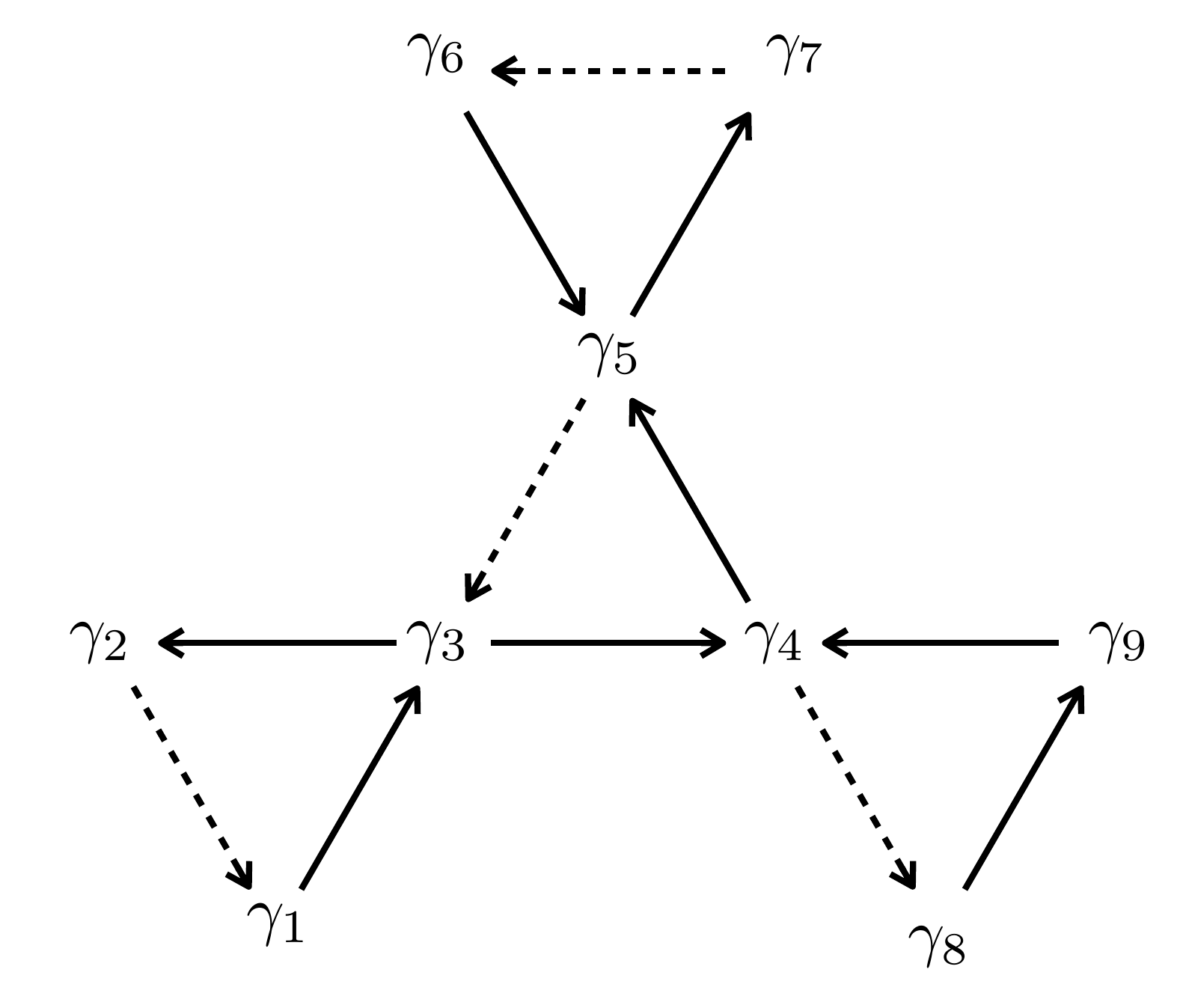}
\caption{Example of a quiver with several 3-cycles appearing in the triangulation of a polygon ($n=12$). The 3-cycles form a tree-like structure. To establish a partial ordering of the nodes, one arrow is removed from each 3-cycle (dashed arrows). The removed arrows are chosen to have disjoint endpoints.}
\label{fig:3-cycles}
\end{figure}

Motivated by the preceding discussion, we now present a general procedure for associating a poset of states to an arbitrary triangulation of a polygon. For quivers arising from polygon triangulations, the most complicated cyclic substructures are tree-like configurations of corner-sharing 3-cycles, as illustrated in Figure~\ref{fig:3-cycles}. More elaborate cases, such as closed chains of 3-cycles or longer oriented cycles, are instead associated with punctures. Such configurations will appear later, when we study correlation functions on the sphere.

The tree-like structure ensures that the elementary wall-crossing operation discussed for the $A_3$ quiver can be applied systematically to any polygon triangulation.
For each 3-cycle, we select a pair of nodes---equivalently, an arrow in the quiver---on which to perform the wall-crossing transformation. These choices are made so that no node is selected more than once, ensuring that the transformations act on disconnected subgraphs. As a result, the pairings between composite states vanish, hence no arrows are generated between them.

\begin{figure}[h]
\centering
\includegraphics[width=13cm]{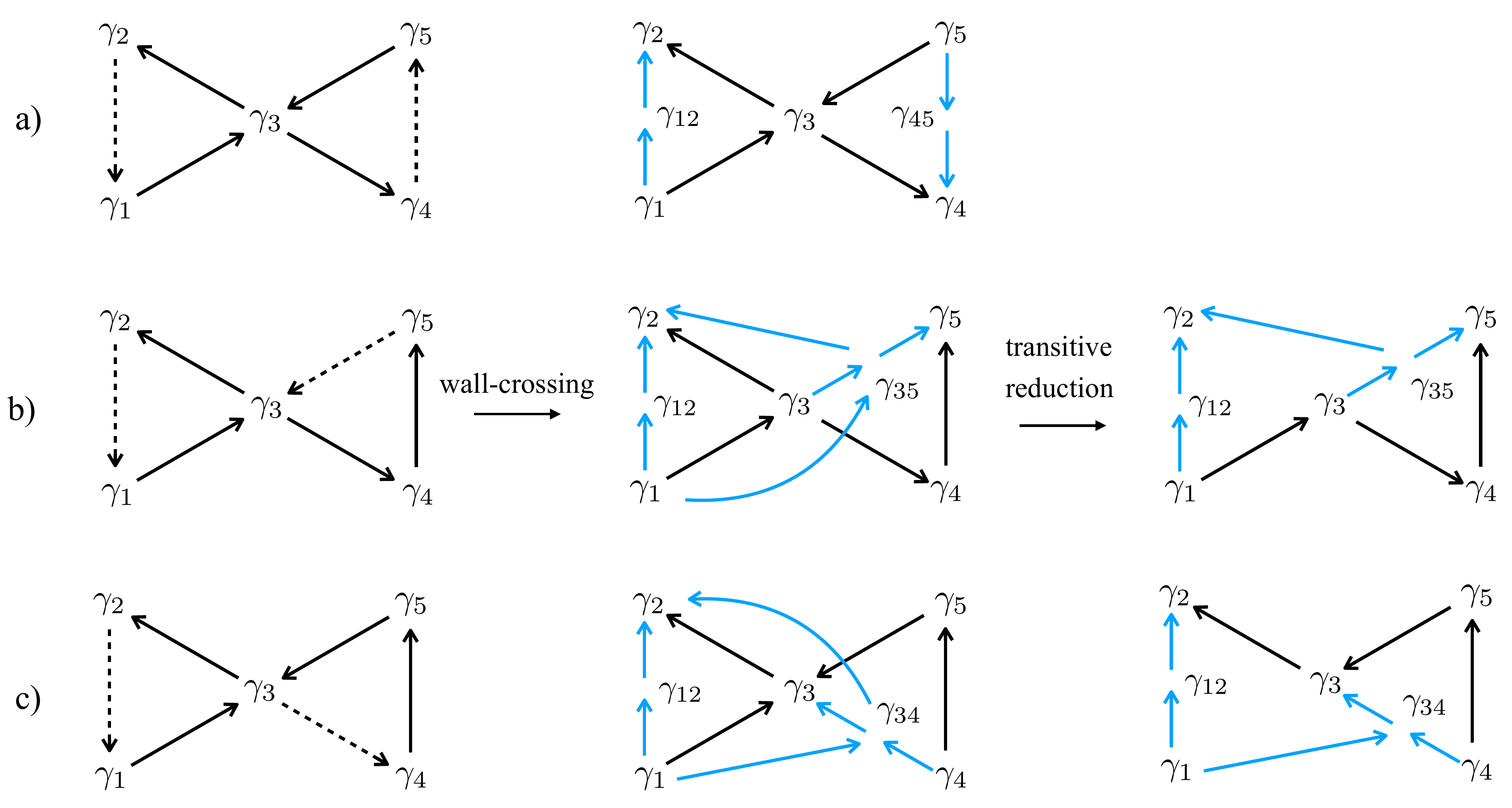}
\caption{The three basic wall-crossing transformations for two 3-cycles. {\bf{Left}}: Dotted arrows indicate the choice of edges on which the wall-crossing is performed. {\bf{Middle:}} Blue arrows indicate the connections involving the composite states after wall-crossing. {\bf{Right:}} Transitive reduction (removal of redundant arrows) of the directed acyclic graph obtained after wall-crossing. The resulting graphs encode the corresponding posets.}
\label{fig:DAG}
\end{figure}

The resulting transformation proceeds as follows: 
\begin{enumerate}
\item For each 3-cycle, remove the selected arrow.
\item Perform the wall-crossing transformation, then reintroduce each removed arrow with reversed orientation and a composite node inserted along it.
\item For each composite node $\gamma_{ij} = \gamma_{i}+\gamma_{j}$, add arrows between $\gamma_{ij}$ and the remaining nodes $\{\gamma\}$, oriented so that an arrow points from $\alpha$ to $\beta$ whenever $\langle \beta, \alpha \rangle = +1$.
\end{enumerate}
This construction produces a directed acyclic graph, and hence a partial order on the set of states. Figure~\ref{fig:DAG} illustrates the mechanism for the case of two 3-cycles and the corresponding three basic choices of non-overlapping wall-crossing transformations. The middle panel shows the connection pattern between nodes, with the newly formed connections highlighted in blue. These connections correspond to non-vanishing pairings \(\langle \gamma,\gamma' \rangle\), and therefore to terms that appear in the respective TBA equations.

Since we are only interested in the induced partial order, the right panel displays the \emph{transitive reduction} of the graph, obtained by removing redundant arrows~\cite{AhoGareyUllman1972}. In this way, we obtain the Hasse diagram associated with the directed acyclic graph. The net effect is that each 3-cycle is replaced by either an unoriented 4-cycle or an unoriented 5-cycle, depending on the relative configuration.

The resulting poset can then be used to define the TBA equations associated with the triangulation. In practice, it is convenient to choose any total ordering that is compatible with this partial order.%
\footnote{In graph theory, this step corresponds to choosing a \emph{topological ordering}, i.e., a (totally-ordered) linear sequence in which each element appears in a way that respects the partial order encoded by the directed acyclic graph.} For the three cases shown in Figure~\ref{fig:DAG}, one such choice is $\{\gamma_{2}\succ \gamma_{12} \succ \cB' \succ \gamma_{1}\}$, where $\cB'$ denotes an ordering of the 3-cycle on the right,
\be\label{eq:B-prime-abc}
\cB'_{a} = \{\gamma_{4} \succ \gamma_{45} \succ \gamma_{3} \succ \gamma_{5}\}\, ,
\ee
with $\gamma_{45} = \gamma_{4}+\gamma_{5}$. The orderings $\cB'_{b}$ and $\cB'_{c}$ are defined analogously by cyclic permutations of the indices $4\rightarrow 5\rightarrow 3\rightarrow 4$.
Other compatible orderings can be obtained by permuting elements of $\cB$ with vanishing pairings. For example, the condition $\gamma_{12} \succ \cB'$ may be relaxed, since $\gamma_{12}$ has zero pairing with all elements of $\cB'$, i.e.~$\langle \gamma_{12}, \gamma'\rangle = 0, \forall \gamma'\in \cB'$. More generally, it is only the poset itself---namely, the equivalence class of compatible total orderings---that is relevant for defining the TBA equations.

In this form, the construction extends recursively to larger configurations.
\begin{figure}[h]
\centering
\includegraphics[width=9cm]{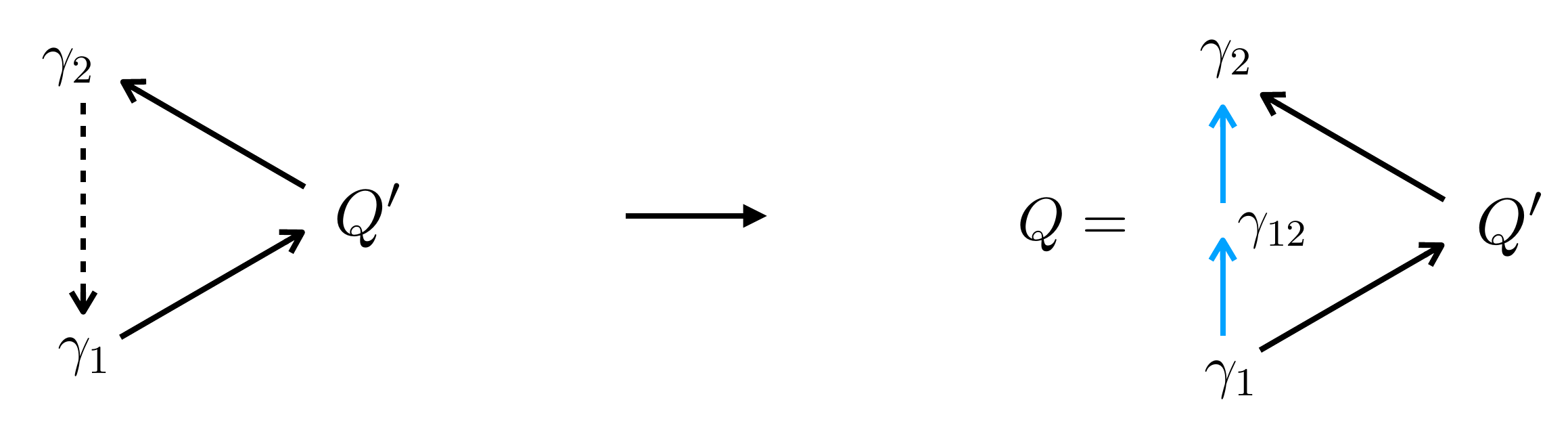}
\caption{Recursive structure of a tree-like quiver of the type shown in Figure~\ref{fig:3-cycles}. By applying wall-crossing transformations successively to the leaves of the tree graph, one constructs an acyclic graph that admits an ordering satisfying the recurrence relation~\eqref{eq:Brec}. Wall-crossing transformations inside $Q'$ can introduce additional blue arrows between $Q'$ and $\gamma_1$ or $\gamma_2$, as illustrated in Figure~\ref{fig:DAG}. These arrows are important for determining the full poset structure but do not affect the compatibility of the chosen ordering.  }
\label{fig:RecurrenceQ}
\end{figure}
Starting from a tree of 3-cycles $Q$,
\be\label{eq:Qrec}
Q = Q' \rightarrow \gamma_{2} \rightarrow \gamma_{1} \rightarrow  Q'\, ,
\ee
as in Figure~\ref{fig:RecurrenceQ}, one removes the leaf $\gamma_{1} \rightarrow \gamma_{2}$ using the ordering
\be\label{eq:Brec}
\cB_{Q} = \{\gamma_{2}\succ \gamma_{12} \succ \cB_{Q'} \succ \gamma_{1}\}\, .
\ee
Iterating this procedure on the leaf-stripped quiver $Q'$ eventually eliminates all pairs to be wall-crossed, yielding a total ordering of the full system.

The resulting poset is minimal in the sense that a single composite state is introduced to resolve the ordering constraint within each 3-cycle. As illustrated in Figure~\ref{fig:DAG}, the construction is not unique: any choice obtained by deleting non-overlapping arrows provides an equally valid starting point for defining the TBA system. Although the equivalence of these constructions is not immediately obvious, it ultimately follows from wall-crossing. We return to this point below.

\section{$\chi$-system formulation}\label{sec:chi-system-lin-quivers}

As discussed above, a consistent ordering of states for a general triangulation can be constructed via wall-crossing transformations. The resulting equations, however, are not unique, and establishing their equivalence becomes increasingly cumbersome for large quivers.

A more invariant formulation is obtained by following the framework of Gaiotto, Moore, and Neitzke, developed for the TBA equations arising from the Hitchin-system description of 4D $\mathcal{N}=2$ theories~\cite{Gaiotto:2010okc}. The emphasis is on the discontinuities of the TBA equations, which induce the action of Kontsevich–Soibelman (KS) transformations~\cite{Kontsevich:2008fj} on a set of $\chi$-functions serving as Fock-Goncharov coordinates on the Hitchin moduli space. In the GMN framework, these transformations occur across rays encoding the spectrum of BPS states in a given chamber---i.e.~a region of the parameter space where the spectrum is constant. While the spectrum itself changes between chambers due to wall-crossing, the total discontinuity remains the same, providing an intrinsic description of the system.%
\footnote{This formalism is closely related to the exact WKB analysis of Schr\"odinger-like equations. In that setting, the role of the $\chi$–variables is played by the Voros symbols \cite{iwaki2014}, and the KS transformations are realized as Stokes automorphisms describing how these symbols jump when a Stokes line is crossed.}

The same structure underlies our equations, with $\cB$ playing the role of the chamber in the $\mathcal{N}=2$ setting. The system is governed by the same KS transformations acting on suitably defined $\chi$-functions. We explain below how these functions map to the $Y$-functions constructed earlier and, following~\cite{Gaiotto:2010okc}, derive the $\chi$-system form of the TBA equations from the spectrum generator, which encodes the total discontinuity.

This formulation bypasses the need to construct a specific poset/chamber and can instead be read off directly from the geometric data of the triangulation. It also makes clear that our TBA equations are essentially equivalent to those of GMN, as they share the same discontinuity structure. As discussed in Appendix~\ref{app:comparison-GMN}, the differences are mainly cosmetic: after suitable changes of variables and redefinitions, the two formulations coincide, both for AdS and for the sphere considered separately.

\subsection{From $Y$-functions to $\chi$-functions}\label{sec:KS-transformation}

To handle the singularity structure more efficiently, it is convenient, following~\cite{Gaiotto:2010okc}, to introduce piecewise continuous functions $X_{\gamma}$ that solve the same TBA equations everywhere in the fundamental strip except along the integration contours. That is, we define
\be\label{eq:def-chi}
\log{X_{\gamma}(\theta)} = \log{\sfchi_{\gamma}(\theta)} +\sum_{\gamma'\, \in\, \mathcal{B}}\langle \gamma, \gamma'\rangle \left[G * \log(1+X_{\gamma'}(\theta))+\bar G * \log(1+\bar X_{\gamma'}(\theta))\right]\,,
\ee
and similarly for $\bar{X}_{\gamma}$, for \textit{all} $\theta \notin\{\cC_{\gamma'}, \gamma' \neq \gamma\}$, with the convolutions defined along the same contours as before.

To make the structure more transparent, one may further separate the contours by assigning them a finite but arbitrary spacing, thereby opening finite domains between consecutive contours. This deformation does not affect the equations, provided the contours remain within the fundamental strip and assuming the integrals remain convergent. It also singles out two distinguished domains, denoted by $\mathscr{D}_{\pm}$, located above and below all the contours.

So defined, $X_{\gamma}$ is continuous in each domain separately, with jumps across the contours, unlike the $Y$-functions which are defined globally by analytic continuation. Since $X_{\gamma}$ and $Y_{\gamma}$ satisfy the same TBA equation along $\cC_{\gamma}$, they coincide in a neighborhood of that contour. In other domains, however, these functions generally differ. In particular, the analytic continuation of $Y_{\gamma}$ to the upper domain $\mathscr{D}_{+}$ does not agree with the value of $X_{\gamma}$ obtained there using~\eqref{eq:def-chi}. We denote this upper determination by $\chi_{\gamma}$. Like the $Y$-functions, the $\chi$-functions can then be extended to the full strip by analytic continuation.

\begin{figure}[h]
    \centering
    \includegraphics[width=0.5\linewidth]{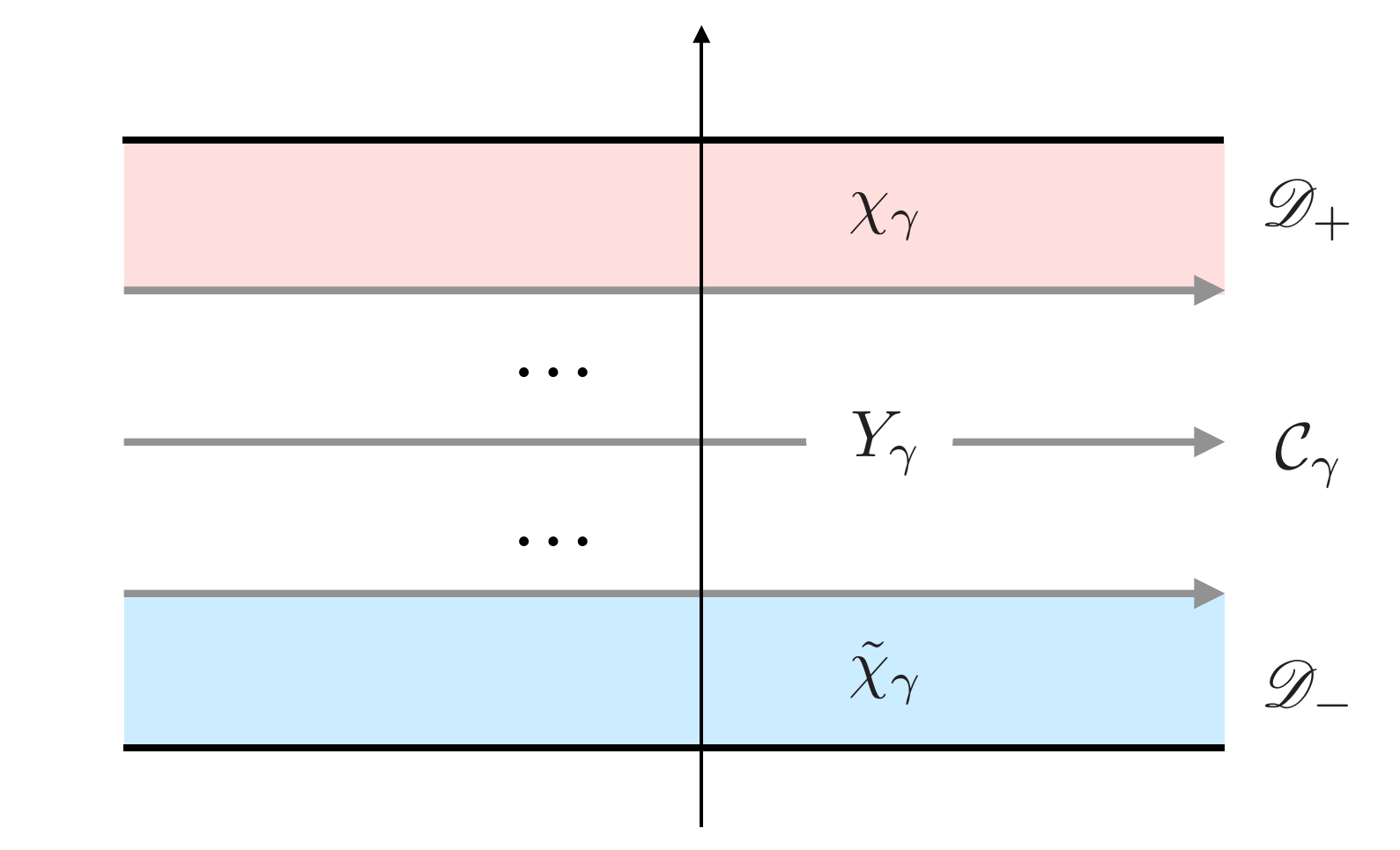}
    \caption{Sketch of the integration contours and domains, with $\mathscr{D}_{\pm}$ located at the top and bottom of the fundamental strip. The function $Y_{\gamma}$ is initially defined in a neighborhood of its contour $\cC_{\gamma}$ and then analytically continued. The function $\chi_{\gamma}$ is defined as the solution to the TBA equations in the upper domain, and similarly for $\tilde{\chi}_{\gamma}$ in the lower domain.}
    \label{fig:KS_jumps}
\end{figure}

Thus, the analytic $Y$- and $\chi$-functions can be viewed as different determinations of the same underlying discontinuous functions $\{X_{\gamma}\}$,
\be\label{eq:main-determinations}
{\chi}_\gamma \equiv X_{\gamma}\big|_{\mathscr{D}_{+}}\,,\qquad
{Y}_\gamma \equiv X_{\gamma}\big|_{\cC_\gamma}\,  ,
\ee
as illustrated in Figure~\ref{fig:KS_jumps}.
Since the original TBA equations are written in terms of the $Y$-variables, the hexagonalization procedure effectively selects this determination. As we will see, however, the $\chi$-determination leads to a more geometric formulation of these equations, in line with the string-theory description~\cite{Caetano:2012ac}.

A useful property of the $\chi$-functions, and more generally of the $X$-functions, is the multiplicative relation
\be\label{eq:multiplication-rule}
\chi_{\gamma+\gamma^{\prime}}=\chi_{\gamma}\chi_{\gamma^{\prime}}\,, 
\ee
which follows from the linearity in $\gamma$ of the TBA interactions~\eqref{eq:def-chi}, together with the same property for the driving terms $\sfchi_{\gamma}$. Bound-state fusion is thus encoded simply as multiplication in the $\chi$-variables, in contrast with the more involved composition rules for the $Y$-variables (see e.g.~\eqref{eq:newcompositeYfunction} for $A_2$).

This structure mirrors the behavior of cross ratios, $z_{\gamma+\gamma'} = z_{\gamma}z_{\gamma'}$, and reflects the Fock-Goncharov nature of the $\chi$-functions~\cite{Gaiotto:2010okc,Caetano:2012ac}. These functions also satisfy the inversion property, $\chi_{-\gamma}=1/\chi_{\gamma}$, where $-\gamma$ denotes the negative of the state $\gamma$ in the GMN framework. Although such states play no role in the present analysis, they are useful when discussing general properties of the $\chi$-functions, in particular their transformations under flips. We refer the reader to Appendix~\ref{app:comparison-GMN} for a discussion of the mutation properties of the $\chi$-functions, as well as a proof of the resulting flip invariance of the correlation functions.

We can now focus on the value of the discontinuities
 of the functions $\{X_\gamma\}$. Crossing the contour $\cC_{\gamma'}$ relates the value of $X_{\gamma}$ \textit{below} the contour to its value \textit{above} via the KS transformation
\begin{align}
\label{KSdefinition}
\cK_{\gamma'}: \qquad X_{\gamma} \,\,\, \mapsto \,\,\, X_{\gamma} (1+X_{\gamma'})^{\langle \gamma', \gamma\rangle}\, .
\end{align}
This follows from the residue analysis of~\eqref{eq:def-chi} and becomes trivial when $\langle \gamma, \gamma'\rangle = 0$, i.e.~when there is no mutual pole in the equations. More generally, the KS action extends to any rational function $F(\{X_{\gamma}\})$ via
\be\label{eq:extension-KS}
\cK_{\gamma^{\prime}} F(\{X_\gamma\})=F(\{\cK_{\gamma^{\prime}} X_{\gamma}\})\,,
\ee
enabling to write successive crossings as ordered products of KS factors. For instance, crossing a second contour $\cC_{\gamma''}$ located just below $\cC_{\gamma'}$ yields $\cK_{\gamma'}\cK_{\gamma''}X_{\gamma}$, and so on.%
\footnote{This ordering may appear counterintuitive, since one first crosses $\gamma'$ and then $\gamma''$. However, it correctly reflects the implementation of the KS action and its extension in~\eqref{eq:extension-KS}. In particular, using~\eqref{eq:extension-KS}, one may equivalently write \(\cK_{\gamma'}\cK_{\gamma''}X_{\gamma}= \cK_{\gamma''}(\cK_{\gamma'}(X_{\gamma}))\), as expected.
}

Lastly, this structure allows one to express the $Y$-functions in terms of the $\chi$-functions in~\eqref{eq:main-determinations},
\be
\label{Y-chi-sys}
Y_{\gamma}
   = \ordprod_{\gamma' \succ \gamma}
      \mathcal{K}_{\gamma'}\ \chi_\gamma\, ,
\ee
where the product is ordered from the largest to the smallest element in the chamber $\cB$. This relation can be inverted to give
\begin{align}
\label{eq:chi-Y-sys}
\chi_\gamma=Y_\gamma\,\prod_{\gamma'\succ\gamma}(1+Y_{\gamma'})^{\langle\gamma,\gamma'\rangle}\,.
\end{align}
As in the derivation of the $Y$-system in Section~\ref{sec:Y-sys}, this result follows from analytically continuing the $Y$-functions to the upper domain $\mathscr{D}_{+}$, collecting residues at the poles encountered when crossing the contours $\cC_{\gamma'}$ with $\gamma'\succ \gamma$. Alternatively, it can be derived algebraically from~\eqref{Y-chi-sys} by computing
\be
\label{eq:chi-Y-inv}
\ordprod_{\gamma'\succ\gamma}\left[\cK_{\gamma'}(1+\chi_{\gamma'})^{\langle\gamma,\gamma'\rangle}\right]\chi_\gamma
\ee
in two different ways. The left-hand side of~\eqref{eq:chi-Y-sys} is obtained by commuting $\chi_\gamma$ to the left through the KS operators, canceling all factors $(1+\chi_{\gamma'})$. The right-hand side follows by moving these factors to the left, where they are converted into $(1+Y_{\gamma'})$ via~\eqref{Y-chi-sys}.

\subsection{Spectrum generator and KS wall-crossing formula}\label{sec:chi-system}

As mentioned above, is also useful to consider the total jump obtained by traversing all contours in \(\mathcal{B}\). The corresponding ordered product of KS factors defines the
spectrum generator
\be
\label{eq:specgen}
\boldsymbol{S} = \ordprod_{\gamma\, \in\, \mathcal{B}} \mathcal{K}_\gamma\, ,
\ee
which encodes the full BPS spectrum in the GMN framework~\cite{Gaiotto:2010okc,Gaiotto:2009hg}. Its main property is invariance under wall-crossing. While permuting the integration contours may introduce new domains through composite $Y$-functions (see Figure~\ref{fig:crossing-contours}), the total discontinuity---and hence $\boldsymbol{S}$---remains unchanged. As such, $\boldsymbol{S}$ depends only on the quiver, or equivalently the triangulation, and not on a specific choice of chamber.

This invariance is encoded in the Kontsevich–Soibelman wall-crossing formula. In the simplest $A_2$ case, it reduces to the so-called pentagon identity~\cite{Kontsevich:2008fj,Gaiotto:2010okc,Gaiotto:2009hg}
\be\label{eq:pent-id}
\cK_{\gamma_{i}}\, \cK_{\gamma_{j}}
=
\cK_{\gamma_{j}}\,
\cK_{\gamma_{i}+\gamma_{j}}\,
\cK_{\gamma_{i}}\, ,
\qquad
\text{if}\qquad
\langle \gamma_{i}, \gamma_{j} \rangle = +1\, ,
\ee
which follows directly from the definition of the KS transformations~\eqref{KSdefinition}. If instead $\langle \gamma_{i}, \gamma_{j} \rangle = 0$, the operators commute
\be\label{eq:noWC}
\cK_{\gamma_{i}}\, \cK_{\gamma_{j}}
= 
\cK_{\gamma_{j}}\, \cK_{\gamma_{i}}\, ,
\ee
reflecting the absence of interactions. More general pairings ($|\langle \gamma_i, \gamma_j \rangle|>1$) lead to more involved relations. However, these do not arise for polygons and are typically associated with punctured geometries. Thus, for polygons, all chambers can be generated purely by reordering the KS factors in $\boldsymbol{S}$ using the two basic relations~\eqref{eq:pent-id}--\eqref{eq:noWC}.

As an application, we can confirm that our construction for the cyclic $A_3$ quiver is indeed minimal by inspecting all possible chambers. In fact, it is known that, up to cyclic relabeling, only two types of chambers can arise in this case~\cite{Alim:2011kw,Gaiotto:2010okc}, namely
\be\label{eq:S-4and5-states}
\boldsymbol{S} = \cK_{\gamma_{2}}\cK_{\gamma_{12}}\cK_{\gamma_{3}} \cK_{\gamma_{1}} =\cK_{\gamma_{2}}\cK_{\gamma_{12}}\cK_{\gamma_{1}} \cK_{\gamma_{13}} \cK_{\gamma_{3}}\, ,
\ee
with $\gamma_{12} = \gamma_{1}+\gamma_{2}$ and $\gamma_{13} = \gamma_{1}+\gamma_{3}$. The first decomposition corresponds to a minimal (four-state) chamber, while the second describes a maximal (five-state) chamber. Using the pentagon identity, one also verifies that $\boldsymbol{S}$ is invariant under cyclic permutations of $\gamma_{1,2,3}$, as expected for this quiver.%
\footnote{For instance, applying $\cK_{\gamma_{2}}\cK_{\gamma_{12}}\cK_{\gamma_{1}} = \cK_{\gamma_{1}}\cK_{\gamma_{2}}$ to the final term in~\eqref{eq:S-4and5-states}, and using that $\cK_{\gamma_{2}}\cK_{\gamma_{13}} = \cK_{\gamma_{13}}\cK_{\gamma_2}$, yields a cyclic image of the middle term in~\eqref{eq:S-4and5-states}.} 

Similarly, the three orderings in~\eqref{eq:B-prime-abc} yield equivalent spectrum generators,
\be\label{eq:Sabc}
\boldsymbol{S}'_a = \boldsymbol{S}'_b = \boldsymbol{S}'_c\, .
\ee
This extends to higher quivers via the recursion relation for polygons~\eqref{eq:Qrec}, which can now be expressed compactly as
\be\label{eq:recursion-S}
\boldsymbol{S} = \cK_{\gamma_{2}}\cK_{\gamma_{12}} \boldsymbol{S}' \cK_{\gamma_{1}} = \cK_{\gamma_{2}} \boldsymbol{S}' \cK_{\gamma_{12}} \cK_{\gamma_{1}}\, ,
\ee
where $\boldsymbol{S}'$ is the spectrum generator for the leaf-stripped quiver $Q'$, and $\boldsymbol{S}$ that of the full quiver $Q$. In the second expression, we used that the composite charge $\gamma_{12}$ decouples from all elements in $Q'$, allowing the KS factors to be reordered. This recursively establishes the equivalence of the various minimal chambers.

\subsection{$\chi$-system}\label{sec:chi-system}
 
These properties make $\boldsymbol{S}$ a natural starting point for deriving a more invariant form of the TBA equations. Acting with $\boldsymbol{S}$ on $\chi_\gamma$ yields the determination of $X_\gamma$ in the lower domain $\mathscr{D}_{-}$,
\be\label{eq:tilde-chi}
\tilde{\chi}_\gamma \equiv X_{\gamma}\big|_{\mathscr{D}_{-}} = \boldsymbol{S}(\chi_\gamma)\,  ,
\ee
as illustrated in Figure~\ref{fig:KS_jumps}. A key feature of this lower determination is that it is simply related to the upper one through the shift $\theta\rightarrow \theta+i\pi$. Denoting by $\bar{\chi}_\gamma^{++}$ the analytic continuation of $\bar{\chi}_\gamma$ under this shift, one finds
\be\label{eq:tilde-chi-chi++}
\tilde{\chi}_{\gamma} = 1/\bar{\chi}^{++}_{\gamma}\, ,
\ee
and similarly with $\chi\leftrightarrow \bar{\chi}$. As a result, the $\chi$-functions functions satisfy the $\chi$-system
\be\label{eq:chi-syste-S}
\chi_\gamma^{++}\boldsymbol{S}(\bar{\chi}_\gamma)= 1\,, \qquad \bar{\chi}_\gamma ^{++}\boldsymbol{S}({ \chi}_\gamma)=1\,.
\ee
As a cross-check, we verify its equivalence with the $Y$-system using~\eqref{eq:chi-Y-sys}, together with%
\footnote{This follows from computing $\ordprod_{\gamma'\in \cB}\, \big( \mathcal{K}_{\gamma'}\, (1+\chi _{\gamma'})^{\langle\gamma,\gamma'\rangle}\big)\ \chi_\gamma$ using the same method as that employed to derive~\eqref{eq:chi-Y-sys} from~\eqref{Y-chi-sys}.}
\be\label{eq:Ychisysgen}
\boldsymbol{S}(\chi_\gamma) =\chi_\gamma \prod_{\gamma'\, \in\, \cB}\,(1+Y_{\gamma'})^{\langle\gamma',\gamma\rangle}\, ,
\ee
and similarly for the barred functions.

A remarkable result of~\cite{Gaiotto:2010okc,Gaiotto:2014bza} is that $\boldsymbol{S}$ can be computed directly from the geometric data of the triangulation, independently of the chosen chamber. We present here the relevant expressions for polygons, which correspond to Argyres–Douglas theories in the terminology of~\cite{Gaiotto:2010okc} (see Section 11 therein). The case of closed geometries is presented in Section~\ref{sec:closed}.

\begin{figure}[h]
\centering
\includegraphics[width=8cm]{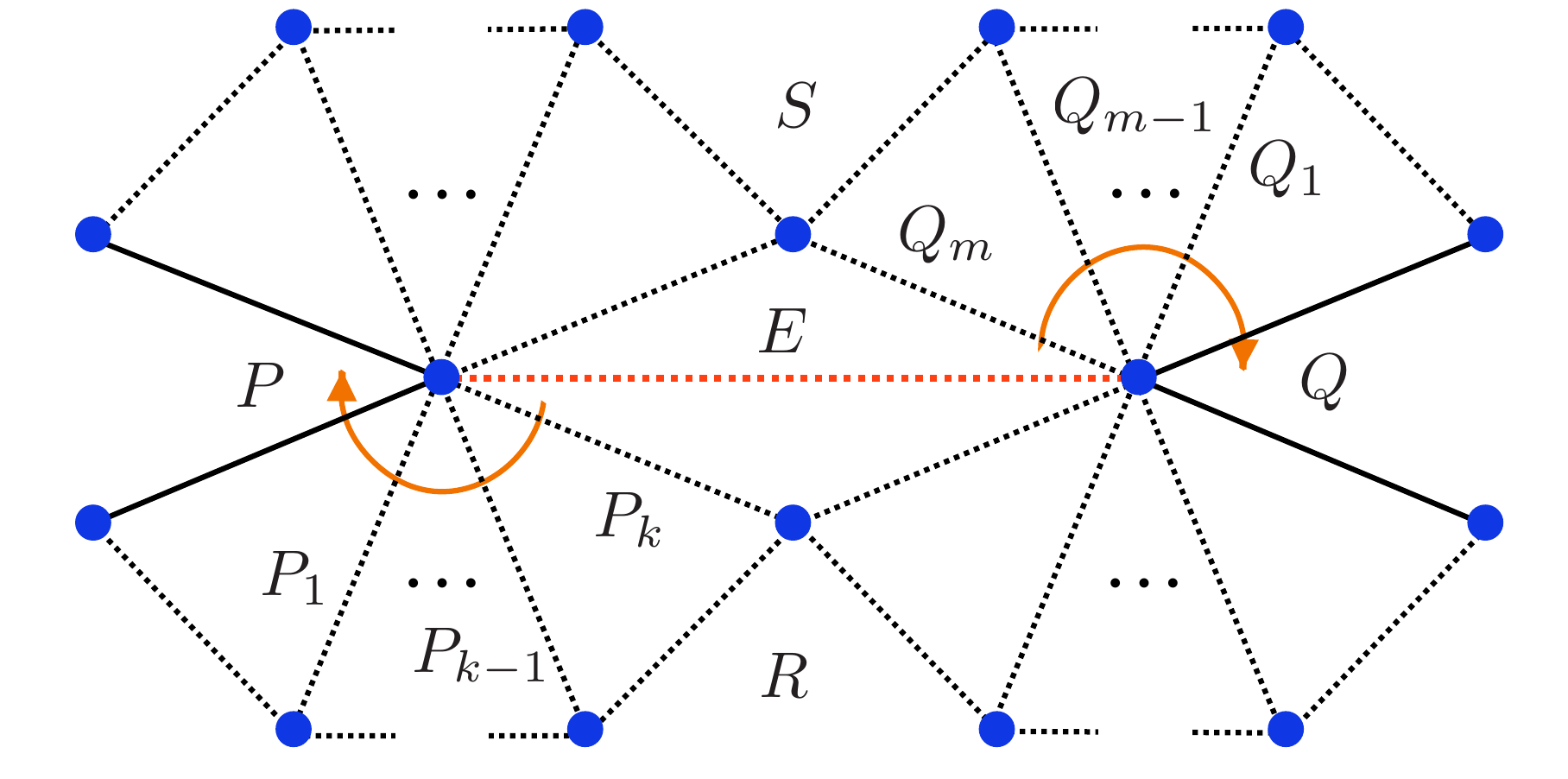}
\caption{Geometrical illustration of~\eqref{eq:A-to-chi} for $A_E\equiv A_{PQ}$ associated with the edge $E$ between the points $P$ and $Q$ (in red). The quadrilateral framing the edge $E$ is $PRQS$. The dotted lines denote internal edges and bold lines external ones. Orange arrows indicate quiver orientation; note that our convention differs from~\cite{Gaiotto:2009hg} by a clockwise orientation of the arrows. }
\label{fig:A-chi}
\end{figure}

To simplify the notation, we denote by $\chi_{E}$ the function associated with the charge $\gamma_E$ of the edge $E$, and write $\langle E, E'\rangle =\langle \gamma_{E}, \gamma_{E'}\rangle$. For the configuration in Figure \ref{fig:A-chi}, we associate to each edge $E$ the quantity
\begin{align} \label{eq:A-to-chi}
A_{E} &\equiv A_{PQ}=\\\nonumber (1+\chi_{P_k}(1+\chi_{P_{k-1}}(\ldots(1+\chi_{P_1})\ldots)))\ 
\chi_{E}&\ (1+\chi_{Q_m}(1+\chi_{Q_{m-1}}(\ldots(1+\chi_{Q_1})\ldots)))\, .
\end{align}
In these terms, the $\chi$-system \eqref{eq:chi-syste-S} can be rewritten as
\be\label{eq:chisysA}
\bar \chi^{++}_E\, \chi_E = \frac{\chi_E}{\boldsymbol{S}({\chi}_E)} =   \frac{(1+ A_{PS})(1+ A_{QR})}{(1+ A_{PR})(1+ A_{QS})}=\prod_{E'}(1+A_{E'})^{\langle E,E'\rangle}\,,
\ee
with similar equations under $(\chi_E,A_E)\leftrightarrow(\bar\chi_E,\bar A_E)$, in both the AdS and sphere sectors. The identification of the edges surrounding $E$ is given in Figure~\ref{fig:A-chi}.

A derivation of~(\ref{eq:chisysA}) in our setup can be done using the identity
\be\label{eq:A-to-Y}
1+A_{E} = 
\prod_{\gamma\, \ni\, \gamma_E} (1+Y_{\gamma})=\prod_\gamma (1+Y_{\gamma})^{\delta_{\gamma,E}}\qquad (\textrm{for polygons})\,,
\ee
where $\gamma=\sum_{E}\delta_{\gamma,E}\,\gamma_E$, $\gamma_E\in\gamma$ if $\delta_{\gamma,E}\neq 0$, and $\delta_{\gamma,E}\in \{0,1\}$ for polygons. Substituting this relation in~\eqref{eq:specgen} yields~\eqref{eq:chisysA}, using
\be
\prod_{\gamma'}(1+Y_{\gamma'})^{\langle E,\gamma'\rangle}=\prod_{E',\gamma'}(1+Y_{\gamma'})^{\langle E,E'\rangle \delta_{\gamma',E'}}=
\prod_{E'}(1+A_{E'})^{\langle E,E'\rangle}\,.
\ee
The identity~\eqref{eq:A-to-Y} is immediate in the canonical poset (minimal chamber) of a linear quiver, where each product reduces to a single factor $A_{E} = Y_{\gamma_{E}}$. It is also preserved under the basic wall-crossing transformation~\eqref{eq:pent-id}: the left-hand side of~\eqref{eq:A-to-Y} remains unchanged, while the right-hand side transforms as (see~\eqref{YtoYprime})
\be
\label{eq:WCij}
\begin{aligned}
(1+Y_{\gamma_{i}}) &= (1+Y'_{\gamma_{i}}) (1+Y'_{\gamma_{i}+\gamma_{j}})\, , \\
(1+Y_{\gamma_{j}}) &= (1+Y'_{\gamma_{j}}) (1+Y'_{\gamma_{i}+\gamma_{j}})\, ,
\end{aligned}
\ee
with all other $Y$-functions unchanged, $Y'_{\gamma_k}=Y_{\gamma_k}$ for $\gamma_k\neq \gamma_i,\gamma_j$. Iterating this transformation extends the validity of~\eqref{eq:A-to-Y} to any chamber of a linear quiver.

The general proof for quivers with cycles uses the recursion $\boldsymbol{S} = \cK_{\gamma_{2}}\cK_{\gamma_{1}+\gamma_{2}}\boldsymbol{S}'\cK_{\gamma_{1}}$. We outline it below for the interested reader. 

\paragraph{Proof of~\eqref{eq:A-to-Y}.} Using the notations from Figure~\ref{fig:RecurrenceQ}, assume that the identity holds for the subquiver $Q'$ associated with $\boldsymbol{S}'$ and let $\gamma_{3}$ denote the state associated with the node in $Q'$ that forms a 3-cycle with nodes $1$ and $2$. We are going to show that~\eqref{eq:A-to-Y} holds for all the nodes in the new quiver $Q$.

From the geometrical definition of the $A$-functions in~\eqref{eq:A-to-chi}, we have 
\be\label{eq:geomA}
A_{1} = \chi_{\gamma_1}\left(1+\frac{A_{3}}{1+\chi_{\gamma_{2}}}\right)\, , \quad A_{2} = \chi_{\gamma_{2}}(1+\chi_{\gamma_{1}})\,,\quad A_E=A_E^{Q'}\big|_{\chi_{\gamma_{3}}\rightarrow \chi_{\gamma_{3}}(1+\chi_{\gamma_{2}})}\,,
\ee
where $A_E^{Q'}$ denotes the restriction of $A_E$ to $Q'$, obtained by setting $\chi_{\gamma_1}=\chi_{\gamma_2}=0$. The KS transformations immediately give
\be
Y_{\gamma_{2}} = \chi_{\gamma_{2}}\, , \qquad Y_{\gamma_{12}} = \cK_{\gamma_{2}}\chi_{\gamma_{12}} = \chi_{\gamma_{1}}\chi_{\gamma_{2}} (1+\chi_{\gamma_{2}})^{-1}\, ,
\ee
using $\gamma_{12} = \gamma_{1}+\gamma_{2}$ and $\langle \gamma_{1}, \gamma_{2}\rangle = 1$. This implies
\be
(1+Y_{\gamma_{2}}) (1+Y_{\gamma_{12}}) = 1+\chi_{\gamma_{2}}(1+\chi_{\gamma_{1}})=(1+A_2)\,,
\ee
and therefore proves~\eqref{eq:A-to-Y} for node $2$. Next, we observe that the KS transformations associated with the states $\gamma_2$ and $\gamma_{12}$ act on $\boldsymbol{S}'$ only through $\gamma_{3}$, via
\be
\label{eq:replacegamma3}
\chi_{\gamma_{3}} \rightarrow \cK_{\gamma_{2}} \cK_{\gamma_{12}} \chi_{\gamma_{3}} = \chi_{\gamma_{3}} (1+\chi_{\gamma_{2}})\, .
\ee
This allows one to rewrite the last equality in~\eqref{eq:geomA} as
\be
A_E=\cK_{\gamma_{2}} \cK_{\gamma_{12}}\, A_E^{Q'}\,,
\ee
for all $E$ in $Q'$. Together with $Y_{\gamma} = \cK_{\gamma_{1}}\cK_{\gamma_{12}}Y^{Q'}_{\gamma}$, this implies that the relations between $A$ and $Y$ are preserved when embedding $Q'$ in the larger quiver $Q$, and establishes the relation~\eqref{eq:A-to-Y} for all nodes inside $Q^\prime$.

Finally, to prove the identity for node $1$, suppose we choose an ordering for $\boldsymbol{S}'$ in which $\gamma_3$ does not appear in any composite state, such that
\be\label{eq:A3Y3}
A_3=Y_{\gamma_{3}}=\ordprod_{\gamma\succ\gamma_3}\cK_\gamma\, \chi_{\gamma_3}\, .
\ee
This means that  $\cK_{\gamma_{3}}$ is the only operator in $\boldsymbol{S}'$ acting nontrivially on $\chi_{\gamma_{1}}$, with $\cK_{\gamma_{3}}\chi_{\gamma_1} = \chi_{\gamma_1}(1+\chi_{\gamma_3})$. This implies
 \be
Y_{\gamma_{1}} = \cK_{\gamma_{2}}\cK_{\gamma_{12}} \boldsymbol{S}' \chi_{\gamma_{1}} = \ordprod_{\gamma\succ\gamma_3}\cK_\gamma \cK_{\gamma_{3}}\chi_{\gamma_{1}}= (1+A_{3})\left(\cK_{\gamma_{2}}\cK_{\gamma_{12}} \chi_{\gamma_{1}}\right)  = (1+A_{3})\frac{\chi_{\gamma_{1}}}{1+A_{2}}\, ,
\ee
and reproduces the sought-off relation
\be
(1+Y_{\gamma_{1}}) (1+Y_{\gamma_{12}}) = 1 + \chi_ {\gamma_ 1}\left(1+\frac{A_3}{1+\chi_{\gamma_2}}\right)=1+A_1\,.
\ee
Since any other chamber can be reached by a sequence of wall-crossing transformations starting from the present one, this concludes the proof for polygons.

\subsection{Integral equations}\label{sec:integral-equations}

The $\chi$-system has a simpler difference structure than the corresponding $Y$-system, since each equation involves only a single shifted $\chi$-function. This makes it straightforward to recast it as a set of integral equations using the standard inversion procedure~\cite{Alday:2010vh,Caetano:2012ac}. Taking the logarithm of~\eqref{eq:chisysA} and performing a Fourier transform, assuming analyticity in the fundamental strip, yields the integral equations
\be\label{eq:chi-system-integral-form}
\log{\chi_{E}(\theta)} = \log{\sfchi_{E}(\theta)} +\sum_{E'}\langle E, E'\rangle \left[\GMN*\log{(1+A_{E'})} + \bGMN*\log{(1+\bar{A}_{E'})}\right]\, ,
\ee
together with the analogous equation obtained by exchanging $\chi_E \leftrightarrow\bar \chi_E$.

These equations generalize those studied in~\cite{Caetano:2012ac} (see also~\cite{Gaiotto:2014bza}) for correlation functions in 1d kinematics, which correspond in our notation to the identification $\bar{\chi} = \chi$ (see Section~\ref{sec:closed}). In this formulation, no contour ordering is required: all convolution contours can be chosen with imaginary parts slightly below that of $\theta$ (i.e.~using $G(\theta-\theta'+i0)$ in \eqref{eq:chi-system-integral-form}).

The inversion procedure leaves the driving term $\boldsymbol{\chi}_{E}$ undetermined. It can be any zero-mode solution of the homogeneous equation
\be
\sfchi^{++}_{E} \bsfchi_{E} = 1\, ,
\ee
and similarly for $\sfchi_{E} \leftrightarrow \bsfchi_{E}$. Consistency with the previous TBA equations requires
\be\label{eq:chi-zero-mode}
\boldsymbol{\chi}_{E} = C_{E}+Z_{E} e^{\theta}+\bar{Z}_{E} e^{-\theta}\,,
\ee
where $Z_{E}$ and $\bar{Z}_{E}$ are flavor independent.

The hexagon data---bridge lengths and cross ratios---are encoded in these zero modes and can be extracted by evaluating the $\chi$-functions at special values of $\theta$. In particular, at $\theta = i\pi/2$, the $\chi$-functions are directly related to cross-ratios. In the AdS sector,
\be\label{eq:chi-ads-special-point}
\chi_{E}^{\AdS}(i\pi/2) = z_{E}\, , \qquad \bar{\chi}_{E}^{\AdS}(i\pi/2) = \bar{z}_{E}\, ,
\ee
where $z_{E}$ is the cross ratio associated with edge $E$. In the sphere sector, the result depends on the dressing of scalar excitations,
\be\label{eq:chi-sph-special-point}
\chi_{E}^{\Sph}(i\pi/2) = \alpha_{E}\, \qU_{E}^{-(1-\qw)}\, , \qquad \bar{\chi}_{E}^{\Sph}(i\pi/2) = \bar{\alpha}_{E}\,  \qU^{-(1+\qw)}_{E}\, 
\ee
with $\qU$ defined in~\eqref{eq:logU-TBA}. The bridge length follows from the derivative at $\theta = i\pi/2$; for instance,
\be\label{eq:chi-derivative}
\frac{\partial}{\partial \theta} \, \log{\bar{\chi}^{\Sph}_{E}}(i\pi/2) = -il_{E}\, , \qquad \textrm{for}\qquad \qw = +1\, ,
\ee
and similarly for the opposite dressing under $\bar{\chi}^{\Sph} \leftrightarrow \chi^{\Sph}$.

In summary, the $\chi$-system~\eqref{eq:chi-syste-S}, together with the zero mode~\eqref{eq:chi-zero-mode}, is fully equivalent to the TBA equations derived from the hexagon formalism. The shifts in bridge lengths and cross-ratios in the TBA precisely reproduce the boundary conditions~\eqref{eq:chi-ads-special-point}--\eqref{eq:chi-derivative}.

An alternative expression for the free energy can also be written in terms of $\chi$-functions,
\be\label{eq:area-A-form}
\cA = \sum_{E} \str\int \frac{d\theta}{2\pi} (Z_{E}\, e^{\theta}+\bar{Z}_{E}\, e^{-\theta})\,  \log{\left(1+A_{E}\right)} + \sum_{E} l_{E} \,\kappa_{E}\, ,
\ee
where
\be\label{eq:kappaE}
\kappa_{E} =  \str\int \frac{d\theta}{2\pi \cosh{\theta}}  \log{\left(1+A_{E}\right)}\, .
\ee
For linear quivers with the canonical ordering, one has $A_{E} = Y_{\gamma_E}$, and the above expression reduces to the standard TBA result~\eqref{nicer-action}. The equivalence with the TBA formula, however, holds for arbitrary triangulations. The proof relies on~(\ref{eq:A-to-Y}), together with the additivity properties of the bridge length and central charge,
\be
l_{\gamma+\gamma'} = l_{\gamma}+l_{\gamma'}\, , \qquad Z_{\gamma+\gamma'} = Z_{\gamma}+Z_{\gamma'}\, ,
\ee
and similarly with $Z\rightarrow \bar{Z}$.

\section{Closed string correlation functions}\label{sec:closed}

In this section, we extend our analysis to punctured spheres, which correspond to actual single-trace correlators of the theory.
The presence of punctures introduces additional complications due to cycles that encircle them. Magnons propagating along cuts around a puncture can generate wrapping divergences that cannot be removed by contour deformation, and instead give rise to various finite-size corrections, see e.g.~\cite{Basso:2017muf,Basso:2022nny,Ferrando:2025qkr}.

For half-BPS operators, however, the wrapping contributions cancel once we sum over magnon flavors. This can be verified explicitly in the classical limit considered here. In this limit, a magnon wrapping a puncture $P$ carries a charge
\be
\gamma_{P} = \sum_{E:P} \gamma_{E}\, ,
\ee
where the sum runs over all edges $E$ of the triangulation ending on $P$. This state interacts trivially with any external state $\gamma'$, since $\langle \gamma_{P}, \gamma'\rangle = 0, \forall \gamma'$. Furthermore, when $P$ corresponds to a half-BPS operator, the product $\prod_{E:P} \cY^{I}_{\gamma_{E}}$ is independent of the flavor index $I$~\cite{Fleury:2016ykk}. Thus, the state $\gamma_P$ decouples, and its contribution vanishes upon taking the supertrace.

As a result, the TBA equations~\eqref{eq:tbafirstform} and the free energy~\eqref{eq:actionasrogers} therefore remain unchanged in the presence of half-BPS punctures and, following the previous discussion, admit the same $\chi$-system as in the GMN formalism. Below, we highlight the minor modifications of this formalism associated with punctures, referring to~\cite{Gaiotto:2010okc} for a more complete discussion.

The quivers associated with triangulations of punctured geometries are in general more intricate, typically featuring multiple cycles, which makes the construction of a consistent poset more delicate. We illustrate this in the case of the four-point function, using wall-crossing transformations on $A_3$ subquivers to construct a consistent ordering. More general methods, such as the mutation method~\cite{Alim:2011ae,Alim:2011kw} reviewed in Appendix~\ref{app:mutation-method}, can be used to address this problem in full generality.

We then analyze the effect of summing over surface moduli, which corresponds to integrating over bridge lengths in the hexagon formalism. This step enables a precise comparison with the equations derived by Caetano and Toledo from the string-theory perspective.

\subsection{$\chi$-system for punctured geometries}

\begin{figure}[h]
\centering
\includegraphics[width=8cm]{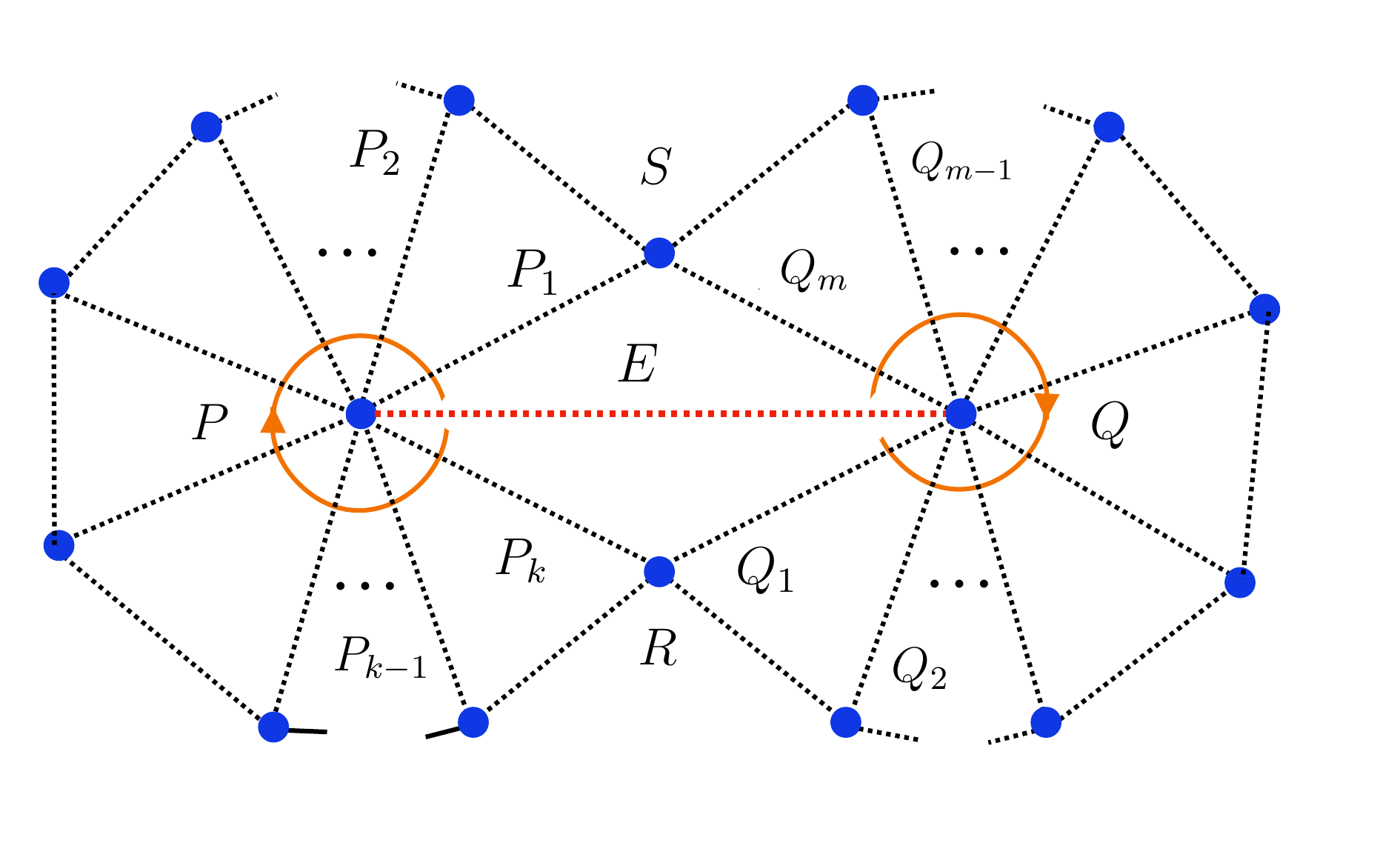}
\caption{The analogue of Figure~\ref{fig:A-chi} for punctures.}
\label{fig:A-chi-Closed}
\end{figure}

Under the conditions described above, the construction developed in the previous sections extends to punctured spheres. The $Y$-system and the $\chi$-system take the same form as in~\eqref{YsysAdSSph} and~\eqref{eq:chisysA}, respectively. The $A$-functions, however, are modified to incorporate the so-called monodromy factors, defined by%
\footnote{As such, these monodromy factors appear somewhat \emph{ad hoc} in our construction. In the GMN approach, however, they arise in a more natural way.}
\be
\mu_{P} = \prod_{E:P} \chi_{E}\, ,
\ee
for each puncture $P$. The precise relation reads~\cite{Gaiotto:2010okc,Caetano:2012ac}
\be\label{eq:APQ-sphere}
A_{PQ} = \frac{A_{PQ}|_{\eqref{eq:A-to-chi}}}{(1-\mu_{P})(1-\mu_{Q})}\, ,
\ee 
where the numerator is defined as in~\eqref{eq:A-to-chi}, with nested products taken over all edges ending on $P$ and $Q$, respectively, as illustrated in Figure~\ref{fig:A-chi-Closed}. Observe that in the polygon case the $\chi$-functions associated with the boundary can be set to zero, which effectively eliminates the monodromy factors.

The general relation between the $Y$-functions and the $\chi$-functions remains unchanged. However, a modification analogous to~\eqref{eq:APQ-sphere} is required for the relation~\eqref{eq:A-to-Y} between the $Y$- and $A$-functions. It takes the form
\be
\label{eq:A-to-Y-closed}
(1-\mu_{P})(1-\mu_{Q}) (1+A_{PQ}) = \prod_{\gamma' \, \ni\,  \gamma_{PQ}  }(1+Y_{\gamma'})\, , \qquad (\textrm{for finite chambers})\, ,
\ee
where the product runs over all $\gamma'$ containing the excitation on the edge $PQ$ in its linear expansion.

As in the polygon case, this relation can be used to connect the $Y$- and $\chi$-systems, and is also needed to express the free energy in terms of the $A$-variables. It is remarkable that the free energy ultimately takes the same form as before in these variables. This is not entirely trivial in the presence of the $\mu$-factors in~\eqref{eq:A-to-Y-closed}. Reintroducing the flavor index $I$ and repeating the analysis for polygons described around eq.~\eqref{eq:area-A-form}, one finds additional contributions of the form
\be\label{eq:free-energy-puncture}
\sum_{I}(-1)^{I}\log{(1-\mu^{I}_{P})}
\ee
for each puncture $P$. At first sight, this suggests that punctures may contribute nontrivially to the free energy, in contrast with previous expectations.

However, the monodromy factors are in fact flavor-independent. This follows from the $\chi$-system, which implies that $\mu_{P}$ is a zero-mode solution completely fixed by the boundary conditions at the puncture. More precisely,
\be
\mu_{P}^{I}(\theta) = (-1)^{|P|} e^{-\cL_{P} \cosh{\theta}}\, , \qquad \forall I\, ,
\ee
where $\cL_{P}$ is the classical R-charge at $P$, and $|P|$ denotes the number of edges ending on $P$ in the triangulation. As a result, the additional contributions~\eqref{eq:free-energy-puncture} cancel in the supertrace, as expected for half-BPS operators, and the free energy reduces to the form given in~\eqref{eq:area-A-form}.

A few remarks on the formula~\eqref{eq:A-to-Y-closed} are in order. Unlike in the polygon case, we do not have a general proof of this relation, though we have verified it in numerous examples, including mixed correlators involving polygons with punctures and the triangulation of the four-point function discussed in the next section.

The formula~\eqref{eq:A-to-Y-closed} is not expected to hold in complete generality, but only when the chamber used to define the $Y$-functions is finite. Indeed, punctured spheres---and more generally Riemann surfaces---may admit chambers with infinitely many elements~\cite{Gaiotto:2010okc,Alim:2011ae}. In such cases, higher composites of the form $\gamma = \sum_{E} \delta_{E,\gamma} \gamma_{E}$ with occupation numbers $\delta_{E,\gamma} > 1$, can arise, requiring modifications of equation~\eqref{eq:A-to-Y-closed}.

Moreover, there are marginal cases where the formula must be adjusted. A simple example is the three-point function, where the quiver consists of three disconnected nodes, so all pairings vanish identically. In this case, the $Y$-functions reduce to $Y_{PQ} = \chi_{PQ}$, with $P, Q = 1,2,3$ labeling the three operators. Computing the corresponding $A$-functions shows that~\eqref{eq:A-to-Y-closed} must be modified by a common factor $(1+\chi_{12}\chi_{23}\chi_{13})$.

This modification is, however, immaterial in the present context. For structure constants of half-BPS operators, the $Y$-functions are flavor-independent, and their contributions to the free energy cancel upon summing over all flavors, in agreement with non-renormalization theorems. In other words, a relation of the form~\eqref{eq:A-to-Y-closed} is not required to compute the free energy in this case.

To remain on the safe side, we therefore assume that equation~\eqref{eq:A-to-Y-closed} holds for finite chambers and connected quivers. It would be interesting to delineate more precisely the domain of validity of~\eqref{eq:A-to-Y-closed}, building on the general formalism of~\cite{Gaiotto:2010okc}. 

Finally, let us recall that the geometric form of the $\chi$-system and the definition~\eqref{eq:APQ-sphere} are independent of the chamber type and therefore universal. Moreover, for a large class of geometries and triangulations---including all punctured spheres---it should always be possible to find a finite chamber~\cite{Alim:2011ae,Alim:2011kw}. In all such cases, equation~\eqref{eq:A-to-Y-closed} is expected to apply, up to the marginal situations discussed above.

\subsection{Four-point function in the Caetano--Toledo triangulation}

Let us now turn to the particular case of the four-point function. In the GMN framework, this corresponds to $N=2$ SQCD with $N_f = 4$ fundamental hypermultiplets (see section 10.7 and 11.6 in \cite{Gaiotto:2009hg}).

\begin{figure}[h]
\centering
\includegraphics[width=10cm]{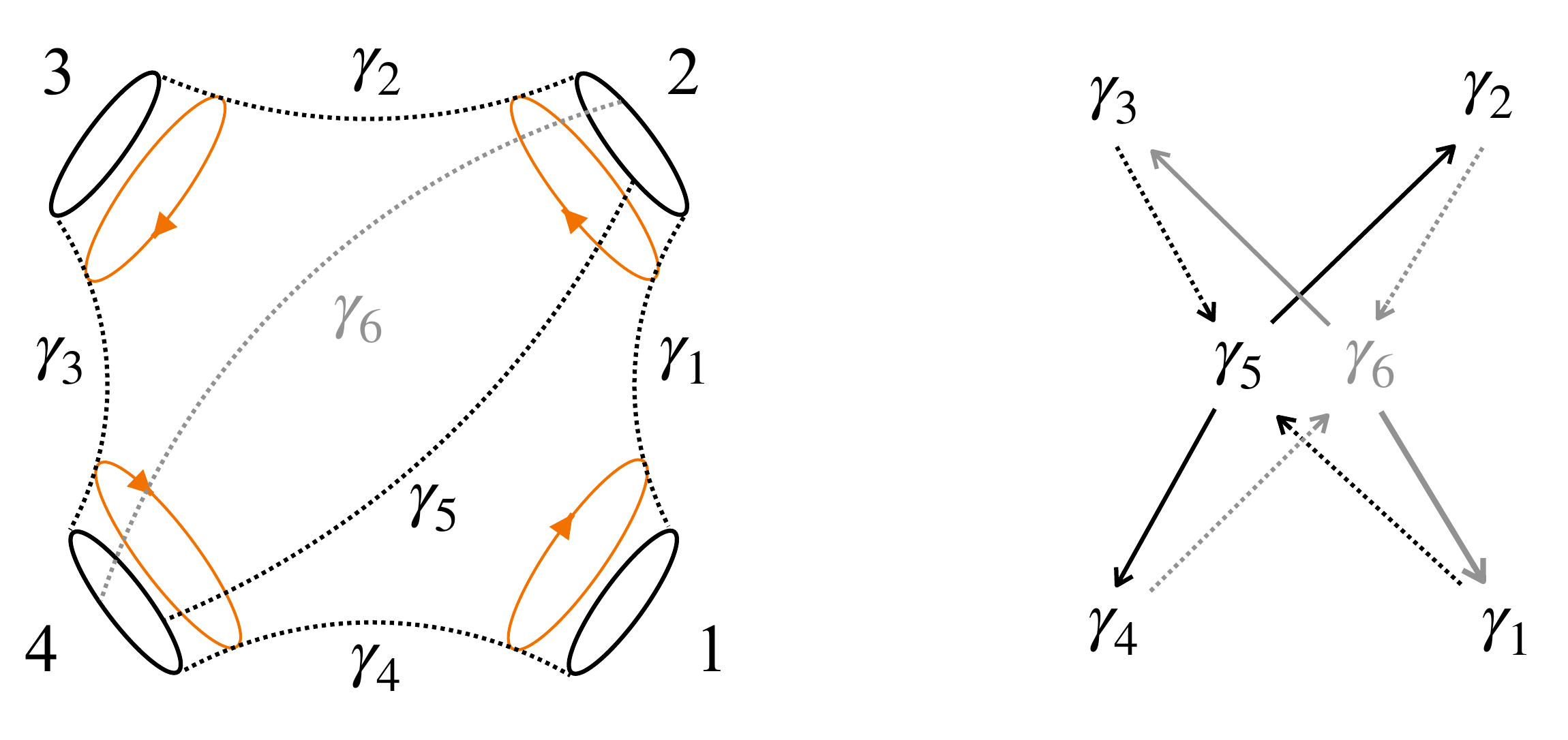}
\caption{{\bf Left:} Triangulation of the four-point function used in Figure 9 of \cite{Caetano:2012ac}. The edge on the back of the sphere and its associated charge are represented with a lighter color. The identification of the charges  to the edges is given in \eqref{charges_edges}.  {\bf Right:} The associated quiver, after removing the two-cycles involving $\gamma_{2,3}$ and $\gamma_{1,4}$. Wall crossing transformations are performed on the dotted lines, according to equation~\eqref{eq:WC3bs} and Figure~\ref{fig:WC3bs}.}\label{fig:4pt_triangulation}
\end{figure}

Following~\cite{Caetano:2012ac}, we consider the triangulation shown in Figure~\ref{fig:4pt_triangulation}, with edges
\be
E_{ij}\in \{E_{12}, E_{23}, E_{34}, E_{14}, E_{\hat{24}}, E_{24}\}\, ,
\ee
connecting operators $i$ and $j$. In this triangulation, some quadrilaterals are degenerate, in the sense that two edges, $E_{24}$ and $E_{\hat{24}}$ connect the same pair of operators. As a consequence, several cross ratios are frozen to fixed values,
\be
\label{eq:framecr}
z_{12}=z_{23}=z_{34}=z_{14}=\alpha_{12}=\alpha_{23}=\alpha_{34}=\alpha_{14}=-1 \, ,
\ee
and similarly for $z_{ij}\to \bar z_{ij}$ and $\alpha_{ij} \to \bar \alpha_{ij}$. Fixing, without loss of generality, $\textbf{x}_1=-z,\, \textbf{x}_2=\infty,\,\textbf{x}_3=1,\,\textbf{x}_4=0$, one is left with a single independent cross-ratio,
\be
z_{24}=1/z_{\hat{24}}=z\, ,
\ee
as expected. The same applies to the remaining variables upon replacing $z_{ij}\to \bar z_{ij},\alpha_{ij}, \bar \alpha_{ij}$

\begin{figure}[h]
\centering
\includegraphics[width=13cm]{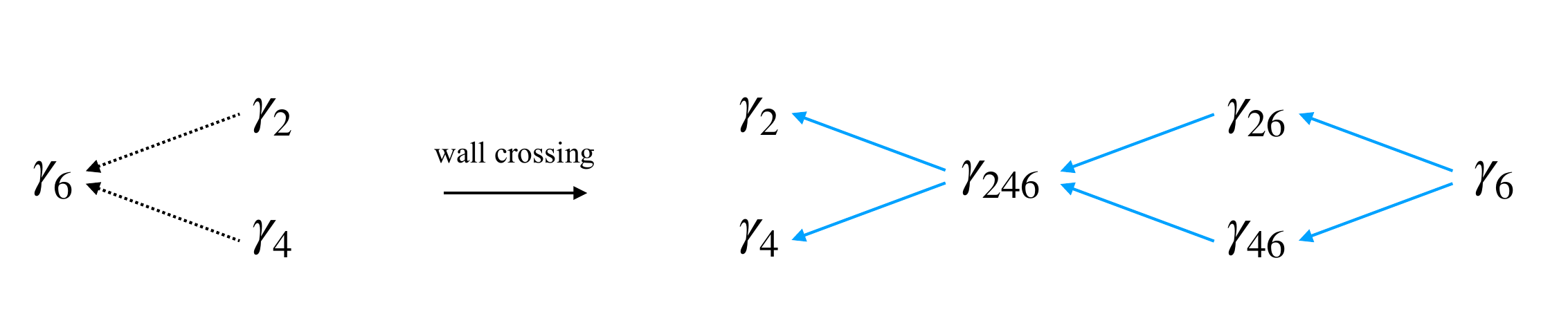}
\caption{Wall-crossing transformation on the $A_{3}$ quiver associated with the three adjacent edges $E_{23}$, $E_{14}$ and $E_{24}$, involving the charges $\gamma_2$, $\gamma_4$ and $\gamma_6$. Blue arrows indicate the new ordering, cf.~\eqref{eq:WC3bs}. An analogous transformation is performed for the remaining charges $\gamma_1$, $\gamma_3$ and $\gamma_5$.}
\label{fig:WC3bs}
\end{figure}

\paragraph{Chamber and states.} A finite chamber for this triangulation can be constructed by performing a pair of $A_{3}$ wall-crossing transformations, as illustrated in Figure~\ref{fig:WC3bs}, or equivalently by applying the more general mutation method~\cite{Alim:2011ae} (see Section 4.1 therein). The resulting chamber $\{\gamma\}$ consists of 12 states, six of which correspond directly to the edges of the triangulation, which we denote as
\be
\label{charges_edges}
\gamma_{1} = \gamma_{E_{12}}\, ,\,\, \gamma_{2} = \gamma_{E_{23}} ,\,\, \gamma_{3} = \gamma_{E_{34}}\, , \,\, \gamma_{4} = \gamma_{E_{14}}\, ,\,\, \gamma_{5} = \gamma_{E_{\hat{24}}}\, ,\,\, \gamma_{6} = \gamma_{E_{24}}\, .
\ee
The exchange matrix has the following non-vanishing entries,
\be
\langle \gamma_{1,3}, \gamma_{6} \rangle = \langle \gamma_{5}, \gamma_{1,3} \rangle = \langle \gamma_{2,4}, \gamma_{5} \rangle = \langle \gamma_{6}, \gamma_{2,4} \rangle = 1\, .
\ee
Performing wall-crossing transformations for the triple $(\gamma_2, \gamma_4,\gamma_6)$ and repeatedly using the pentagon identity, one finds (see Figure~\ref{fig:WC3bs})
\be
\label{eq:WC3bs}
\cK_{\gamma_6}\cK_{\gamma_2}\cK_{\gamma_4}=\cK_{\gamma_2}\cK_{\gamma_4}\cK_{\gamma_{246}}\cK_{\gamma_{46}}\cK_{\gamma_{26}}\cK_{\gamma_6}\,,
\ee
and similarly for the triple $(\gamma_1, \gamma_3,\gamma_5)$. This generates six additional states
\be
\begin{aligned}
&\gamma_{15} = \gamma_{1} + \gamma_{5}\, ,\,\, \gamma_{35} = \gamma_{3} + \gamma_{5}\, , \,\, \gamma_{135} = \gamma_{1}+\gamma_{3}+\gamma_{5}\, , \\
&\gamma_{26} = \gamma_{2} + \gamma_{6}\, ,\,\, \gamma_{46} = \gamma_{4} + \gamma_{6}\, , \,\, \gamma_{246} = \gamma_{2}+\gamma_{4}+\gamma_{6}\, .
\end{aligned}
\ee
The ordering of integration contours is encoded in the spectrum generator~\cite{Gaiotto:2009hg}
\be\label{eq:S-for-four}
\boldsymbol{S} = \cK_{\gamma_{1}} \cK_{\gamma_{2}} \cK_{\gamma_{3}} \cK_{\gamma_{4}} \cK_{\gamma_{135}} \cK_{\gamma_{246}} \cK_{\gamma_{15}} \cK_{\gamma_{26}} \cK_{\gamma_{35}} \cK_{\gamma_{46}} \cK_{\gamma_{5}} \cK_{\gamma_{6}}\, .
\ee
Acting with the KS transformations in this sequence yields the relations between $Y$- and $\chi$-functions, 
\begin{align}
&Y_{\gamma_1} = \chi_{\gamma_1}\,,\quad Y_{\gamma_2} = \chi_{\gamma_2}\,,\quad Y_{\gamma_3} = \chi_{\gamma_3}\,,\quad Y_{\gamma_4} = \chi_{\gamma_4}\,,\\ \nonumber
&\frac{Y_{\gamma_{135}}}{\chi_{\gamma_1}\chi_{\gamma_3}\chi_{\gamma_5}}=\frac{\chi_{\gamma_2}\chi_{\gamma_4}\chi_{\gamma_6}}{Y_{\gamma_{246}}}=\frac{(1+\chi_{\gamma_2})(1+\chi_{\gamma_4})}{(1+\chi_{\gamma_1})(1+\chi_{\gamma_3})}\,,\\\nonumber
&\frac{Y_{\gamma_{46}}}{\chi_{\gamma_4}\chi_{\gamma_6}}=\frac{Y_{\gamma_{26}}}{\chi_{\gamma_2}\chi_{\gamma_6}}=\frac{\chi_{\gamma_3}\chi_{\gamma_5}}
{Y_{\gamma_{35}}}=\frac{\chi_{\gamma_1}\chi_{\gamma_5}}{Y_{\gamma_{15}}}
=\frac{\chi_{\gamma_1}\chi_{\gamma_3}\chi_{\gamma_5}}{Y_{\gamma_{135}}}\frac{(1+ Y_{\gamma_{135}})}{(1+Y_{\gamma_{246}})}\,,\\ \nonumber 
&\frac{Y_{\gamma_{5}}}{\chi_{\gamma_5}}=\frac{\chi_{\gamma_6}}{Y_{\gamma_{6}}}=\frac{Y_{\gamma_{135}}}{\chi_{\gamma_1}\chi_{\gamma_3}\chi_{\gamma_5}}\frac{(1+Y_{\gamma_{26}})(1+Y_{\gamma_{46}})(1+Y_{\gamma_{246}})^2}{(1+Y_{\gamma_{15}})(1+Y_{\gamma_{35}})(1+Y_{\gamma_{135}})^2}\,.
\end{align}
In writing them, we have used expressions from previous lines to keep the notation compact.

\paragraph{$\chi$-system.} 
One can finally derive the $\chi$-system directly from the action of the spectrum generator, using the general formula~\eqref{eq:chisysA} and the correspondence~\eqref{charges_edges}. The result is (with $\chi_{ij} = \chi_{E_{ij}}$ and $A_{ij} = A_{E_{ij}}$)
\be\label{eq:chi-sys-4pt}
\begin{aligned}
&\bar{\chi}^{++}_{24} \chi_{24} = \left(\bar{\chi}_{\hat{24}}^{++}\chi_{\hat{24}}\right)^{-1} = \frac{(1+A_{23})(1+A_{14})}{(1+A_{34})(1+A_{12})}\, ,\\
&\bar{\chi}_{12}^{++} \chi_{12} = \left(\bar{\chi}_{14}^{++}\chi_{14}\right)^{-1} = \bar{\chi}_{34}^{++} \chi_{34} = \left(\bar{\chi}_{23}^{++}\chi_{23}\right)^{-1} = \frac{(1+A_{24})}{(1+A_{\hat{24}})}\, ,
\end{aligned}
\ee
together with the relations obtained by exchanging $\chi \leftrightarrow \bar{\chi}$ and $A\leftrightarrow \bar{A}$. The $A$-functions are given by
\be\label{eq:A-sys-4pt}
\begin{aligned}
{(1-\mu_{2})(1-\mu_{4})}\,A_{24} &= {\chi_{24}\,(1+\chi_{12}(1+\chi_{\hat{24}}(1+\chi_{23})))\,(1+\chi_{34}(1+\chi_{\hat{24}}(1+\chi_{14})))}\, , \\
{(1-\mu_{2})(1-\mu_{4})}\,A_{\hat{24}} &= {\chi_{\hat{24}}\,(1+\chi_{23}(1+\chi_{24}(1+\chi_{12})))\,(1+\chi_{14}(1+\chi_{24}(1+\chi_{34})))}\, , \\
{(1-\mu_{1})(1-\mu_{2})} \,A_{12} &= {\chi_{12}\,(1+\chi_{14})\,(1+\chi_{\hat{24}}(1+\chi_{23}(1+\chi_{24})))}\, , \\
{(1-\mu_{2})(1-\mu_{3})}\,A_{23} &= {\chi_{23}\,(1+\chi_{34})\,(1+\chi_{24}(1+\chi_{12}(1+\chi_{\hat{24}})))}\, , \\
{(1-\mu_{3})(1-\mu_{4})}\,A_{34} &= {\chi_{34}\,(1+\chi_{23})\,(1+\chi_{\hat{24}}(1+\chi_{14}(1+\chi_{24})))}\, , \\
{(1-\mu_{1})(1-\mu_{4})}\, A_{14} &= {\chi_{14}\,(1+\chi_{12})\,(1+\chi_{24}(1+\chi_{34}(1+\chi_{\hat{24}})))}\,,
\end{aligned}
\ee 
while the monodromy factors read
\be\label{eq-mu-4pt}
\mu_{1} = \chi_{12}\chi_{14}\, , \qquad \mu_{2} = \chi_{12}\chi_{24}\chi_{23}\chi_{\hat{24}}\, , \qquad \mu_{3} = \chi_{23}\chi_{34}\, , \qquad \mu_{4} = \chi_{14}\chi_{24}\chi_{34}\chi_{\hat{24}}\, .
\ee
As a consistency check, one verifies agreement with the geometric formula~\eqref{eq:APQ-sphere}, as well as the validity of~\eqref{eq:A-to-Y-closed}.

The system~\eqref{eq:chi-sys-4pt}--\eqref{eq-mu-4pt} applies equally to AdS and sphere $\chi$-functions. In the special case $\chi = \bar{\chi}$, it reduces to the system derived in~\cite{Caetano:2012ac} for four-point functions in $AdS_{2}$. We have adopted the same edge notation to facilitate comparison. These finite-difference equations can be recast into integral form using the procedure described in Section~\ref{sec:integral-equations}. We will analyze them in more detail below and compare with the results of~\cite{Caetano:2012ac}. 

Finally, although the action of the spectrum generator is independent of the choice of chamber, it is worth emphasizing that---unlike in the polygon case---not all chambers can be reached by permuting the KS factors in~\eqref{eq:S-for-four} via the pentagon identity. In the presence of multiple punctures, there exist chambers with infinitely many states, which require new types of wall-crossing transformations.

In the language of $\mathcal{N}=2$ wall-crossing, see discussions in~\cite{Gaiotto:2010okc,Alim:2011ae,Alim:2011kw}, this phenomenon is associated with the appearance of vector multiplets in the BPS spectrum. In the present context, these correspond to states wrapping cycles around two or more punctures. Such states are expected to play an important role in OPE limits, where punctures approach one another.

\subsection{Sum over bridge lengths}\label{sec:sum-over-moduli}

To make precise contact with string theory, we must also sum over the bridge lengths assigned to the edges of the triangulation, which in the hexagonalization procedure are interpreted as the moduli of the geometry. This leads to 
\be
\qZ [\{ d_{E}\}] = \sum_{\ell_{E}} \left(d_{E}\right)^{\ell_{E}}\, \qZ [\{\ell_{E}\}]\, ,
\ee
where $d_{E} = y_{E}^2/x_{E}^2$ denotes the scalar propagator (see eq.~\eqref{eq:superdistance}).

In the classical limit $\ell_{E} \rightarrow \infty, \epsilon\rightarrow 0$, with $l_{E} = \epsilon \ell_{E}$ held fixed, the set of allowed lengths becomes dense. The discrete sum may then be replaced by an integral, which can be evaluated using a saddle-point approximation in the $l_E$ variables,
\be
\qZ[\{ d_{E}\}] \sim e^{-\cF/\epsilon}\, ,
\ee
where $\cF$ is given by
\be\label{eq:free-energy-closed-string}
\cF = \cA - \sum_{E}l_{E} \log{d_{E}}\, ,
\ee
evaluated at the saddle-point configuration $l_{E} = l^{\textrm{saddle}}_{E}$.

To derive the saddle-point equations for the bridge lengths, we must determine how $\cA$ varies under changes of the external parameters. Remarkably, this variation is entirely governed by the driving terms $\cY$ appearing in the TBA equations,
\be\label{eq:vary-A}
\delta \cA = -\sum_{I,E} (-1)^{F_{I}}\int \frac{d\theta}{2\pi \cosh^{2}{\theta}} \delta \log{\cY^{I}_{E}(\theta)} \times \log{(1+A^{I}_{E}(\theta))}\, .
\ee
This relation follows from evaluating the variation of the free energy density using the TBA equations~\eqref{eq:tbafirstform}, as detailed in Appendix~\ref{app:simpler-action}.

The bridge lengths cannot be varied independently, since they are constrained by charge conservation conditions at each puncture,
\be\label{eq:charge-cons}
\sum_{E:P} l_E = \cL_P\, .
\ee
The saddle-point analysis therefore naturally involves Lagrange multipliers. Rather than introducing them explicitly, one may instead exploit the fact that, if \(E\) is a diagonal of a quadrilateral as in Figure~\ref{fig:A-chi-Closed}, the lengths of the adjacent edges can be varied in such a way that the R-charges at its vertices remain fixed. This variation is given by
\be
\delta_{E} l_{E'} = \langle E, E'\rangle \, ,
\ee
and, under it, one finds
\be\label{delta-E-cY}
\delta_{E} \log{\cY^{I}_{E'}} = -\langle E, E'\rangle \cosh{\theta}\, , \qquad \textrm{for all flavors}\,\, I\, .
\ee

Combining~\eqref{eq:vary-A} and~\eqref{delta-E-cY} leads to the saddle-point equations
\be
\label{eq:U-cstr}
0=\delta_{E} \cF = \sum_{E'}\langle E,E' \rangle(\kappa_{E'}-\log d_{E'})=-\log{\qU^2_{E}} \, , \qquad \forall E\, ,
\ee
or equivalently,
\be
0 = \sigma^{\Sph}_{E}-\sigma^{\AdS}_{E} + \sum_{E'} \langle E, E'\rangle \int\frac{d\theta'}{4\pi \cosh{\theta'}} \str \, L_{E'}(\theta')\, , \qquad \forall E\, .
\ee
Similarly, $\log{\qU_{\gamma}}=0$ for all $\gamma$, by linear extension to any chamber. Here we used the definition
\be
\log z_E=\sum_{E'}\langle E,E' \rangle \log \textbf{x}_{E'}\, ,
\ee
and analogously for $\bar z_E,\,\alpha_E,\,\bar\alpha_E$, to rewrite
\be
\sum_{E'}\langle E,E' \rangle \log d_{E'}=\log{\left(\frac{\alpha_E \bar \alpha_E}{z_E \bar z_E}\right)}  = \sigma_{E}^{\Sph}-\sigma_{E}^{\AdS} = \log{U^2_{E}}\, ,
\ee
together with the definition \eqref{eq:logU-TBA}.

Not all of the equations~\eqref{eq:U-cstr} are independent. Indeed, taking the product of $\qU_{E}$ over all edges ending on $P$ automatically yields
\be
\qU_{P} = \prod_{E:P}\qU_{E} = 1\, ,
\ee
even without imposing~\eqref{eq:U-cstr} on the individual factors. However, since each puncture comes with the conservation equation~\eqref{eq:charge-cons}, the total number of independent constraints matches the number of edges in the triangulation, namely $3(n-2)$ for an $n$-punctured sphere. As this coincides with the number of bridge lengths, it follows that, for generic kinematics, all bridge lengths are completely fixed by~\eqref{eq:U-cstr} and~\eqref{eq:charge-cons}.

These constraints have important consequences, bringing the system closer to the structure expected from string theory. First, they restore a full symmetry between the AdS and sphere sectors. Indeed, as can be seen from~\eqref{eq:constant-C}, on the support of~\eqref{eq:U-cstr} the extra term $\propto \qw \log{\qU_{\gamma}}$ in the TBA equations for the sphere $Y$-functions vanishes. As a result, AdS and sphere $Y$-functions satisfy identical TBA equations, differing only in the choice of the kinematical angles $\phi_{\gamma}$ in~\eqref{eq:constant-C}. It is also worth noting that this reduces the coupling between the two sectors, which is then entirely encoded in their common central charges.

An equivalent way to see this symmetry restoration is to examine the values of the $\chi$-functions at the special point $\theta = i\pi/2$ in~\eqref{eq:chi-ads-special-point} and~\eqref{eq:chi-sph-special-point}. As discussed earlier, these values control the zero modes $\sfchi_{E}$ and are the only potential source of discrepancy between the two sectors. Using~\eqref{eq:U-cstr}, they reduce (for any dressing $\qw$) to
\be
\begin{aligned}
\chi_{E}^{\AdS}(i\pi/2) &= z_{E}\, , \qquad \bar{\chi}_{E}^{\AdS}(i\pi/2) = \bar{z}_{E}\, ,\\
\chi_{E}^{\Sph}(i\pi/2) &= \alpha_{E}\, , \qquad \bar{\chi}_{E}^{\Sph}(i\pi/2) = \bar{\alpha}_{E}\, ,
\end{aligned}
\ee
showing that the AdS and sphere $\chi$-functions coincide up to an exchange of kinematic data, $(z_{E}, \bar{z}_{E}) \leftrightarrow (\alpha_{E}, \bar{\alpha}_{E})$.

The second important consequence is that the bridge lengths $l_{E}$ can be completely eliminated in favor of the central charges $Z_{E}$ and $\bar{Z}_{E}$. This is manifest at the level of the TBA equations, which can already be written entirely in these variables (see~\eqref{eq:ZplusZbar}). It is less immediate for the free energy $\cF$ (see~\eqref{eq:free-energy-closed-string} and~\eqref{eq:area-A-form}), where the lengths enter explicitly through
\be\label{eq:l-term}
\sum_{E}l_{E} (\kappa_{E}-\log{d_{E}})\, .
\ee
However, on the support of~\eqref{eq:U-cstr}, this term is invariant under shifts of the form
\be
l_{E} \rightarrow l_{E} + \sum_{E'}\langle E, E'\rangle f_{E'}\, ,
\ee
since
\be
\sum_{E, E'} \langle E, E'\rangle (\kappa_{E}-\log{d_{E}}) f_{E'} = \sum_{E'} f_{E'}\log{\qU^2_{E'}} = 0\, .
\ee
With a suitable choice of $f_{E'}$, one can set $l_{E} \rightarrow -(Z_{E}+\bar{Z}_{E})$ in~\eqref{eq:l-term}, thus eliminating the bridge lengths entirely. A similar rewriting applies to the conservation equation~\eqref{eq:charge-cons}, which becomes
\be
\sum_{E:P} (Z_{E}+\bar{Z}_{E}) = -\cL_{P}\, .
\ee
As a by-product, this procedure removes any explicit dependence on the dressing parameter $\qw$, which is absorbed into $Z_{E}+\bar{Z}_{E}$ (see eq.~\eqref{eq:ZplusZbar}).

In conclusion, when $\qU_{E} = 1$ for all $E$, neither the TBA equations nor the free energy $\cF$ depends explicitly on $\qw$ or $l_{E}$: the two dressings become equivalent, and the description reduces entirely to central charges, in agreement with the string theory formulation~\cite{Caetano:2012ac}. The bridge lengths then reappear only as derived quantities, obtained from derivatives of the sphere $\chi$-functions at $\theta = i\pi/2$, as in~\eqref{eq:chi-derivative}.

\subsection{Comparison with string theory results in $AdS_{2}\times S^{1}$}

In this section we compare our final equations with those of Caetano and Toledo \cite{Caetano:2012ac} for strings in $AdS_{2}\times S^{1}$. To make contact with this setup, we must reduce the kinematics in both the AdS and sphere sectors.

We begin with the reduction to $AdS_{2}\times S^{2}$, where all insertion points are aligned along a line in both sectors. In terms of cross ratios, this corresponds to the symmetric configuration
\be
z_E=\bar z_E\,, \qquad \alpha_E = \bar \alpha_E\,,
\ee
for each edge $E$. As a consequence, the $\chi$- and $\bar{\chi}$-functions obey identical boundary conditions at $\theta = i\pi/2$. Since the $\chi$-system is symmetric under the exchange $\chi \leftrightarrow \bar{\chi}$, these equalities can be lifted to the functions themselves (assuming the solution is unique)
\be\label{eq:chi=chibar}
\chi_{E}(\theta) = \bar{\chi}_{E}(\theta) \, .
\ee
It is important here to impose the constraint $\qU_E = 1$ derived earlier. Indeed, evaluating the above relation at $\theta=i\pi/2$ in the sphere sector using~\eqref{eq:chi-sph-special-point} implies
\be
\alpha_E = \bar{\alpha}_E\, \qU_E^{-2\qw}\, ,
\ee
and therefore requires $\qU_E = 1$.

The TBA equations~\eqref{eq:TBA-GMN} can then be simplified using~\eqref{eq:chi=chibar}. They take the form
\be\label{eq:integrate-chi-system-1d}
\log{\chi^{\AdS}_{E}(\theta)} = C^{\AdS}_E+ Z_{E} e^{\theta}+\bar{Z}_{E} e^{-\theta} + \sum_{E'}\langle E, E'\rangle \, \GMN_{\textrm{1d}} * \log{(1+A^{\AdS}_{E'})}\, , 
\ee
where
\be\label{eq:1d-GMN}
\GMN_{\textrm{1d}}(\theta) = \GMN(\theta) + \bGMN(\theta) = \frac{i}{\sinh{\theta}}\, ,
\ee
and similarly for the sphere sector. As discussed earlier, in this integrated form of the $\chi$-system there is no ordering issue, and the contour of integration may be chosen to lie along the real $\theta$-axis for all $E$, using $\theta\rightarrow \theta+i0$ in~\eqref{eq:1d-GMN}.

\begin{figure}[h]
    \centering
    \includegraphics[width=0.8\linewidth]{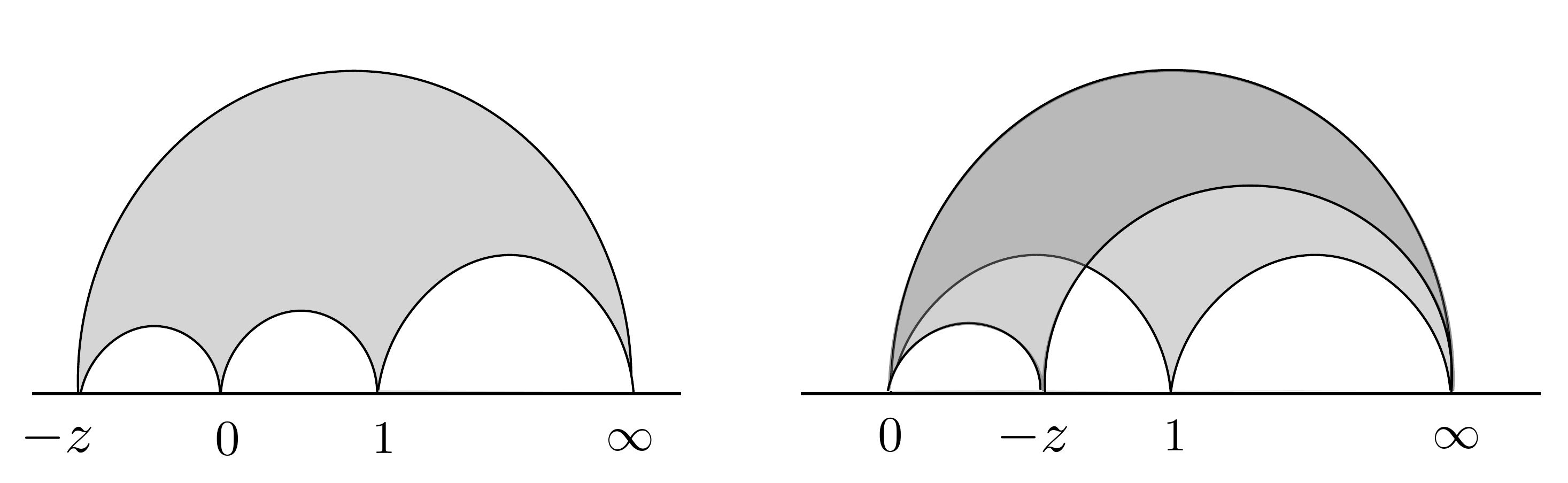}
    \caption{Cartoon of a minimal surface in $AdS_{2}$ corresponding to a folded closed string ending at four points along a line at the boundary~\cite{Caetano:2012ac}. \textbf{Left:} For $z>0$, the AdS phase $\phi^{\AdS} = 0$ and the string spans a nontrivial area. \textbf{Right:} For $z<0$, where $\phi^{\AdS} = \pi$, the string develops additional folds and is expected to degenerate into a set of geodesics, according to the geodesic hypothesis~\cite{Sikorowski:2015wzd}.}
    \label{fig:AdS2string}
\end{figure}

Condition~\eqref{eq:chi=chibar} also restricts the integration constants to
\be
C_{E} = \phi_{E} = ik\pi\, ,
\ee
with $k=0$ or $k=1$, both in the AdS and sphere sectors. These constants encode the sign of the corresponding cross ratios, and therefore the ordering of the points along the line, as illustrated in Figure~\ref{fig:AdS2string} for the spacetime cross ratios.

These considerations are generic for any triangulation in the restricted one-dimensional kinematics. In the particular triangulation considered here, $C_E^{\AdS}=C_E^{\Sph}=i\pi$ for the edges on the external frame, $E\in\{E_{12},\,E_{23},\,E_{34},\,E_{14}\}$, since the corresponding cross ratios are fixed to $-1$ (see~\eqref{eq:framecr}). The only cross ratios left unfixed are the diagonal ones,
\be
z_{24}=1/z_{\hat{24}}=z\,,\qquad \alpha_{24}=1/\alpha_{\hat{24}}=\alpha\,,
\ee
with the corresponding constants being $0$ or $i\pi$, depending on the sign of $z$ and $\alpha$, respectively.

We now further restrict the sphere kinematics to $S^1$. On the gauge theory side, this amounts to choosing operators $\mathcal{O}_i$, with $i=1,2,3,4$, such that there are no contractions between $(2,4)$ or $(1,3)$, i.e.~$y_{1}\cdot y_{3} = y_{2}\cdot y_{4} = 0$. The corresponding graphs therefore have non-zero bridge lengths along the edges of the frame $1234$, and vanishing lengths along the diagonals.

In terms of cross-ratios, this kinematics corresponds to
\be
\alpha = \bar{\alpha} = -1 \qquad \Rightarrow \qquad C_{24}^{\Sph} = C_{\hat{24}}^{\Sph} = i\pi\,.
\ee
Remarkably, this choice admits an exact constant solution for the diagonal $\chi$-functions,
\be\label{eq:s1-reduction}
\chi^{\Sph}_{24}(\theta) = \chi^{\Sph}_{\hat{24}}(\theta) = -1\, , \qquad \textrm{for all}\,\, \theta\, .
\ee
This solution was first identified in~\cite{Caetano:2012ac,Sikorowski:2015wzd} as a limit of the AdS $\chi$-system. It applies equally well in the sphere sector since the two systems are formally the same. It implies a complete trivialization of the sphere $A$-functions,
\be\label{eq:sphere-A-0d}
A^{\Sph}_{12},\, A^{\Sph}_{23}, \,A^{\Sph}_{34},\, A^{\Sph}_{41} \rightarrow 0\, , \qquad A^{\Sph}_{24},\, A^{\Sph}_{\hat{24}} \rightarrow -1\, ,
\ee
as one can verify by substituting~\eqref{eq:s1-reduction} into the $\chi$-system equations~\eqref{eq:chi-sys-4pt}.

An important consequence is that the central charges vanish on the diagonals,
\be\label{eq:vanish-Z}
Z_{24} = \bar{Z}_{24}  = Z_{\hat{24}} =  \bar{Z}_{\hat{24}} = 0\, .
\ee
This reflects the vanishing bridge lengths and follows from the diagonal integral equations when $A_{ii+1} = 0$,
\be
0 = \log{(-\chi_{24})} = Z_{24} e^{\theta}+\bar{Z}_{24} e^{-\theta}\, ,
\ee
and similarly for $\log{(-\chi_{\hat{24}})}$. Another important consequence is that the frame edges do not contribute to the sphere free energy.%
\footnote{The absence of contributions does not mean that the functions $\chi_{ii+1}$ are trivial. However, since the corresponding $A$'s vanish, these edges decouple from the sphere free energy.}

The diagonal contribution to the sphere free energy can also be evaluated explicitly, although some care is needed. Setting $A^{\Sph}_{24} = A^{\Sph}_{\hat{24}} = -1$ produces logarithmic divergences that compensate the vanishing central charges~\eqref{eq:vanish-Z}. Introducing a regulator as in~\cite{Caetano:2012ac}, one finds that the net sphere contribution reduces to a constant $2\pi/3$,%
\footnote{More precisely, setting $Z_{24} = \bar{Z}_{24} = -\varepsilon/2$ with $\varepsilon$ a small regulating length, the leading contribution as $\varepsilon \rightarrow 0$ comes from large rapidities, $|\theta| \approx -\log{\varepsilon}$. In this regime the function $\chi_{24}$ is exponentially small and dominated by the driving term of the TBA equation, so that $A^{\Sph}_{24} \approx \chi^{\Sph}_{24} \approx -e^{-\varepsilon \cosh{\theta}}$. Substituting into the free energy then yields $\lim_{\varepsilon \rightarrow 0}\int \frac{d\theta}{\pi} (-\varepsilon \cosh{\theta}) \log{(1-e^{-\varepsilon \cosh{\theta}})} = \pi/3$, giving a contribution $\pi/3$ from edge $24$, and similarly from $\hat{24}$.} yielding
\be
\cF = \frac{2\pi}{3} -\sum_{E} \int \frac{d\theta}{\pi} (Z_{E}\, e^{\theta}+\bar{Z}_{E} \, e^{-\theta}) \log{(1+A^{\AdS}_{E})} + 2\sum_{E}(Z_{E}+\bar{Z}_{E})(\kappa^{\AdS}_{E}-\log{x_{E}})\, ,
\ee
with $x_{E} = \textbf{x}_{E} =\bar{\textbf{x}}_{E}$ and
\be
\kappa^{\AdS}_{E} =  \int \frac{d\theta}{2\pi \cosh{\theta}}  \log{(1+A^{\AdS}_{E})}\, .
\ee
In summary, in the $S^1$ kinematics the sphere sector becomes trivial (up to a constant shift to the free energy) leaving us with a single copy of the $\chi$-system with $\chi = \bar{\chi}$ for the AdS modes, in perfect agreement with the string theory analysis~\cite{Caetano:2012ac,Toledo:PhDthesis,Sikorowski:2015wzd}.

An important input carried over from the sphere analysis is the vanishing of diagonal central charges (see eq.~\eqref{eq:vanish-Z}). In string theory, this follows from the worldsheet Virasoro constraints relating the sphere and AdS stress tensors, suggesting that the equality of central charges across sectors in the TBA equations is a general consequence of these constraints.

At this stage, the analysis parallels the string theory case. In particular, there remain two possible choices for the constant
\be
C_{24}^{\AdS} = C_{\hat{24}}^{\AdS} = i\phi^{\AdS}\, , \qquad \textrm{with}\qquad \phi^{\AdS} = 0 \,\, \textrm{or} \,\, \pi\, ,
\ee
corresponding to the two possible orderings of points on the line (see Figure~\ref{fig:AdS2string}). To conclude this section, let us briefly discuss each case in turn.

\paragraph{Case $\phi = 0$.} In this case additional symmetries reduce the system of equations. In particular, the condition~\eqref{eq:vanish-Z} implies
\be\label{eq:chi24}
\chi_{24} = 1/\chi_{\hat{24}} = \exp{\int \frac{id\theta'}{2\pi \sinh{(\theta-\theta'+i0)}} \log{\frac{(1+A_{14}(\theta'))(1+A_{23}(\theta'))}{(1+A_{12}(\theta'))(1+A_{34}(\theta'))}}}\, ,
\ee
showing that there is only one independent diagonal $\chi$-function in this setup. This relation also constrains the $\chi$-functions along the frame edges. The monodromy constraints at each puncture~\eqref{eq-mu-4pt} simplify to
\be
\mu_{i} = \chi_{i-1, i}\chi_{i, i+1} = e^{-\cL_{i}\cosh{\theta}}\, , \qquad i =1,2,3,4,
\ee
with periodicity $i \equiv i+4$, and where $\cL_i$ denotes the R-charge at point $i$. Hence, knowing a single $\chi_{i, i+1}$ (say $\chi_{12}$) determines all others. One then obtains
\be\label{eq:chi12}
\log{(-\chi_{12})} = Z_{12} e^{\theta} + \bar{Z}_{12} e^{-\theta} + \int \frac{id\theta'}{2\pi \sinh{(\theta-\theta'+i0)}} \log{\frac{(1+A_{24}(\theta'))}{(1+A_{\hat{24}}(\theta'))}}\, .
\ee
Since $\bar{Z}_{12} = Z_{12}^*$, the system of equations~\eqref{eq:chi12} and~\eqref{eq:chi24} is characterized by two real parameters. One combination, $Z_{12}-\bar{Z}_{12}$, is fixed by imposing $\chi_{12} = -1$ at $\theta = i\pi/2$; the other, $Z_{12}+\bar{Z}_{12}$, is chosen so that $\chi_{24} = z = e^{-\sigma^{\AdS}}$ at the same point.%
\footnote{The string theory analysis~\cite{Sikorowski:2015wzd} also suggests that $Z_{12} = \bar{Z}_{12}$, as a consequence of the reality condition imposed on the stress tensor in the string worldsheet theory. As emphasized above, this condition translates into an integral equation obtained by evaluating~\eqref{eq:chi12} at $\theta = i\pi/2$ using $\chi_{12} = -1$. Although it is conceivable that this condition indeed holds, we have not been able to provide a proof of it.}

Constructing a general analytic solution may only be feasible numerically. Exact solutions, however, can be found in specific cases such as $\sigma^{\AdS} = 0$ with symmetric charges $\cL_i$, as discussed in detail in~\cite{Sikorowski:2015wzd,Caetano:2012ac}.

\paragraph{Case $\phi = \pi$.}This case is more striking: the equations become identical to those of the sphere sector. In particular,
\be
C_{E}^{\AdS} = C_{E}^{\Sph}\, ,
\ee
for any edge $E$ (either along the frame or its diagonals). Since the central charges also coincide in the two sectors, all driving terms are identical. Assuming the solution is unique for a given set of driving terms, it follows that the AdS and sphere solutions coincide,
\be
\chi_{E}^{\AdS} = \chi_{E}^{\Sph}\, .
\ee
This corresponds to a ``supersymmetric" configuration in which the geometry in AdS and on the sphere is identical. In this case $\cA$ vanishes identically and the correlator reduces to its free-theory value. This conclusion is in fact not restricted to the one-dimensional kinematics considered here: the same reasoning applies in general kinematics whenever the constants coincide in the two sectors for all edges.

Although this result may appear surprising, it is consistent with findings from string theory and is related to the geodesic hypothesis formulated in~\cite{Sikorowski:2015wzd}. The latter proposes that folded string solutions for the four-point function in $AdS_{2}\times S^1$ degenerate into a set of disconnected geodesics when $z<0$. In other words, for such configurations, the string does not span a finite worldsheet, and its free energy (area) $\cF$ reduces to that of the free gauge theory, in agreement with our ``supersymmetric'' argument.

\section{Conclusion and outlook}\label{sec:conclusions}

In this work, we derived a TBA description of correlation functions of half-BPS operators from hexagonalization in the classical strong-coupling limit of planar $\mathcal{N}=4$ SYM theory, focusing on two-dimensional kinematics. By combining standard integrability techniques with the wall-crossing framework of Gaiotto, Moore, and Neitzke~\cite{Gaiotto:2010okc,Gaiotto:2009hg}, we showed that the problem admits a natural reformulation in terms of $Y$- and $\chi$-systems. Our analysis applies to both open and closed geometries. In the four-point case, we recover, in the $AdS_2\times S^1$ limit, the $\chi$-system previously obtained from semiclassical string theory~\cite{Caetano:2012ac}.

Our results open several directions for future investigation. First of all, from a practical perspective, it would be interesting to study the TBA equations numerically and to test the saddle-point approximation over moduli space performed in Section~\ref{sec:closed}, which played an important role in achieving agreement with string-theory results. This should clarify how bridge lengths---and the underlying triangulation---are dynamically fixed in terms of cross-ratios. A related question is whether the free energy varies smoothly with these cross-ratios~\cite{Caetano:2012ac,Sikorowski:2015wzd}, or if phase transitions arise when the saddle point shifts between triangulations.

On the string-theory side, our findings call for a direct extension of minimal-surface techniques to $AdS_{3}\times S^{3}$ \cite{Erkan2026}. The emerging picture suggests a factorization into two $SU(2)$ Hitchin systems, one for the AdS sector and one for the sphere, sharing the same central charges (stress tensors). Understanding whether a similar structure persists in more general kinematics, and how it relates to higher-rank Hitchin systems~\cite{Alday:2009dv,Alday:2010vh,Gaiotto:2012rg,Gaiotto:2012snakes}, is an important open problem. On the hexagon side, it would entail exploring how to efficiently handle the matrix part of the hexagon sums in these more general settings.

An important ingredient for a complete comparison with the classical string analysis is the construction of flat sections of the auxiliary linear problem associated with the Hitchin equations. In the minimal-surface/GMN framework~\cite{Gaiotto:2009hg,Caetano:2012ac}, these sections serve as the fundamental building blocks. They determine the underlying triangulation through WKB analysis and define the $\chi$-functions in terms of Fock–Goncharov (cluster) coordinates. In this formulation, the $\chi$-system arises more naturally and follows directly from the analytical properties of the sections, in contrast with the more indirect TBA derivation adopted here.
In the integrability formalism, the auxiliary linear problem can be derived from the Lax connection associated with the string sigma model~\cite{Bena:2003wd}; obtaining it directly within the hexagonalization formalism would provide a better understanding of how these geometric structures enter the TBA in this particular framework. 

Throughout this paper, we have restricted attention to half-BPS operators. A natural generalization would be to consider operators dual to spinning strings, whose non-trivial spectral data are encoded, at strong coupling, in their quasi-momenta~\cite{Kazakov:2004qf}. An important question is whether the $\chi
$-system is sufficiently universal to accommodate this generalization. If so, we expect the quasi-momenta of the operator insertions to enter the $\chi$-system through the driving terms.
The available results for the classical three-point functions of spinning strings~\cite{Janik_2011,Kazama_2011,Kazama_2012} can serve as a guide. However, for higher-point functions of non-BPS operators,  where it is necessary to sum over bridge lengths, disentangling the contributions to the driving terms might be more challenging. Minimization over the bridge lengths may also bring some important simplification, in analogy to what we have seen in this work.

Another challenge when considering non-BPS operators in this formalism is the handling of wrapping corrections, which are built into the QSC formalism. Once these problems are solved, we expect to be able to take into account in an unified manner both the spectral data, encoded in the quasi-momenta and the $Q$-functions, and the geometric data encoded in the cross ratios. Alternatively, further insight can be gained from the Separation of Variables (SoV) method, used in~\cite{Kazama:2016cfl} to retrieve the semiclassical three-point function at weak coupling. Recent progress on the SoV representation of the three-point functions at weak coupling, see e.g.~\cite{Bercini:2022jxh,Bargheer:2025kli,Bargheer:2026qsov} and references therein, may shed additional light on the connection between correlation functions and the spectral problem via $Q$-functions.

A further natural extension of the present classical analysis is to higher-genus surfaces relevant for non-planar corrections~\cite{Bargheer:2017nne,Eden:2017ozn,Bargheer:2018jvq,Bargheer:2019kxb}. In the \emph{Handling Handles} framework, this involves polygonization, hexagonalization, and sprinkling, i.e., summing over tree-level Wick-contraction graphs and their discrete moduli, decomposing the resulting geometry into hexagons, and including mirror-state contributions along the bridges. If our methods for closed geometries extend to higher genus, the last two steps may already be largely under control in the classical regime. The main open problems are the treatment of more general triangulations, a more systematic implementation of the moduli sum, and the resolution of Riemann-surface degenerations through stratification. These advances would bring the present framework closer to a genuine non-planar extension.

Finally, going beyond the classical limit remains an outstanding problem. While some quantum corrections should be captured by modifications of the TBA kernels and by fluctuations around the path-integral saddle point, others---such as flavor loops in the hexagon matrix part---lie outside the present framework. A key challenge is to understand how to incorporate these contributions efficiently. Another open question concerns whether, and in what way, wall-crossing transformations extend to loop corrections and, conversely, whether these transformations can be used to constrain the overall structure of such corrections.

\section*{Acknowledgments}

We thank Antoine Bourget, Gwena\"el Ferrando, Thiago Fleury, Philippe Di Francesco, Rinat Kedem, Shota Komatsu, Gregory Korchemsky, Gabriel Lefundes, Juan Maldacena and Alexander Tumanov for interesting discussions.
EK thanks Ivan Ip for sharing his unpublished lecture notes on cluster algebras.
DS thanks the Laboratoire de Physique de l’\'Ecole Normale Sup\'erieure for hospitality during the final stage of this work, and the Erwin Schrödinger International Institute for Mathematics and Physics (ESI), University of Vienna, for the opportunity to participate in the 2026 Thematic Programme “Amplitudes and Algebraic Geometry,” where part of this work was carried out, and for their support.
We acknowledge support from the French National Research Agency through the research grant “Observables” (ANR-24-CE31-7996) and from the CNRS through the IRP grant NP-Strong.

\appendix

\section{Octagon at strong coupling}
\label{app:cauchy}

In this appendix, we review the strong-coupling computation of the octagon using an approach similar to~\cite{Jiang:2016ulr,Bargheer:2019exp}. Recall that
\be
\label{octagonSC}
\bbO =\sum_{N\, =\, 0}^\infty \,\frac{1}{N!}
\sum_{\a} \int d\u\, \mu_{\a}(\u) \,T_{\a} (\u;\qg)\, ,
\ee
where we adopt the notation of Section~\ref{sec:Octonecut}.
The weight $T_{\a}(\u;\qg)$ is defined as the average over the two dressings,
\begin{align}
\label{Ttwodress}
T_{\textbf{a}}(\u;\qg)&=\frac{1}{2}(T^{+}_{\textbf{a}}(\u;\qg)+T^{-}_{\textbf{a}}(\u;\qg))\,,
\end{align}
so that the octagon can be written as
\begin{align}
\label{octagonFDlogtot}
\bbO\equiv\frac{1}{2}\left(\bbO^+ + \bbO^-\right)\,.
\end{align}
We are interested in the strong-coupling regime, where the measure $H_{a_i,a_j}(u_i,u_j)$ and $\mu_a(u)$ are given in \eqref{eq:mu2}. Defining shifted rapidities $u_j^{[\pm a_j]}=u_i\pm ia_j\e/2$, the $N$-magnon integrand can be written as a Cauchy determinant,
\begin{align}
 \mu_\a(\u) \,T^{\pm}_\a (\u)\,=\frac{\epsilon^N}{i^N}\det \frac{T^{\pm}_{a_i}(u_i)}{u_i^{[-a_i]}-u_j^{[+a_j]}}=\frac{\epsilon^N}{i^N}\det \frac{T^{\pm}_{a_i}(u_i)}{u_i^{[-2a_i]}-u_j}\,,
\end{align}
where, in the second equality, we shifted $u_j\to u_j-ia_j\e/2$ in the denominator and neglected the corresponding shift in the numerator.

Taking the logarithm of $\mathbb{O}^{\pm}$ in \eqref{octagonFDlogtot} isolates the connected contributions. These correspond to the longest cycles in the cycle expansion of the determinant at each $N$. In the presence of bound states, the result can be expressed in terms of a generalized Fredholm determinant,
\begin{align}
\label{octagonFD}
\mathbb{O}^\pm=\Det(1+\sum_{a>0}\KK^\pm_a)\equiv\Det(1+\KK^\pm)\;, \qquad \KK^\pm_a(u,v)=\frac{1}{2\pi}\frac{T^\pm_{a}(u)}{u^{[-2a]}-v}\,,
\end{align}
so that the connected expansion reads
\begin{align}
\label{octagonFDlog}
\log \mathbb{O}^\pm&=\sum_{N>0}\frac{(-1)^{N+1}}{N}\Tr (\KK^\pm)^N\,.
\end{align}
The sum in~\eqref{octagonFD} runs over bound states of arbitrary length. The traces in~\eqref{octagonFDlog} can be evaluated by residues,
\begin{align}
\label{GN}
\Tr (\KK^\pm)^N &=\frac{1}{i^N}\sum_{\a} \int\,\frac{du_1}{2\pi }\ldots\frac{du_N}{2\pi }\,\frac{T^\pm_{a_1}}{u_1^{[-2a_1]}-u_2}\ldots \frac{T^\pm_{a_N}}{u_N^{[-2a_N]}-u_1}\nonumber\\
 &=\sum_\a\int \,\frac{du_1}{2\pi i}\,\frac{du_2}{2\pi i}\,\frac{1}{u_1^{[-2a_1]}-u_2}\,\frac{1}{u_2^{[-2a_2-\ldots-2a_N]}-u_1}\, T^\pm_\a
=\frac{1}{\epsilon} \int \frac{du}{2\pi }\,\sum_\a \frac{1}{|\a|} \,T^\pm_\a \nonumber\\
&=\frac{1}{\epsilon} \int \frac{du}{2\pi }\,\int_0^1\frac{dt}{t} \sum_\a t^{|\a|}\,T^\pm_\a=\frac{1}{\epsilon} \int \frac{du}{2\pi }\,\int_0^1\frac{dt}{t}\, (G^{\pm}(t)-1)^N\,.
\end{align}
Here $|\a|=\sum_j a_j$, and $G^{\pm}(t)$ is the generating function for the characters,
\begin{align}
G^{\pm}(t)=\frac{(1+t\,\cY_{\pm}^{\Sph})(1+t\,\bar \cY_{\pm}^{\Sph})}{(1+t\,\cY^{\AdS})(1+t\,\cY^{\AdS})}=1+\sum_{a\geq 1} t^a\,T^{\pm}_a\,,
\end{align}
such that for fundamental particles $T_1^\pm=\cY_\pm^{\Sph}+\bar \cY_\pm^{\Sph}-\cY^{\AdS}-\bar \cY^{\AdS}$. Writing the overall factor in $T^\pm_1$ as $\textrm{q}=\exp{i\sigma^{\AdS} \cP-\ell \cE}$ and comparing with~\eqref{eq:gentrans}, the $\cY$-functions can be identified as~\eqref{lYAdS} and~\eqref{lYSph},
\begin{align}
\label{cYvalues} \cY_{\pm}^{\Sph}=\textrm{q}\exp{{\pm\varphi}+i\phi^{\Sph}}&\,,\quad \bar\cY_{\pm}^{\Sph}=\textrm{q}\exp{{\mp\varphi}-i\phi^{\Sph}}\,, \nonumber \\\cY^{\AdS}=\textrm{q}\exp{i\phi^{\AdS}}&\,,\quad
    \bar\cY^{\AdS}=\textrm{q}\exp{-i\phi^{\AdS}}\,.
\end{align}
We can now compute the octagon at leading order in $\epsilon$. Substituting the last line of~\eqref{GN} in \eqref{octagonFDlog}, we obtain
\begin{align}
\label{OctLi4}
\log \mathbb{O}^{\pm}&=-\frac{1}{\epsilon}\int \,\frac{du}{2\pi }\int_0^1 \frac{dt}{t} \sum_{N>0}\frac{(1-G^\pm(t))^N}{N}=\frac{1}{\epsilon}\int\frac{du}{2\pi}\int_0^1 \frac{dt}{t}\log G^\pm(t)\nonumber\\ 
& =-\frac{1}{\epsilon}\int\frac{du}{2\pi}\Big[\Li(-\cY_{\pm}^{\Sph})+ \Li(-\bar\cY_{\pm}^{\Sph})-\Li(-\cY^{\AdS})-\Li(-\bar\cY^{\AdS})\Big]\,,
\end{align}
in agreement with~\eqref{eq:oct-sc} after the change of variables $u=\tanh\theta$. Re-expanding in terms of the original bound state index $a$ one obtains
\begin{align}
\label{OcctLiweights}
\log \mathbb{O}^{\pm}=\frac{1}{\epsilon}\sum_{a> 0}\sum_I(-1)^{F_I+1}\int\frac{du}{2\pi} \frac{(-\cY_\pm^I)^a}{a^2}\,.
\end{align}

The interpretation of this result is that, if we work at fixed total weight $a$ of the bound states and arbitrary number $N$ of bound states, the net effect of the clustering procedure is to combine the $N$ bound states into stacks of $a$ fundamental magnons of each flavor, 
\begin{align}
\label{dressing trees}
{\rm weights} =\frac{T^\pm_a}{a}\quad \rightarrow \quad 
\sum_I(-1)^{F_I+1}\frac{(-\cY_\pm^I)^a}{a^2}\,.
\end{align}

\section{Classical $R$-matrix}
\label{app:r-matrix}

In this appendix, we present the expressions for the fundamental $R$-matrix at strong coupling. We focus on the diagonal components $\cR_{IJ}^{IJ}(u,v)$, with \(I,J \in \{\psi_{1}, \psi_{2} \,|\, \phi_{1}, \phi_{2}\}\), which are summarized in Table~\ref{table:diagonal}.
\begin{table}[h!]
\centering
\setlength{\arrayrulewidth}{1pt}
\renewcommand{\arraystretch}{2}
\begin{tabular}{ |c|c|c|c|c|  }
 \hline 
\diagbox[width=4em,height=4em]{$I$}{$J$}& $\psi_1$ & $\psi_2$ &$\phi_1$&$\phi_2$\\
 \hline
 $\psi_1$   & $-D(u,v)$   &$-\frac{D(u,v)-E(u,v)}{2}$&  $L(u,v)\xi(u)^{\qw}$& $L(u,v)\frac{1}{\xi(u)^{\qw}}$\\
 \hline
 $\psi_2$&   $-\frac{D(u,v)-E(u,v)}{2}$  & $-D(u,v)$   &$L(u,v)\xi(u)^{\qw}$&$L(u,v)\frac{1}{\xi(u)^{\qw}}$\\
 \hline
$\phi_1$ &$G(u,v)\frac{1}{\xi(v)^{\qw}}$ & $G(u,v)\frac{1}{\xi(v)^{\qw}}$& $A(u,v)\left(\frac{\xi(u)}{\xi(v)}\right)^{\qw}$&$\frac{A(u,v)-B(u,v)}{2}\frac{1}{\left(\xi(u)\xi(v)\right)^{\qw}}$\\
\hline
$\phi_2$   &$G(u, v)\xi(v)^{\qw}$ & $G(u,v)\xi(v)^{\qw}$&  $\frac{A(u,v)-B(u,v)}{2}\left(\xi(u)\xi(v)\right)^{\qw} $&$A(u,v)\left(\frac{\xi(v)}{\xi(u)}\right)^{\qw}$\\
 \hline
\end{tabular}
\caption{Diagonal components of the fundamental $R$-matrix.}\label{table:diagonal}
\end{table}
We follow the conventions of~\cite{Caetano:2016keh}, where the notation $D_{12}=D(u,v)$, etc., is used instead. The signs accompanying the fermionic components in the table account for the grading factor included in the $R$-matrix. Additional $\xi$-factors arise from the dressing procedure~\cite{Fleury:2017eph}, where $\qw=\pm 1$ and
\be
\xi(u)=\left(\frac{x^{+}}{x^{-}}\right)^{1/2}
=1+\epsilon\,\frac{i x}{x^2-1}+\mathcal{O}\!\left(\epsilon^2\right)\,,
\ee
with $x^{\pm}=x\!\left(u\pm i\epsilon/2\right)$ and $x=x(u)$ denoting the Zhukovsky variable.

Including the dynamical part of the hexagon form factor and expanding at strong coupling, one finds that all diagonal elements asymptote to unity in the limit $\epsilon\rightarrow0$. The leading corrections take the form
\be
\frac{\cR_{IJ}^{IJ}(u, v)}{h(v, u)} = 1 + \epsilon K^{IJ}(u, v) + \mathcal{O}\left(\epsilon^2\right)\, ,
\ee
where the kernel satisfies the antisymmetry property
\be\label{app:antisymmetry-K}
K^{IJ}(u, v) = -K^{JI}(v, u)\, .
\ee
From Table~\ref{table:diagonal}, we observe that the kernel is symmetric under exchange of $\psi_{1}$ and $\psi_{2}$, while exchanging $\phi_{1}$ and $\phi_{2}$ interchanges the two dressing choices. It is therefore sufficient to present the expressions for $\qw = +1$. The components independent of the dressing are
\be
\begin{aligned}
&K^{\psi_{1}\psi_{1}}(u, v) = \frac{2ixy (xy-1)}{(x^2-1)(x-y)(y^2-1)} = K(\theta, \theta') \, , \\
&K^{\psi_{1}\psi_{2}}(u, v) = \frac{2ixy(x-y)}{(x^2-1)(xy-1)(y^2-1)} = \bar{K}(\theta, \theta')  \, , 
\end{aligned}
\ee
where, in the last equalities, we have transformed to rapidity variables $\theta$ and $\theta'$ using~(\ref{eq:x-to-theta}). This yields
\be
K(\theta, \theta')=\frac{i}{2}
\cosh\theta \,
\cosh\theta'
\coth\!\left(\frac{\theta-\theta'}{2}\right)\,,\quad\bar{K}(\theta, \theta')=\frac{i}{2}
\cosh\theta \,
\cosh\theta'
\tanh\!\left(\frac{\theta-\theta'}{2}\right)\, .
\ee
For the mixed AdS–sphere kernels, we obtain
\be
\begin{aligned}
&K^{\psi_{1}\phi_{1}}(u, v)   =  \frac{2ix y^2}{(x^2-1)(y^2-1)} = k_{+}(\theta, \theta') \, , \\
&K^{\psi_{1}\phi_{2}}(u, v)   = \frac{2ix}{(x^2-1)(y^2-1)} = k_{-}(\theta, \theta') \, ,
\end{aligned}
\ee
with
\be
k_{\pm}(\theta, \theta') = -\frac{i}{2}\cosh\theta\, (\sinh\theta'\mp i)\, .
\ee
We observe that the dependence on $\theta$ and $\theta'$ is completely factorized. Moreover, since these kernels are proportional to $\cosh\theta$, the corresponding contributions in the TBA equations can be absorbed into the redefinition of the bridge lengths in~\eqref{eq:delta-l}. Similarly, using~\eqref{app:antisymmetry-K}, the sphere-AdS kernels take the form
\be
K^{\phi_{1}\psi_{1}}(u, v)=-k_{+}(\theta^{\prime}, \theta)\,,\quad
K^{\phi_{2}\psi_{1}}(u, v)=-k_{-}(\theta^{\prime}, \theta)\,.
\ee
Their overall proportionality to $\cosh\theta'$ implies that these contributions can instead be absorbed into the shift $\delta \log U$ in~(\ref{eq:delta-logU}).

The most intricate expressions are found in the scalar sector. They take the form
\be
\begin{aligned}
&K^{\phi_{1}\phi_{1}}(u, v)  = -\frac{2 i x y (x^2-x y+y^2-1)}
{(x^2-1)(y^2-1)(x-y)} = - K(\theta,\theta') - k_{+}(\theta', \theta) + k_{+}(\theta, \theta') \, , \\
&K^{\phi_{1}\phi_{2}}(u, v)  = -\frac{2 i x (x^2 y^2-x y-y^2+1)}
{(x^2-1)(y^2-1)(x y-1)} = - \bar{K}(\theta, \theta') - k_{+}(\theta', \theta) + k_{-}(\theta, \theta')\, , \\
&K^{\phi_{2}\phi_{2}}(u, v)  = -\frac{2 i (x^2 (y^2-1)+x y-y^2)}
{(x^2-1)(y^2-1)(x-y)}= - K(\theta,\theta') - k_{-}(\theta', \theta) + k_{-}(\theta, \theta')\,  . \\
\end{aligned}
\ee
The remaining components can be obtained using the symmetry properties of the kernels discussed above. Finally, the kernels satisfy the crossing symmetry relations
\be\label{eq:kernel-crossing-sym}
K^{IJ}(\theta+i\pi, \theta') = K^{J\bar{I}}(\theta', \theta) = - K^{\bar{I}J}(\theta, \theta')\, , 
\ee
where $\bar{I}$ is defined as $\bar{\psi}_{1} = \psi_{2}, \bar{\phi}_{1} = \phi_{2}$, and similarly under $1\leftrightarrow 2$. These relations play a key role in the derivation of the $Y$-system relations~\eqref{YsysAdSSph}.

\section{TBA equations and tree expansion}
\label{app:TBAtree}

As alluded to in the main text, the strong coupling expansion in Section~\ref{sec:EFD} can be reformulated as an expansion over trees. In Section~\ref{sec:Diagonal}, this tree expansion was used to show that only the diagonal part of the scattering matrix contributes in the strong-coupling limit. 
Here, we outline the main ingredients for deriving the TBA equations from~\eqref{eq:simplifiedNcutpartitionfunction} within this framework.

Each tree is constituted by vertices and branches, and they contain no cycles. The power counting in $\epsilon$ is exactly the same as in ~\eqref{eq:tree_count}. Here, after taking the diagonal approximation for the scattering matrices and performing clustering, vertices correspond to stacks of $a$ fundamental magnons with weight
\be
\label{eq:vertextree}
\textrm{vertex}_{\, (\gamma,I, a,\theta)}\ \to \ (-1)^{F_I+1} \frac{[-\cY^I_{\gamma}(\theta)]^a}{a^2}\equiv(-1)^{F_I}\,\cY_{\gamma,a}^I(\theta)\, ,
\ee
where $\gamma$ labels the edge, $I$ the flavor index, and $a$ the stack index. Branches connect vertices and they are assigned the weight
\be
\textrm{branch}_{\,( \gamma,I, a,\theta) \to (\gamma',J,b,\theta')}\ \to -ab\,K_{\gamma \gamma'}^{IJ}(\theta,\theta')=-ab\,\langle  \gamma,\gamma'\rangle \,K^{IJ}(\theta,\theta')\,.
\ee
The expansion of the $\NN$-cut partition function \eqref{eq:simplifiedNcutpartitionfunction} takes the form of a sum over forests, where each connected component is a tree.

We define a rooted tree $(-1)^{F_I}\,Y^I_{\gamma,a}(\theta)$ as a tree whose root is the vertex $(\gamma,I,a,\theta)$, with an arbitrary number of attached branches. At each vertex other than the root we sum over  $(\gamma',J,b,\theta')$ .
Rooted trees satisfy a Schwinger-Dyson-type equation, reflecting the fact that any tree can be constructed recursively from smaller trees~\cite{Kostov:2017vwz}. In the present context, this equation takes the form
\be
\label{TBAbsTree}
Y^I_{\gamma,a}(\theta)=\cY^I_{\gamma,a}(\theta)
\exp\Big[a \sum_{\gamma',J,b} (-1)^{F_J+1}b\,K_{\gamma\gamma'}^{IJ}\star Y^J_{\gamma',b}(\theta)\Big]\,.
\ee
The equations for $a>1$ can be derived from the one for $a=1$, corresponding to the fundamental $Y$-functions $Y^I_{\gamma,1}(\theta)\equiv Y^I_{\gamma}(\theta)$, by setting
\be
Y^I_{\gamma,a}(\theta)=-\frac{[-Y_{\gamma}^I(\theta)]^a}{a^2}\,.
\ee
This relation is consistent with the definition of the bare weights $\cY^I_{\gamma,a}$ for the stack of $a$ magnons in~\eqref{eq:vertextree}, and ensures that the sum over $b$ in the exponent of \eqref{TBAbsTree} reconstructs the expansion of $\log{(1+Y_{\gamma'}^J)}$.

The fundamental $Y$-functions then satisfy
\be
\label{TBAfdtTree}
Y^I_{\gamma}(\theta)=\cY^I_{\gamma}(\theta)
\exp\Big[ \sum_{\gamma',J} (-1)^{F_J+1}\,K_{\gamma\gamma'}^{IJ}\star\log \left(1+ Y^J_{\gamma'}\right)(\theta)\Big]\,.
\ee
which is equivalent to the TBA equation \eqref{eq:tbafirstform}.

To obtain the expression for the action, one observes that summing (and integrating) over the variables associated with the root $Y^I_{\gamma,a}$ effectively removes the root, thereby turning rooted trees into unrooted ones. The combinatorial counting of such trees differs from that in~\cite{Kostov:2017vwz}, and leads to the appearance of the Rogers dilogarithm~\eqref{eq:RogerDi}.
A related tree representation for the solution to the GMN equations was discussed in~\cite{neitzke2013}.

\section{Free energy in various forms}\label{app:simpler-action}

In this appendix, we discuss various rewritings of the free energy~$\cA$ and compute its variation.

Let us start by giving the main steps that allow one to recast the free energy into the form~\eqref{nicer-action}. The starting point is the general formula~\eqref{eq:actionasrogers}. Performing an integration by parts on the integrals containing the dilogarithms, we obtain
\begin{align}
\label{eq:dilog-app}
& \int \frac{d\theta}{2\pi \cosh^2{\theta}} \textrm{Li}_{2}(-Y^{I}_{\gamma}(\theta))  =  \int \frac{d\theta}{2\pi}\, \textrm{tanh}{\,\theta}\, L_{\gamma}^{I}(\theta) \, \partial_{\theta} \log{Y^{I}_{\gamma}(\theta)} \\ \nonumber
 = & \int \frac{d\theta}{2\pi} \, \textrm{tanh}{\,\theta} \, L_{\gamma}^{I}(\theta) \, \partial_{\theta}(Z_{\gamma} e^{\theta}+\bar{Z}_{\gamma} e^{-\theta})  - \sum_{\gamma',J}(-1)^{F_{J}} \int \frac{d\theta}{2\pi} \, \textrm{tanh}{\,\theta}\, L_{\gamma}^{I}(\theta)\, \partial_{\theta} G_{\gamma \gamma'}^{IJ}* L_{\gamma'}^{J} (\theta)\, .
\end{align}
In the last step, we used the TBA equations~\eqref{eq:TBA-GMN}, rewritten in the more compact form
\be
\log{Y_{\gamma}^{I}(\theta)} = C^{I}_{\gamma} + Z_{\gamma} e^{\theta} + \bar{Z}_{\gamma} e^{-\theta} +\sum_{ \gamma', J} \GMN_{\gamma \gamma'}^{I J}* L_{\gamma'}^{J}(\theta)\, ,
\ee
where $G_{\gamma \gamma'}^{IJ}$ is a block-diagonal kernel defined analogously to~\eqref{eq:Kerblock}, with the replacement $\qK, \bqK\rightarrow \GMN, \bGMN$. Summing~\eqref{eq:dilog-app} over $(\gamma, I)$ with weight $(-1)^{F_{I}}$ and using the symmetry of the integrals and sums, we can replace
\be
\textrm{tanh}{\,\theta} \,\partial_{\theta} \GMN_{\gamma \gamma'}^{IJ}(\theta-\theta') \,\, \rightarrow \,\,\frac{1}{2} (\textrm{tanh}{\,\theta}\, \partial_{\theta}+\textrm{tanh}{\,\theta'}\, \partial_{\theta'}) \GMN_{\gamma \gamma'}^{IJ}(\theta-\theta') = -\frac{\tilde{\qK}_{\gamma \gamma'}^{IJ}(\theta, \theta')}{2\cosh^{2}{\theta} \cosh^{2}{\theta'}} \, ,
\ee
with $\tilde{\qK}_{\gamma \gamma'}^{IJ}$ given in~\eqref{eq:Kerblock}. This identity, together with the TBA equations~\eqref{eq:tbahexagon}, allows us to rewrite the last term in the final line in~\eqref{eq:dilog-app} into the form
\be
-\frac{1}{2} \sum_{\gamma,I} (-1)^{F_{I}} \int \frac{d\theta}{2\pi \cosh^2{\theta}} \log{\left(Y_{\gamma}^{I}(\theta)/\tilde{\cY}_{\gamma}^{I}(\theta)\right)} L_{\gamma}^{I}(\theta)\, ,
\ee
where the $\tilde{\cY}$'s denote the modified driving terms, see eqs.~\eqref{eq:shifted-driving-terms},~\eqref{eq:delta-l},~\eqref{eq:delta-logU} and~\eqref{eq:kappa}. Adding the remaining contributions to the free energy and using the relation
\be
\textrm{tanh}{\, \theta}\, \partial_{\theta} (Z_{\gamma} e^{\theta}+\bar{Z}_{\gamma}e^{-\theta}) = Z_{\gamma} e^{\theta} + \bar{Z}_{\gamma} e^{-\theta} -\frac{Z_{\gamma}+\bar{Z}_{\gamma}}{\cosh{\theta}}\, ,
\ee
we find
\begin{align}
\nonumber
\cA = \sum_{\gamma,I} (-1)^{F_{I}} &\int \frac{d\theta}{2\pi} (Z_{\gamma} e^{\theta}+\bar{Z}_{\gamma}e^{-\theta}) L_{\gamma}^{I}(\theta)\,\,\, - \sum_{\gamma,I} (-1)^{F_{I}} \int \frac{d\theta}{2\pi \cosh{\theta}}(Z_{\gamma}+\bar{Z}_{\gamma}) L_{\gamma}^{I}(\theta) \\
&\,\,\, + \frac{1}{2}\sum_{\gamma,I} (-1)^{F_{I}} \int \frac{d\theta}{2\pi\cosh^2{\theta}} \log{\left(\tilde{\cY}_{\gamma}^{I}(\theta)/\cY_{\gamma}^{I}(\theta)\right)} L_{\gamma}^{I}(\theta)\, .
\end{align}
Evaluating the last term in this equality using
\begin{align}
\nonumber
&\frac{1}{2}\sum_{\gamma,I} (-1)^{F_{I}} \int \frac{d\theta}{2\pi\cosh^2{\theta}} \log{\left(\tilde{\cY}_{\gamma}^{I}(\theta)/\cY_{\gamma}^{I}(\theta)\right)} L_{\gamma}^{I}(\theta)  = -\frac{1}{2} \sum_{\gamma,I}(-1)^{F_{I}} \delta l_{\gamma} \int \frac{d\theta}{2\pi \cosh{\theta}} L_{\gamma}^{I}(\theta)\\ \nonumber
& \!\!+ \frac{1}{2}\sum_{\gamma} \delta \log{U_{\gamma}} \int \frac{d\theta}{2\pi \cosh^2{\theta}} \left[\left(\qw+i\sinh{\theta}\right)L_{\gamma}^{\Sph}(\theta) + \left(-\qw+i\sinh{\theta}\right)\bar{L}_{\gamma}^{\Sph}(\theta) \right]\!  =\! -\!\sum_{\gamma} \delta l_{\gamma}\kappa_{\gamma}\, ,
\end{align}
and recalling that $Z_{\gamma}+\bar{Z}_{\gamma} = -l_{\gamma}-\delta l_{\gamma}$, we verify eq.~\eqref{nicer-action}.

Another expression for the free energy can be obtained by changing variables from the $Y$-functions to $\chi$-functions, as previously done in~(\ref{eq:area-A-form}). An alternative representation may also be derived starting from~\eqref{eq:actionasrogers}. Performing an integration by parts, we obtain
\be
\cA = -\frac{1}{2}\int \frac{\tanh \theta}{2\pi}  \, df \, ,
\ee
where
\be\label{eq:action-Y}
df \equiv 2 \sum_{\gamma}\str\,  d R (-Y_{\gamma})=\sum_{\gamma}\str 
\Big[\log Y_\gamma\wedge \log (1+Y_\gamma)-d\,\left(\log \cY_\gamma \,\log(1+Y_\gamma)\right)\Big]\, ,
\ee
with the shorthand notation
\be\label{eq:wedge}
g\wedge h\equiv g\,dh-dg\,h=-h\wedge g\, .
\ee
This expression can be rewritten in terms of the $\chi$- and $A$-functions as
\begin{align}
\label{eq:action-chi-A}
df  =  
\sum_E \str\Big[& \log \chi_E\wedge \log(1+A_E) - d\left(\log \cY_E\,\log(1+A_E)\right)\Big]
\nonumber\\
&+\frac{1}{2}\sum_{E,E'}\langle E,E'\rangle \, \str\Big[ \log (1+A_E)\wedge \log(1+A_{E'})\Big]\,,
\end{align}
where $E$ and $E'$ run over the edges in the triangulation. This result follows from~\eqref{eq:chi-Y-sys} and~\eqref{eq:A-to-Y}, together with the decompositions
\be\label{eq:decompositions}
\log \cY_\gamma=\sum_E \delta_{\gamma, E} \log \cY_E\, , \qquad \log \chi_\gamma=\sum_E \delta_{\gamma, E} \log \chi_E\, ,
\ee
where $\delta_{\gamma,E}=1$ if $\gamma_E\in\gamma$, and zero otherwise. We also use the notation $\langle E,E' \rangle\equiv \langle \gamma_E,\gamma_{E'} \rangle$. 

Finally, let us provide a proof of formula~\eqref{eq:vary-A} giving the variation of the free energy with respect to external parameters. When varying the free energy~\eqref{eq:actionasrogers},
\be\label{eq:delta-A-app}
\delta \cA = \sum_{\gamma, I} (-1)^{F_{I}} \int\frac{d\theta}{2\pi \cosh^2{\theta}} \delta R(Y^{I}_{\gamma})\, ,
\ee
we adopt the notation~\eqref{eq:wedge}, replacing $d\to \delta$, to write
\begin{align}
\label{eq:var-R}
\delta R(Y^{I}_{\gamma}) 
& 
= \frac{1}{2} \left( \log{(Y^{I}_{\gamma}/\cY^{I}_{\gamma})}\, \delta L^{I}_{\gamma} - \delta \log{(Y^{I}_{\gamma}/\cY^{I}_{\gamma})} \, L_{\gamma}^{I}\right) -\delta \log{\cY^{I}_{\gamma}} \, L_{\gamma}^{I}\,  \\ \nonumber
& 
=\frac{1}{2}\log(Y^{I}_{\gamma}/\cY^{I}_{\gamma})\wedge L^{I}_{\gamma}- \delta \log{\cY^{I}_{\gamma}} \, L_{\gamma}^{I}\,.
\end{align}
The final term reproduces precisely the contribution appearing in~\eqref{eq:vary-A} after summing over $\gamma \in \cB$ and converting the sum into a sum over edges $E$ in the triangulation. This conversion uses~\eqref{eq:A-to-Y-closed} and the first relation in~\eqref{eq:decompositions}.%
\footnote{We also use that the monodromy $\mu^{I}_{P}$ in~\eqref{eq:A-to-Y-closed} is independent of the flavor $I$ for any puncture $P$, and that the combination $\sum_{I}(-1)^{F_{I}}\log{\cY^{I}_{E}} =0$ for any edge $E$.}

To show that the remaining terms vanish, we substitute them into~\eqref{eq:delta-A-app} and apply the TBA equations~\eqref{eq:tbafirstform}. This yields
\be
\begin{aligned}
&-\frac{1}{2} \sum_{\gamma,\gamma',I,J} (-1)^{F_{I}+F_{J}} \int \frac{d\theta}{2\pi \cosh^{2}{\theta}} \left( \, \qK^{IJ}_{\gamma \gamma'} \star L^{J}_{\gamma'}(\theta)\wedge  L_{\gamma}^{I}(\theta) \right) \\
&= \frac{1}{2} \sum_{\gamma,\gamma',I,J} (-1)^{F_{I}+F_{J}} \int \frac{d\theta\, d\theta'}{(2\pi)^2 \cosh^{2}{\theta} \cosh^{2}{\theta'}} K^{IJ}_{\gamma \gamma'}(\theta, \theta')\, L^{I}_{\gamma}(\theta) \wedge L^{J}_{\gamma'}(\theta')\, .
\end{aligned}
\ee
This integral then vanishes due to the symmetry of the kernel, $\qK^{IJ}_{\gamma \gamma'}(\theta, \theta') = \qK^{JI}_{\gamma' \gamma}(\theta', \theta)$, and the antisymmetry of the wedge product.

\section{Constant $Y$-system solutions}\label{app:constant-Y}

In this appendix, we construct solutions to the TBA equations for polygons in the ``high-temperature'' limit, where the central charges $Z_{E}, \bar{Z}_{E}\rightarrow 0$ for all edges $E$. As for null polygonal Wilson loops~\cite{Alday:2010vh,Alday:2009dv}, in this regime, the $Y$-functions simplify and become independent of the rapidity $\theta$.

In our case, the phenomenon must happen in both the AdS and sphere sectors simultaneously, since both share the same central charges. In particular, since it applies in the sphere sector, derivatives of $\chi^{\Sph}_{E}$ should vanish, meaning that all bridge lengths $l_{E}$ should be zero (see eq.~\eqref{eq:chi-derivative}). Furthermore, the polygon should be regular, that is, dihedrally symmetric, as in the Wilson loops context.

The $Y$-system~\eqref{YsysAdSSph} provides a particularly powerful framework for describing this class of solutions. With vanishing bridge lengths, we could work with any triangulation of the polygon. For simplicity, we work with the fan triangulation shown in the left panel of Figure~\ref{fig:fan}. The \textit{constant} $Y$-system ($Y^{++}_{j} = Y_{j}$) then takes the simple form
\be
\label{eq:Ysyslinearfan}
Y_{j}\bar{Y}_{j} = (1+Y_{j-1})(1+\bar{Y}_{j+1})\, ,\qquad j = 1, \ldots, n-3\,,
\ee
together with analogous relations obtained by exchanging $Y\leftrightarrow \bar{Y}$. Here, $j$ labels the internal edges of the $n$-gon, ordered in the fan triangulation from bottom to top. At the boundaries, we impose $Y_0=Y_{n-2} = 0$.

Both the $Y$-functions for the AdS and sphere sectors solve equations~\eqref{eq:Ysyslinearfan}, differing only in the interpretation of the solution in terms of spacetime and R-space cross ratios, respectively. For simplicity, we restrict to configurations with $\qU_j = 1$ for all $j$. Under this condition, the solutions in the two sectors can be mapped directly by exchange of the corresponding cross ratios, $(\phi^{\AdS}_j, \sigma^{\AdS}_j) \leftrightarrow (\phi^{\Sph}_j, \sigma^{\Sph}_j)$, and moreover all dependence on the dressing parameter $\qw$ drops out. 

\begin{figure}[h]
\centering
\includegraphics[width=12cm]{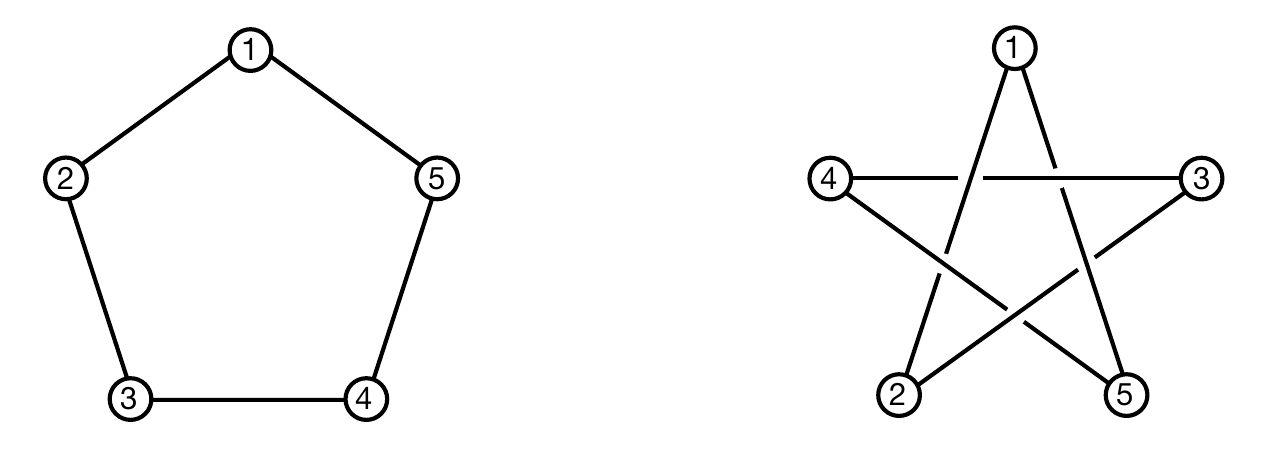}
\caption{Cartoon of regular pentagons for $k=1$ (left) and $k=2$ (right).}
\label{fig:regular}
\end{figure}

\paragraph{Solutions.} The structure of the solutions depends on the parity of $n$. We begin with the odd case, where the solutions are isolated, and then turn to the even case, which supports a richer family of configurations.

For odd $n$, a constant solution requires
\be
Y_j = \bar{Y}_j\, , \qquad j =1, \ldots , n-3\, .
\ee
As discussed before, this condition coincides with the restriction to one-dimensional kinematics, in which all vertices lie on a line (up to conformal transformations). In this case, the problems reduces to the familiar $SU(2)$ constant $Y$-system equations,
\be
Y^2_{j} = (1+Y_{j-1})\,( 1+Y_{j+1})\, ,
\ee
with boundary conditions $Y_0 = Y_{n-2} = 0$. The solution can be conveniently expressed in terms of $Q$-functions,
\be
Y_{j} = Q_{j-1} Q_{j+1}\, ,
\ee
which satisfy the $QQ$-relations
\be
\label{QQsys}
Q_{j}^2 = Q_{j+1} Q_{j-1} + 1\, .
\ee
Setting $Q_{-1}=0$ and $Q_{0}=1$ ensures that $Y_0= 0$, and eq.~\eqref{QQsys} is solved by the character
\begin{align}\label{eq:Q-character}
Q_j=\frac{t^{j+1}-t^{-j-1}}{t-t^{-1}}\,.
\end{align}
Imposing the remaining boundary condition $Y_{n-2}=0$ requires that $Q_{n-1}=0$, which leads to $t^{2n}=1$.
Focusing on distinct, non-degenerate solutions with $Y_j\neq 0$ for all $j=1,\ldots,n-3$, we find that such solutions form a discrete family labeled by an integer $k$ such that
\be
t = e^{\pi i k/n}, \qquad k = 1, \ldots, \lfloor n/2 \rfloor\, ,
\ee
with $k$ coprime with $n$.

As mentioned earlier, the geometric interpretation of these solutions is in terms of regular polygons~\cite{Alday:2010vh}. The case $k=1$  corresponds to the ordinary convex regular $n$-gon, while $k>1$ gives rise to folded regular polygons, in which each edge of the contour passes over $2(k-1)$ other edges. An example for $n=5$ is shown in Figure~\ref{fig:regular}. Relaxing the coprime condition for $k$ produces degenerate solutions in which some $Y$-functions vanish; geometrically, this corresponds to configurations where vertices coincide and the polygon collapses into a union of smaller polygons.

The discrete solutions described above remain valid when $n$ is even. In this case, however, they represent only special points within larger one-parameter families of solutions, corresponding to regular polygons with vertices in two-dimensional kinematics. In this setting, $Y_{2j} = \bar Y_{2j}$, and the $Y$-system reduces to 
\begin{align}
\label{eveny}
Y_{2j-1} \bar{Y}_{2j-1}&=(1+Y_{2j-2})(1+Y_{2j+2})\, , \\ \nonumber
Y_{2j}^2=(1+Y_{2j-1})&(1+\bar Y_{2j+1})=(1+\bar Y_{2j-1})(1+Y_{2j+1})\,.
\end{align}
The solutions depend on a continuous parameter $\zz$, defined by
\begin{align}\label{eq:z-cst-sol}
\zz^2\equiv \frac{1+Y_{2j-1}}{1+\bar Y_{2j-1}}\,, \qquad \forall j = 1, \ldots, (n-2)/2\, .
\end{align}
Using the ansatz
\be
\label{oddnyq}
Y_{j}=\bar Q_{j-1}\,Q_{j+1}\,,
\ee
and analogously for $\bar{Y}$ with $Q\leftrightarrow \bar{Q}$, the system~\eqref{eveny} is solved if the $Q$-functions satisfy
\be
\begin{aligned}
\label{evennq}
1+\bar Q_{2j-2}\,Q_{2j}&=\zz\,\bar Q_{2j-1}\,Q_{2j-1}\,,\\
1+Q_{2j-2}\,\bar  Q_{2j}&=\frac{1}{\zz}\,\bar Q_{2j-1}\,Q_{2j-1}\,,\\
1+\bar Q_{2j-1}\,  Q_{2j+1}&=1+Q_{2j-1}\,  \bar Q_{2j+1}=\bar Q_{2j}\, Q_{2j}\,.
\end{aligned}
\ee
Its general solution is given by
\begin{align}
\label{evennsol}
&Q_{2j}=\zz^{j}\left(\frac{t^{2j+2}-t^{-2j-2}}{t^2-t^{-2}}+\zz\frac{t^{2j}-t^{-2j}}{t^2-t^{-2}}\right)\,,\\
&Q_{2j-1}=\frac{t^{2j}-t^{-2j}}{t^2-t^{-2}}(t+\zz t^{-1})\,,
\end{align}
with the barred expressions following from the replacement $\zz\rightarrow 1/\zz$. The boundary conditions take the same form as before and again require $t = e^{\pi ik/n}$.

When $\zz= 1$, the system reduces to the one-dimensional kinematics previously discussed, with solution given by the character form~\eqref{eq:Q-character}. The case $\zz=-1$ yields another character solution, related to the former by $k\rightarrow n/2-k$. The full solution interpolates continuously between these two special points.

For simplicity, in what follows we restrict to real solutions where $Y$ and $\bar{Y}$ are complex conjugate, and we parametrize $\zz = e^{i\psi}$. We also assume that $k$ is coprime with $n$ and focus on the branch $\psi \in (-\psi_{c}, \psi_{c})$ with $\psi_{c} = (n-2k)\pi/n$. Over this domain, the solution is non-degenerate and its free energy is smooth for any $\psi$.%
\footnote{The endpoints of the domain correspond to $\zz = -t^{\mp 2}$, where all $(n-2)$ odd $Y$-functions equal $-1$ while all even ones vanish.}

To relate the solution parameters to the polygon geometry, we introduce the $\chi$-functions. For the fan triangulation, they take the simple form
\be
\chi_{j} = \frac{Y_{j}}{1+Y_{j-1}}\, .
\ee
Substituting~\eqref{oddnyq} and using~\eqref{evennq}, one finds
\be\label{eq:chi-from-Q}
\chi_{2j-1} = \frac{Q_{2j}}{Q_{2j-2}}\, , \qquad \chi_{2j} = \frac{1}{\zz} \frac{Q_{2j+1}}{Q_{2j-1}}\, ,
\ee
and similarly for barred variables with $\bar{\zz} = 1/\zz$. The $\sigma$- and $\phi$-parameters follow from $\chi_j = e^{i\phi_j-\sigma_j}$ and $\bar{\chi}_j = e^{-i\phi_j-\sigma_j}$. In particular, for $n$ even,
\be\label{eq:psi-to-phi}
\psi = \frac{2(\phi_{1} + \phi_{3} + \ldots + \phi_{n-3})}{n}\, ,
\ee
along with $\sigma_{1}+\sigma_{3}+\ldots + \sigma_{n-3} = 0$. These relations follow from
\be
\chi_{1} \chi_{3}\ldots \chi_{n-3} = \frac{Q_{n-2}}{Q_{0}} = \zz^{n/2}\, .
\ee
using~\eqref{eq:chi-from-Q} and the telescoping property of the $Q$-function products.

This concludes the construction of the constant solutions, which may be applied to either the AdS sector or the sphere sector.

\paragraph{Free energy.} We proceed with the calculation of the free energy. The key quantity is given by
\be
\begin{aligned}\label{eq:cst-An}
A_{n} :=\sum_{j\,=\,1}^{n-3}\int \frac{d\theta}{2\pi \cosh^2{\theta}} \left[\textrm{Li}_{2}(-Y_j) + \frac{1}{2}\log{\left(\frac{Y_{j}}{\cY_j}\right)}\log{(1+Y_{j})}+ \left(Y_j \rightarrow \bar{Y}_j, \cY_j \rightarrow \bar{\cY}_j\right) \right]\, .
\end{aligned}
\ee
For $n$ even, one may eliminate the $\cY$'s from this formula using
\be
\cY_{1}\cY_{3}\ldots \cY_{n-3} = \frac{Y_{1}Y_{3}\ldots Y_{n-3}}{(1+Y_{2})(1+Y_{4})\ldots (1+Y_{n-4})}\, ,
\ee
and analogously for $(Y, \cY)\rightarrow (\bar{Y}, \bar{\cY})$, which follows from the equations above and $l_{j} = 0, \forall j$. It implies
\be
\sum_{j\, =\, 1}^{n-3}\log{\left(\frac{Y_j \bar{\cY}_j}{\bar{Y}_j \cY_j}\right)} \log{\left(\frac{1+Y_{j}}{1+\bar{Y}_j}\right)} = 0\, ,
\ee
using $Y_{2j} = \bar{Y}_{2j}$ and eq.~\eqref{eq:z-cst-sol}. For $n$ odd this equation is automatically satisfied. Substituting this relation into~\eqref{eq:cst-An} and performing the integral over $\theta$ yields the simpler representation
\be
A_{n} = \frac{1}{\pi} \sum_{j\,=\,1}^{n-3} \left[\textrm{Li}_{2}(-Y_{j})+\textrm{Li}_{2}(-\bar{Y}_{j}) + \frac{1}{4}\log{\left(Y_j \bar{Y}_j\right)}\log{(1+Y_j)(1+\bar{Y}_j)}\right]\, .
\ee
Now, the sum over constant $Y$-functions can be evaluated (numerically) and expressed in a compact form.

For the odd-$n$ solution, as well as for the even-$n$ solution with $\zz=1$, the sum can be taken using Rogers dilogarithm identities, familiar from the study of the high-temperature limit of relativistic integrable systems. They yield
\be
A_{n, k} = -\frac{\pi(6  k (k - n) + n (1 + n))}{3n} \, . 
\ee
For even $n$, we must also take into account the dependence on $\zz$. Remarkably, once we set $\zz = e^{i\psi}$, this dependence simplifies and becomes quadratic in $\psi$, as it was the case for polygonal Wilson loops~\cite{Alday:2010vh}. By numerically evaluating the sums for various $n \geqslant 4$ and $k=1, \ldots , (n-2)/2$, and fitting the results, we find
\be
A_{n, k}(\psi) = -\frac{\pi(6  k (k - n) + n (1 + n))}{3n} + \frac{n \psi^2}{2\pi}\, ,
\ee
with $\psi \in (-\psi_{c}, \psi_{c})$. In particular, for $n=4$, only $k=1$ is possible, and the solution takes the simple form
\be
A_{n\, =\, 4} =-\frac{\pi}{6} +\frac{2\psi^2}{\pi}\, ,
\ee
with $\psi\in (-\pi/2, \pi/2).$

To obtain the full result for a constant solution, we include both the AdS and sphere contributions,
\be
\cA_n = A_{n, k}(\psi)^{\Sph} - A_{n, k}(\psi)^{\AdS} \, .
\ee
The parameter $n$ is the same in the two sectors, but the other parameters $k$ and $\psi$ may differ. This gives
\be
\cA_n = \frac{2\pi(k_{\AdS} (k_{\AdS} - n)-k_{\Sph} (k_{\Sph} - n))}{n} + \frac{n (\psi_{\Sph}^2-\psi^2_{\AdS})}{2\pi}\, .
\ee
For $n=4$, one finds simply
\be
\cA_{n\, =\, 4} = \frac{2(\psi^2_{\Sph}-\psi^2_{\AdS})}{\pi}\, .
\ee
This expression gives the free energy of the octagon with constant sources ($\sigma = l =0$)
\be
\cY^{\AdS} = e^{i\phi^{\AdS}}\, , \,\, \bar{\cY}^{\AdS} = e^{-i\phi^{\AdS}}\, , \,\, \cY^{\Sph} = e^{i\phi^{\Sph}}\, , \,\, \bar{\cY}^{\Sph} = e^{-i\phi^{\Sph}}\, ,
\ee
with $\phi^{\AdS} = 2\psi_{\AdS}$ and $\phi^{\Sph} = 2\psi_{\Sph}$.

\section{Non-standard TBA for cyclic quivers}
\label{app:TBAcycles}
\begin{figure}[h]
    \centering
 \includegraphics[width=0.55\linewidth]{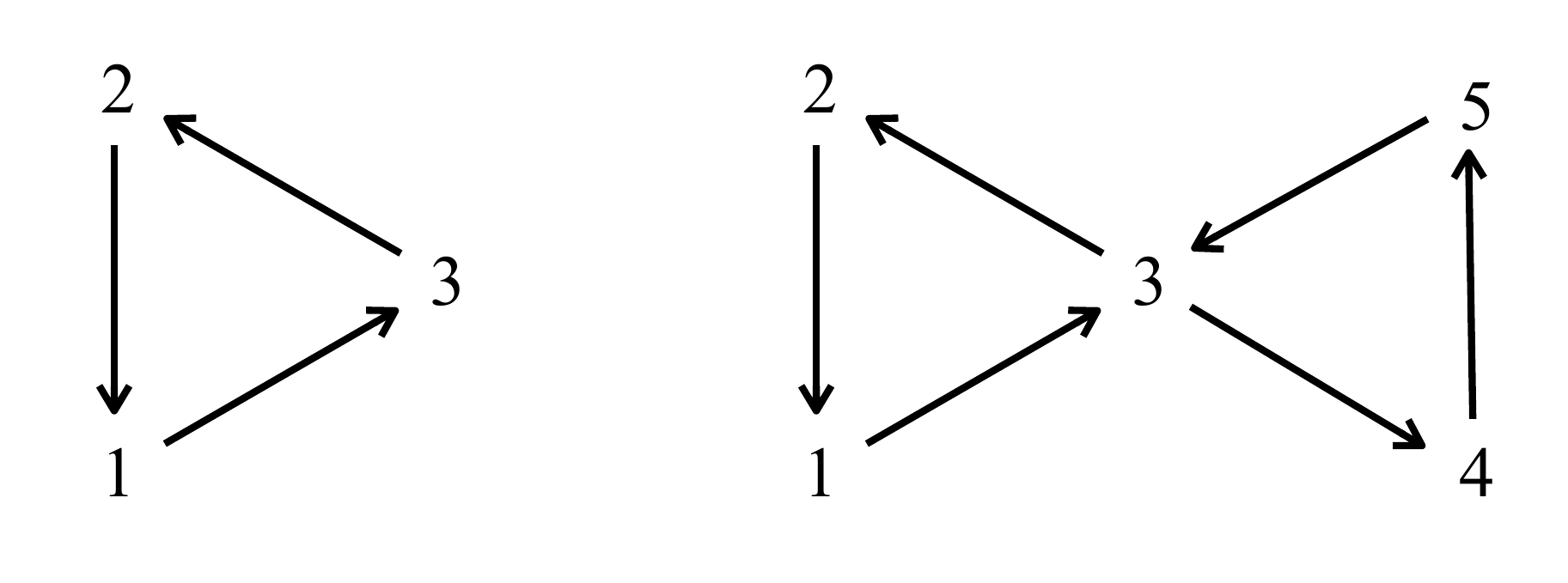}
    \caption{Quivers corresponding to the cyclic configurations considered in this appendix. \textbf{Left:} cyclic $A_3$;  \textbf{right:} $A_5$ with two cycles.
     }
    \label{fig:A3A4A5}
\end{figure}

In this appendix, we analyze the TBA equations for polygon triangulations within a nonstandard framework, where Feynman-like prescriptions~\eqref{eq:Feynman-presc} are employed to regulate the decoupling singularities that arise in the kernels. The cases of interest correspond to quivers with cycles. For clarity, we focus on the simplest triangulations involving cyclic quivers, namely those containing one and two 3-cycles, as illustrated in Figure~\ref{fig:A3A4A5}. These examples correspond to systems with three cuts $(A_3)$ and five cuts $(A_5)$, respectively.

As we will show, reformulating the TBA equations in this Feynman-like framework leads to the appearance of new types of ``composite'' $Y$-functions. These differ from those encountered in the wall-crossing approach: they are associated with nonlocal states supported on 3-cycles and satisfy distinct defining (or composition) relations that express them in terms of the elementary $Y$-functions.

To streamline the notation, we label the edges of the triangulation (equivalently, the nodes of the quiver) by $i$, and denote the corresponding $Y$-functions by
\be
Y_{\gamma_i}\ \rightarrow \ Y_i\,.
\ee
Similarly, we write $Y_{ij\ldots}$ for the composite $Y$-functions associated with states of charges $\gamma_{i}+\gamma_{j}+\ldots\,$. Their zero-mode contributions satisfy the standard multiplicative relation
\be\label{eq:driving-terms-composite}
\cY_{ij...} = \cY_{i}\cY_{j}...\, .
\ee
In this framework, the Feynman prescription assigns to the kernel $K_{\gamma\gamma'}(\theta, \theta')$ an $\pm i0$ deformation of the form $\theta\rightarrow \theta+i\langle \gamma, \gamma'\rangle 0$, where the sign is determined by the pairing. As a result, there is no fixed ordering of integration contours; instead, all contours may be taken along the real rapidity axis. Consequently, the standard wall-crossing formalism and the associated KS transformations are not directly applicable.

Nevertheless, one can derive relations connecting the $Y$-functions in this non-standard framework to the corresponding $\chi$-functions, thereby establishing the equivalence between the two descriptions, as we demonstrate below for the specific cases under consideration.

Aside from the modified contour prescription, the TBA equations for the elementary functions $Y_i$ retain the same structure as in the standard formulation, with the right-hand side involving a sum over a suitable set of states. For brevity, we present only the associated $Y$-system relations. Since these take the same form in both the AdS and sphere sectors, we suppress the flavor index throughout.

\paragraph{Cyclic $A_{3}$.} As discussed in Section~\ref{sec:generaltriangulations}, a peculiarity of using the Feynman prescription to regulate the poles in the kernels of the TBA equations is the appearance of spurious contact terms, which are in tension with the expected decoupling properties of hexagon form factors. The simplest example arises in a three-magnon configuration, with one magnon located on each edge of the inner triangle of the 3-cycle triangulation.

The connected contribution of this configuration to the free energy can be extracted by expanding in the small-$\cY$ regime and isolating terms proportional to $\cY_{1}\cY_{2}\cY_{3}$. Restricting to three identical magnons (say $\psi_1$), this contribution takes the form of a triple integral over the corresponding rapidities,
\be
\mathcal{I}_{123} = \int \prod_{i=1}^{3}\frac{\cY_i(\theta_i)d\theta_{i}}{(2\pi)\cosh^{2}{\theta_{i}}} (K_{12}K_{23}+\textrm{cyclic})\, ,
\ee
where
\be
K_{ij} = \langle \gamma_{i}, \gamma_{j}\rangle K(\theta_{i}, \theta_{j})\, .
\ee
We fix the orientation of the circular quiver such that
\be 
\label{eq:A3quiver}
\langle \gamma_1, \gamma_2\rangle=\langle \gamma_2, \gamma_3\rangle=\langle \gamma_3, \gamma_1\rangle=1\,.
\ee
With the Feynman prescription, each kernel is evaluated with a shifted argument $\theta_{i} \rightarrow \theta_{i}+i\langle \gamma_i, \gamma_j\rangle 0$. While this renders the integrals well-defined, the product of such kernels generates an ultra-local contribution in which all three $Y$-functions are evaluated at the same rapidity,
\be\label{eq:spurion}
\mathcal{I}_{123} = \int \frac{d\theta}{2\pi \cosh^{2}{\theta}} \cY_1(\theta) \cY_2(\theta) \cY_3(\theta) + (\textrm{regular triple integrals})\, ,
\ee
in addition to genuine triple integrals with smooth integrands.

The problem with the ultra-local term is that it does not vanish in the OPE limit (see Figure~\ref{fig:hexagonOPElimit}). In this regime, $\cY_{1}$ and $\cY_{2}$ oscillate rapidly, while their product $\cY_{1}\cY_{2}$ remains of order one. As a result, the smooth triple integrals average to zero, as expected, whereas the ultra-local term persists, in contradiction with the expected OPE factorization.

This issue can be resolved by introducing a counterterm in the form of a new $Y$-function $Y_{123}$, defined by
\be\label{eq:Y123-smallY}
\quad Y_{123}(\theta) \approx -\cY_{1}(\theta) \cY_{2}(\theta) \cY_{3}(\theta)\, ,
\ee
where the minus sign is chosen to cancel the unwanted contribution. Assuming that the overall structure of the TBA equations is preserved, one is naturally led to extend the system by including a composite state $\gamma_{123} = \gamma_{1}+\gamma_{2}+\gamma_{3}$, in addition to $\gamma_{1}, \gamma_{2}, \gamma_{3}$.

Remarkably, under this assumption the associated $Y$-function decouples from the elementary ones. Indeed, $\gamma_{123}$ is a null vector of the 3-cycle pairing, $\langle \gamma, \gamma_{i}\rangle = 0$ for all $i$, implying that $Y_{123}$ does not enter the TBA equations for $Y_{1,2,3}$.

The resulting $Y$-system reads
\begin{align}
\label{eq:YsysA3cycle}
   Y_i^{++}\,\bar Y_i = (1+\bar Y_{i+1})(1+Y_{i-1}^{++})\,,
\end{align}
with cyclic boundary conditions, $Y_{i+3} = Y_{i}$, and similarly for $Y_j\leftrightarrow \bar Y_j$. The decoupling of the composite mode implies a trivial equation for $Y_{123}$,
\be\label{eq:Y123-syst}
Y_{123}^{++}\bar{Y}_{123} = 1\, ,
\ee
(and likewise for $Y\leftrightarrow \bar{Y}$).

The system~\eqref{eq:YsysA3cycle} admits a simple ``integral of motion'', obtained by taking the product of equations~\eqref{eq:YsysA3cycle} over $i$. Together with~\eqref{eq:Y123-syst} and the small-$Y$ asymptotic~\eqref{eq:Y123-smallY}, this allows one to express $Y_{123}$ directly in terms of the elementary $Y$-functions,
\be\label{eq:cubic}
Y_{123} = -\frac{Y_{1}Y_{2}Y_{3}}{(1+Y_{1})(1+Y_{2})(1+Y_{3})} = -\cY_{1} \cY_{2} \cY_{3}\, .
\ee
This provides the simplest example of the composition relation alluded to before.

The consistency of this nonstandard formulation can be checked by comparison with the corresponding $\chi$-system. Defining the $\chi$-functions via~\eqref{eq:chi-Y-sys} and using~\eqref{eq:chi-syste-S}--\eqref{eq:Ychisysgen}, one finds
\be
\label{eq:YtochiA3}
\chi_i=\frac{Y_i}{1+Y_{i-1}}\,, \qquad \chi_i^{++}\bar\chi_i=\frac{(1+\bar Y_{i+1})}{(1+\bar Y_{i-1})}\,,
\ee
with $i=1,2,3$. The composite function is simply $\chi_{123}=-Y_{123}=\chi_1\,\chi_2\,\chi_3$. Inverting the first relation in~\eqref{eq:YtochiA3}, one obtains
\be
\label{eq:chitoYA3}
Y_i=\frac{\chi_i(1+\chi_{i-1}(1+\chi_{i-2}))}{1-\chi_1\,\chi_2\,\chi_3}\,.
\ee
The $A$-variables entering the $\chi$-system (see~\eqref{eq:A-to-chi}) are given in this case by
\be
\label{eq:AA3}
A_i=\chi_i(1+\chi_{i-1})\,, 
\ee
so that 
\be
\label{eq:AYA3}
1+A_i=(1+Y_i)(1+Y_{123})\,.
\ee
Substituting into~\eqref{eq:chitoYA3} reproduces the $\chi$-system, thereby establishing the equivalence between the two formulations.

An additional check follows from comparison with the $Y$-functions obtained via wall crossing. Denoting those by $Y'_1,\ Y'_2,\ Y'_{12}$ and $Y'_3$, one finds
\begin{align}
\label{eq:WCA3}
Y_1'&=\frac{Y_1}{1+Y_2}\,,\quad 
Y_{12}'=\frac{Y_1\,Y_2}{(1+Y_1+Y_2)(1+Y_3)}=-\frac{Y_{123}}{Y'_3}\,,\\ \nonumber
Y_2'&=\frac{Y_2}{1+Y_1}\,, \quad Y'_3=\frac{Y_3\,(1+Y_1+Y_2)}{(1+Y_1)(1+Y_2)}\,.
\end{align}
This mapping can be verified by comparing the corresponding $\chi$-functions,
\begin{align}
\label{eq:YtochiA3WC}
 \chi'_1=\frac{Y'_1(1+Y'_2)\,(1+Y'_{12})}{1+Y'_3}=\chi_1\,, \quad\chi'_2=Y'_2=\chi_2\,,\\\nonumber\chi'_3=\frac{Y_3'}{1+Y_2'}=\chi_3\,, \quad \chi'_{12}=Y'_{12}\,(1+Y_2')=\chi_1\,\chi_2\,.
\end{align}
As a by-product, this provides an alternative proof of the cyclic symmetry of the wall-crossing solution, which is manifest in the present (nonstandard) formulation.

Finally, although the TBA equations themselves are unaffected by the presence of $Y_{123}$, the free energy acquires an additional contribution from the composite mode, which exactly cancels the spurious term in~\eqref{eq:spurion}. The full expression reads
\be
\label{eq:A123}
\cA = \str \int \frac{du(\theta)}{2\pi} \, \big(R(Y_{1})+R(Y_{2})+R(Y_{3})+R(Y_{123})\big)\, .
\ee
One can verify, using the map~\eqref{eq:WCA3} together with Rogers dilogarithm identities, that this agrees with the wall-crossing result,
\be
\cA =\str \int \frac{du(\theta)}{2\pi} \, \big(R(Y'_{1})+R(Y'_{2})+R(Y'_{3})+R(Y'_{12})\big)\,.
\ee

\paragraph{$A_5$ with two cycles.} Following a similar logic, and enforcing consistency with the $\chi$-system, one finds an analogous construction for a quiver consisting of two cycles sharing a common node, as shown in the right panel of Figure~\ref{fig:A3A4A5}. 

To each cycle, we first associate composite functions
\be
Y_{123}=-\frac{Y_1\,Y_2\,Y_3}{(1+Y_1)(1+Y_2)(1+Y_3)}\,,
\qquad Y_{345}=-\frac{Y_3\,Y_4\,Y_5}{(1+Y_3)(1+Y_4)(1+Y_5)}\,,
\ee
as in the previous example. Achieving a complete match, however, requires introducing an additional composite $Y$-function associated with the chain formed by the two cycles, which share node $3$,
\be
Y_{12345}=\frac{Y_{123}\,Y_{345}}{Y_3\,(1+Y_{123})(1+Y_{345})} \,.
\ee
The positive overall sign suggests that this contribution compensates for an over-subtraction and must therefore be reinstated.

The associated $Y$-system for the elementary $Y$-functions can be written compactly in a form analogous to the general expression~\eqref{YsysAdSSph},
\begin{align}
\label{eq:Ysyscomposites}
Y_i^{++}\,\bar{Y}_i =\prod_{\gamma':\langle\gamma',\gamma_i\rangle<0}\left(1+\,\bar{Y}_{\gamma'}\right) \prod_{\gamma':\langle\gamma',\gamma_i\rangle>0}\Big(1+Y_{\gamma'}^{++}\Big)\,.
\end{align}
Here, $\gamma'\in\{\gamma_1,\gamma_2,\gamma_3,\gamma_4,\gamma_5,\gamma_{123},\gamma_{345},\gamma_{12345}\}$, with $\gamma_{123} = \gamma_{1}+\gamma_{2}+\gamma_{3}$, etc., and the products run over $\gamma'$ with negative and positive pairings with $\gamma_i$, respectively. Since the composite $Y$-functions are directly expressed in terms of the elementary ones through the above composition relations, this $Y$-system suffices to determine all $Y$-functions. One may nevertheless derive equations for the composite modes by taking suitable products of the relations above. For instance, one finds
\be
Y_{123}^{++}\bar{Y}_{123} = (1+\bar{Y}_{5})(1+\bar{Y}_{345})(1+Y^{++}_{4})(1+Y^{++}_{345})\, ,
\ee
and
\be
Y_{12345}^{++}\bar{Y}_{12345} = 1\, ,
\ee
and similarly for $Y\leftrightarrow \bar{Y}$.

Although reminiscent of the general form~\eqref{YsysAdSSph}, these relations are not structurally identical. In particular, we observe the presence of interactions between $Y_{123}$ and $Y_{345}$, despite the fact that their pairing vanishes, $\langle \gamma_{123}, \gamma_{345}\rangle =0$. This indicates that, as defined here, the composite modes develop mutual interactions that are not of the same factorized type as those encountered in the wall-crossing framework.
By contrast, the maximal composite $\gamma_{12345}$ behaves more regularly: it decouples from all other modes, in agreement with the fact that $\langle \gamma_{12345}, \gamma\rangle =0$ for all $\gamma$.

Despite this partial departure from the general structure, the correspondence with the $\chi$-system can still be established using
\be\label{eq:chi-function-alternative}
\chi_{i} = \frac{Y_{i}}{\prod_{\gamma':\langle\gamma',\gamma_i\rangle>0}\Big(1+Y_{\gamma'}\Big)}\, ,
\ee
and the relations
\be
1+A_{i} = \prod_{\gamma' \, \ni\, \gamma_i} (1+Y_{\gamma'})\, ,
\ee
where $A_i$ is the $A$-function associated with the edge $i$ of the corresponding triangulation,
\be
\begin{aligned}
&A_{1} = \chi_{1}(1+\chi_{3}(1+\chi_{4}))\, , \quad A_{2} = \chi_{2}(1+\chi_{1})\, , \\
&A_{3} = \chi_{3}(1+\chi_{2})(1+\chi_{4})\, ,\\
&A_{5} = \chi_{5}(1+\chi_{3}(1+\chi_{2}))\, , \quad A_{4} = \chi_{4}(1+\chi_{5})\, .
\end{aligned}
\ee

It is also instructive to compare the above solution with that obtained in Section~\ref{sec:generalsolution} via the wall-crossing method. To this end, consider, for instance, the poset corresponding to case a) in Figure~\ref{fig:DAG}. This configuration includes the elementary states $\gamma_{i}$ with $i=1, \ldots, 5$, as well as the composite states $\gamma_{12} = \gamma_{1}+\gamma_{2}$ and $\gamma_{45} = \gamma_{4}+\gamma_{5}$, and admits the spectrum generator
\be
\boldsymbol{S} = \cK_{\gamma_{2}}\cK_{\gamma_{4}} \cK_{\gamma_{12}}\cK_{\gamma_{45}} \cK_{\gamma_{3}} \cK_{\gamma_{1}}\cK_{\gamma_{5}}\, .
\ee
Denoting the associated $Y$-functions as $Y'_{i}$, $Y'_{12}$ and $Y'_{45}$, one can derive the corresponding $\chi$-functions using the general formula~\eqref{eq:chi-Y-sys}. Comparing these expressions with~\eqref{eq:chi-function-alternative} yields the map
\be
\begin{aligned}
&Y'_{1} = \frac{Y_{1}}{1+Y_{2}}\, , \quad Y'_{12} = \frac{Y_{1}Y_{2}}{(1+Y_{1}+Y_{2})(1+Y_{3})(1+Y_{345})}\, , \quad Y'_{2} = \frac{Y_{2}}{1+Y_{1}}\, , \\
&Y'_{3} = \frac{Y_{3}(1+Y_{1}+Y_{2})(1+Y_{4}+Y_{5})}{(1+Y_{1})(1+Y_{2})(1+Y_{4})(1+Y_{5})}\, , \\
&Y'_{4} = \frac{Y_{4}}{1+Y_{5}}\, , \quad Y'_{45} = \frac{Y_{4}Y_{5}}{(1+Y_{4}+Y_{5})(1+Y_{3})(1+Y_{123})}\, , \quad Y'_{5} = \frac{Y_{5}}{1+Y_{4}}\, . \\
\end{aligned}
\ee
Using these relations, one can verify that the free energy is correctly reproduced by summing over all states $\gamma$ in the non-standard set,
\be
\begin{aligned}
\label{eq:Acomposites}
\cA &= \str \int \frac{du(\theta)}{2\pi} \, \left(\sum_{i\, =\, 1}^{5}R(Y_{i})+R(Y_{123})+R(Y_{345})+R(Y_{12345})\right) \\
& = \str \int \frac{du(\theta)}{2\pi} \, \left(\sum_{i\, =\, 1}^{5}R(Y'_{i})+R(Y'_{12})+R(Y'_{45})\right) \, ,
\end{aligned}
\ee
where the second line corresponds to the free energy obtained via the wall-crossing solution.

Proceeding along the same lines, one finds that similar descriptions exist for higher quivers built from linear chains of 3-cycles. However, an increasing number of composite objects of the type introduced above appears to be required in order to reproduce both the $\chi$-system and the associated free energy.

The general pattern for arbitrary triangulations remains unclear. As emphasized earlier, because the composite $Y$-functions are rigidly expressed in terms of the elementary ones, their $Y$-system relations are less standard and lack the universality of the wall-crossing formulation, making generalizations difficult to anticipate.

Nevertheless, the existence of such a description is intriguing, and it would be interesting to explore whether it can be exploited further in the analytical or numerical study of TBA equations.

\section{Direct comparison with GMN and flip invariance}\label{app:comparison-GMN}

In this appendix, we present a more precise comparison between our equations and those derived by GMN in~\cite{Gaiotto:2014bza}. This analysis also clarifies the origin of the negative-charge states $-\gamma$, which appear as counterparts of the positive states $\gamma \in \mathcal{B}$ in the GMN formalism. We further discuss the nature of the $\chi$-functions and their transformation properties under flips of the triangulation, and we provide a proof of the flip invariance~\cite{Fleury:2017eph} of the free energy at zero bridge length.

\subsection{Comparing the equations}

\begin{figure}[h]
    \centering
    \includegraphics[width=0.5\linewidth]{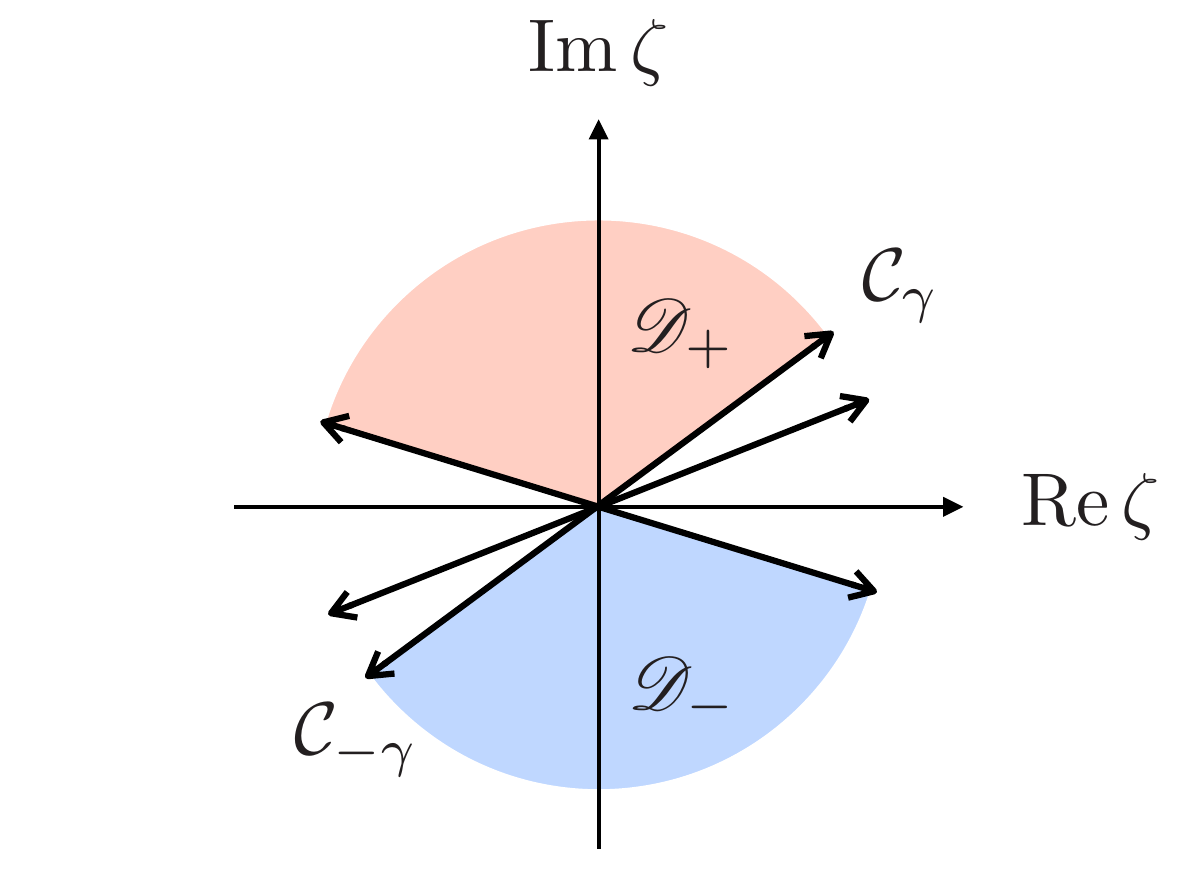}
    \caption{Contours in the $\zeta$-plane. Contours $\cC_{\gamma}$ in the fundamental strip map to rays with $\textrm{arg} \, \zeta_{\gamma} = \textrm{Im}\, \theta_{\gamma}$ in the right half-plane ($\textrm{Re}\, \zeta >0$). In the GMN framework, each such ray is paired with a negative counterpart, $\cC_{-\gamma} = -\cC_{\gamma}$,  extending into the left half-plane ($\textrm{Re}\, \zeta <0$). These negative rays originate from the singularities of the barred kernel $\bar{\GMN}(\theta)$, shifted by $-i\pi$ relative to those of $\GMN(\theta)$.}
    \label{fig:stokes_sectors}
\end{figure}

To facilitate comparison with the GMN formulation, we introduce a convenient parametrization inspired by~\cite{Gaiotto:2010okc}
\be\label{eq:zeta-variable}
    \zeta:=e^{\theta}\, .
\ee
This parametrization provides a useful way to visualize the ordering of contours $\cC_{\gamma}$ in $\cB$. Under this map, each contour $\mathcal{C}_\gamma$ in the fundamental strip $\theta\in (-\pi/2, \pi/2)$ is sent to a ray in the right half of the $\zeta$-plane ($\Re\, \zeta >0$), with the rays ordered according to their arguments, as shown in Figure~\ref{fig:stokes_sectors}. With a slight abuse of notation, we use the same symbol $\cC_{\gamma}$ to denote both the original contour and its image in the $\zeta$-plane. In this representation, $\gamma\succ \gamma'$ corresponds to $\textrm{arg}\, \cC_{\gamma} > \textrm{arg}\, \cC_{\gamma'}$.

In terms of the spectral parameter $\zeta$, the GMN equations take the form~\cite{Gaiotto:2010okc}%
\footnote{Up to a change of variable $\zeta\rightarrow 1/\zeta$.}
\begin{align}\label{eq:GMN-eqs-zeta}
\log X_\gamma(\zeta)
=
Z_\gamma \zeta+ C_\gamma
+
\frac{\bar{Z}_\gamma}{\zeta}
+
\sum_{\gamma'} 
\langle \gamma , \gamma' \rangle
\int_{\mathcal{C}_{\gamma'}}
\frac{\mathrm{d}\zeta'}{4\pi i\zeta'}
\,
\frac{\zeta' + \zeta}{\zeta' - \zeta}
\,
\log\!\left(1+X_{\gamma'}(\zeta')\right)\, .
\end{align}
Here we have suppressed several auxiliary dependencies present in the GMN formulation, including the Coulomb branch parameters $u$ and the overall phase $\vartheta$. In the zero-mode sector, GMN introduce a variable $\theta_\gamma$ associated with the $S^1$-holonomy; here we instead write it as $-iC_\gamma$, in agreement with~(\ref{eq:GMNzeromodes}). Finally, we fix the quadratic refinement of~\cite{Gaiotto:2010okc} to the constant value $-1$, appropriate for a finite chamber of hypermultiplets in $\mathcal{N} = 2$ terminology.

Several features of equation~\eqref{eq:GMN-eqs-zeta} differ from the form of the equations introduced earlier and therefore require clarification.

\paragraph{Negative states.} A first difference concerns the spectrum of charges. In~\eqref{eq:GMN-eqs-zeta} both the label $\gamma$ and the summation over $\gamma'$ run over both the positive charges in the chamber $\mathcal{B}$ and their negatives,
\begin{align}
\gamma, \gamma' \in \mathcal{B} \cup (-\mathcal{B}) .
\end{align}
In the GMN framework, charges $\gamma$ correspond to half-BPS states, while charges $-\gamma$ label their CPT conjugates. The associated functions are not independent: equation~\eqref{eq:GMN-eqs-zeta} implies
\be\label{eq:inversion-law}
X_{-\gamma} = \frac{1}{X_{\gamma}}\, ,
\ee
together with the multiplicative property (see eq.~\eqref{eq:multiplication-rule})
\be
X_{\gamma+\gamma'} = X_{\gamma} X_{\gamma'}\, .
\ee
These relations follow from eq.~\eqref{eq:GMN-eqs-zeta}, together with the linearity in $\gamma$ of $Z_{\gamma}, \bar{Z}_{\gamma}$, and $C_{\gamma}$, and ensure that
\be
X_{\gamma}\in\operatorname{Hom}(\Lambda,\mathbb{C}^{\times})\,,
\ee
i.e., that the $X_{\gamma}$'s form a cluster torus on which KS transformations act naturally.

\paragraph{Integration contours.} A second difference concerns the choice of integration contours. In our construction, these are identified with the $\mathcal{C}_\gamma$ introduced earlier. By contrast, in~\cite{Gaiotto:2010okc} they are defined as
\be\label{eq:C-gamma-GMN}
\mathcal{C}^{(\mathrm{GMN})}_\gamma=\{\zeta : Z_\gamma \zeta <0\} .
\ee
i.e., they are determined by the phases of the central charges.

This choice optimizes the convergence of the integrals near the endpoints $\zeta = 0$ or $\infty$. In this asymptotic regime, $X_{\gamma} \propto e^{Z_{\gamma}\zeta +\bar{Z}_{\gamma}/\zeta}\rightarrow 0$ and the exponential suppression is maximized when
\be\label{eq:optimal}
\textrm{Im}\, (Z_{\gamma} \zeta + \bar{Z}_{\gamma}/\zeta) = 0\, , \qquad \textrm{Re}\, (Z_{\gamma} \zeta + \bar{Z}_{\gamma}/\zeta) < 0\, .
\ee
This condition is equivalent to~\eqref{eq:C-gamma-GMN} when $Z_{\gamma}$ and $\bar{Z}_{\gamma}$ are complex conjugates. Since the choice of contour is kinematical, it deforms as the kinematics are varied.

In our setup, the prescription~\eqref{eq:C-gamma-GMN} is not always the most natural, since $Z_{\gamma}$ and $\bar{Z}_{\gamma}$ are not, in general, complex conjugates. One may nevertheless adopt~\eqref{eq:optimal} to optimize convergence. 

Alternatively, one can fix the contours using the original hexagon weights, which becomes equivalent to~\eqref{eq:optimal} in the regime of large bridge lengths or large $\sigma$'s. This amounts to choosing the imaginary part $\varepsilon_{\gamma}$ of the contour $\cC_{\gamma}$ in the $\theta$-plane according to
\be\label{eq:alternative-contours}
-l_{\gamma} \cosh{\theta} + i \sigma_{\gamma}\sinh{\theta} = -\sqrt{l_{\gamma}^{2}+\sigma_{\gamma}^2}\cosh{(\theta-i\varepsilon_{\gamma})} \qquad \Rightarrow \qquad \textrm{tan}\, \varepsilon_{\gamma} = \frac{\sigma_{\gamma}}{l_{\gamma}}\, ,
\ee
with $\sigma_{\gamma} = \sigma^{\AdS}_{\gamma}$. In our construction, we assumed sufficiently large bridge lengths to ensure convergence of all hexagon integrals and sums. As a result, the contours lie close to one another, corresponding to $\zeta\in \mathbb{R}^{+}$ in the $\zeta$-plane. When the bridge lengths decrease, or in asymptotic limits such as the OPE regime, it is more convenient to fix the contours as in~\eqref{eq:alternative-contours} and track their evolution via analytic continuation of the TBA equations, supplemented by wall-crossing transformations.

\paragraph{Reality condition.} A further difference concerns the reality properties of the equations. In the GMN setup, $Z_{\gamma}$ and $\bar{Z_{\gamma}}$ are complex conjugates and $C_{\gamma}$ is purely imaginary, implying
\be\label{eq:GMN-reality}
\textrm{GMN}: \qquad X_{\gamma}(\zeta)^* = X_{-\gamma}(-1/\zeta^*)\, ,
\ee
where $^{*}$ denotes complex conjugation.

In our case, these conditions are not generally satisfied, even when the kinematical parameters $\sigma_{\gamma}, \phi_{\gamma}$ and $l_{\gamma}$'s are real. The origin of this discrepancy lies in the dressing of scalar excitations, which introduces the terms proportional to $\qw$ (see e.g.~\eqref{eq:constant-C}) and spoils manifest reality. Instead, complex conjugation is accompanied by an exchange of dressing,
\be
Y_{\gamma}(\theta)^*_{\qw} = \bar{Y}_{\gamma}(-\theta^*)_{-\qw}\, ,
\ee
and similarly for the free energy,
\be
\cA^*_{\qw} = \cA_{-\qw}\, .
\ee
Since the hexagon prescription averages over $\qw = \pm$, the final correlator remains real.

\paragraph{Direct comparison.} Having clarified the above differences, we can now directly match the two formulations. It follows directly from the map~\eqref{eq:zeta-variable} that the kernel $\GMN$ coincides with the one in the GMN equations~\eqref{eq:GMN-eqs-zeta},
\be
\frac{d\zeta'}{4\pi i\zeta'} \frac{\zeta'+\zeta}{\zeta'-\zeta} = \frac{d\theta'}{2\pi} G(\theta-\theta')\, .
\ee
To reproduce the second kernel $\bGMN$, one must include the contribution of negative charges. This is achieved by performing the change of variables $\zeta'\rightarrow -\zeta'$ for the sum over $-\gamma'$ in~\eqref{eq:GMN-eqs-zeta}, using that $\cC_{-\gamma'} = -\cC_{\gamma'}$ in the $\zeta$-plane. One then finds agreement with our equations provided that
\be\label{eq:map-to-GMN}
X_{\gamma}(\zeta) = Y_{\gamma}(\theta)\, , \qquad X_{-\gamma}(-\zeta) = \bar{Y}_{\gamma}(\theta)\, ,
\ee
with the identification holding for $\theta\in \cC_{\gamma}$.

In this way, the contribution of the barred functions is naturally interpreted as arising from states of charge $-\gamma$, reproducing the structure of~\eqref{eq:GMN-eqs-zeta}. The negative argument in~\eqref{eq:map-to-GMN} implies that $\cC_{-\gamma}$ in~\eqref{eq:GMN-eqs-zeta} is obtained from $\cC_{\gamma}$ by a shift of $\pi$ in the imaginary $\theta$-direction. This establishes a pairing between contours: each $\mathcal{C}_{\gamma}$ in the fundamental strip is associated with a dual contour $\mathcal{C}_{-\gamma}$ in the \textit{anti-fundamental strip}, obtained by shifting $\theta\rightarrow \theta-i\pi$. The corresponding rays differ by a phase $\pi$ in the $\zeta$-plane, as illustrated in Figure~\ref{fig:stokes_sectors}.

\subsection{$\chi$-functions and mutations}

We can now return to the question of the nature of the $\chi$-functions. In the GMN framework, these arise by construction as cluster $\chi$-coordinates. In our case, this property is not manifest and must be derived. We show below that our $\chi$-functions indeed behave as Fock–Goncharov coordinates, following arguments from~\cite{Gaiotto:2010okc,Alim:2011kw,Alim:2011ae}.

\begin{figure}[h]
    \centering
    \includegraphics[width=0.6\linewidth]{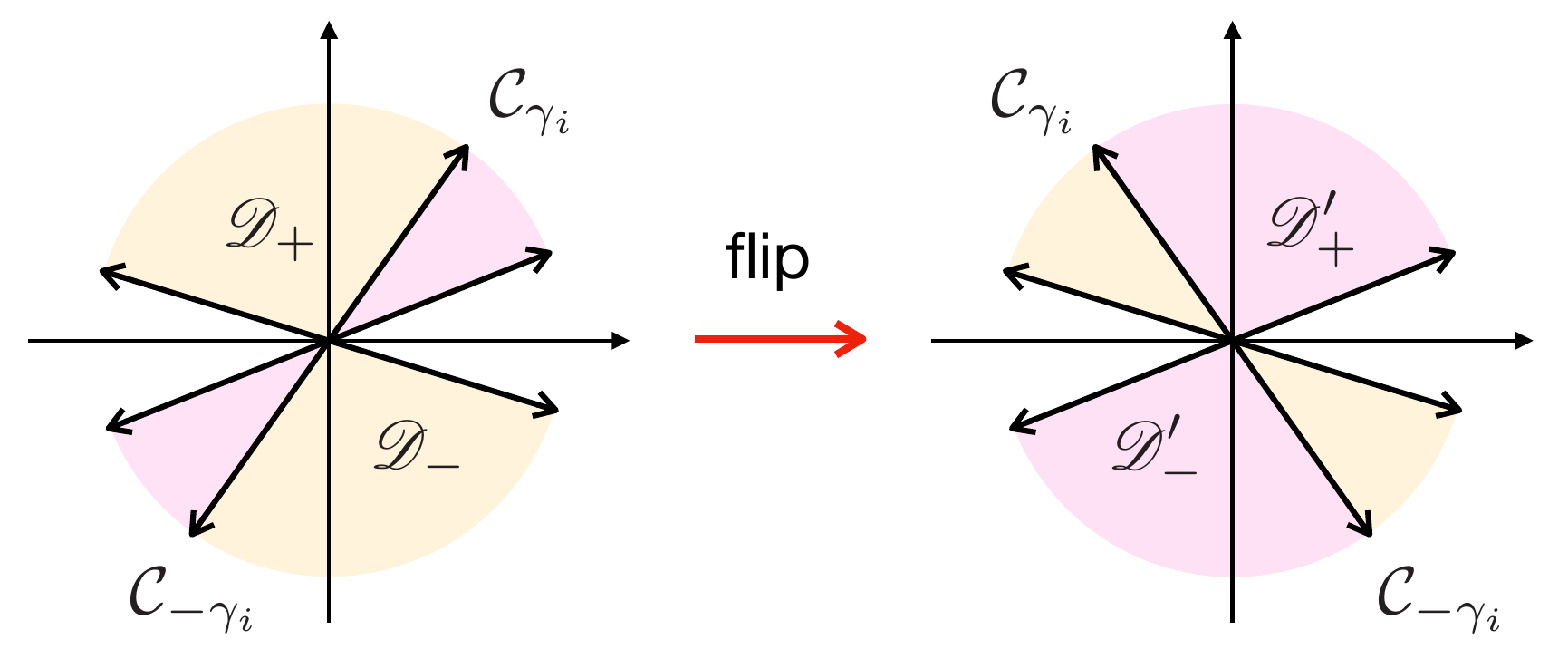}
    \caption{Flip of the topmost contour $\cC_{\gamma_i}$. As this contour exits the fundamental domain $\Re\, \zeta >0$, its companion $\cC_{-\gamma_i} = -\cC_{\gamma_i}$ enters the domain from below, becoming the lowest element in the new contours ordering. The $\chi$-functions in the new domain $\mathscr{D}_+'$ are related to the original ones in $\mathscr{D}_+$ by the action of the KS transformation $\cK_{\gamma_i}$.}
    \label{fig:placeholder}
\end{figure}

To this end, consider a chamber $\mathcal{B}$ and let $\gamma_i$ be the highest element, as illustrated in Figure~\ref{fig:placeholder}. Rotate the ray $\cC_{\gamma_i}$ across the domain $\mathscr{D}_+$ into the anti-fundamental domain $\textrm{Re}\, \zeta <0$, while $\cC_{-\gamma_i}$ simultaneously moves through $\mathscr{D}_-$ into the fundamental domain $\Re\, \zeta >0$.
This yields a new ordering in which $\gamma_i$ is removed and $-\gamma_i$ becomes the lowest element, as illustrated in Figure~\ref{fig:placeholder}.

The resulting set of positive charges is $\{\gamma, \gamma\neq \gamma_i\}\cup \{-\gamma_i\}$. These charges must generate a new positive cone, i.e.~they must be expressible as positive integer combinations of a new basis containing $-\gamma_i$. The only consistent possibility is that the basis transforms according to~\eqref{eq:mutationlatticebasis}, namely via the mutation $\mu_i$ \cite{Alim:2011kw,Alim:2011ae}.

This change of basis induces a transformation $\tau_i$ on the $\chi$-functions, defined on the elementary variables associated with the edges of the triangulation as
\begin{align}
\label{eq:tropic}
    \chi_j\mapsto \tau_i(\chi_j) \qquad \textrm{with}\qquad \tau_i(\chi_j):=
\begin{cases}
\chi_i^{-1} & j=i,\\
\chi_j\,\chi_i^{\max(0,\langle \gamma_j,\gamma_i\rangle)} & j\neq i\,,
\end{cases}
\end{align}
and extended multiplicatively to all $\chi$-functions.

However, this is not the full story. As seen in Figure~\ref{fig:placeholder}, the shift also redefines the upper domain, $\mathscr{D}_+\rightarrow \mathscr{D}'_+$, an effect captured by the KS transformation $\cK_{\gamma_i}$. The full transformation acting on the $\chi$-functions is therefore
\begin{align}
\cK_{\gamma_i}\circ \tau_i=\mu_i\,,
\end{align}
which coincides with the mutation associated with $\gamma_i$, or equivalently the flip of the edge $i$.

This reproduces the Fock–Goncharov decomposition of a mutation into a tropical part, implemented by $\tau_i$ and a non-tropical (dynamical) part, implemented by $\cK_{\gamma_i}$.\footnote{This statement can be made precise using a Hamiltonian approach to cluster algebra mutations; see e.g.~\cite{GekhtmanNakanishiRupel2017}.} The former corresponds to a change of basis in the charge lattice, while the latter encodes the nontrivial transformation of the functions. This establishes that the $\chi$-functions in our construction are indeed cluster $\chi$-variables. 

It is also instructive to describe the corresponding transformation at the level of the $Y$-functions. Since only the contour $\cC_{\gamma_i}$ is moved, the functions $Y_{\gamma}$ with $\gamma\neq \gamma_i$ remain unchanged, up to relabeling induced by the mutation of charges. By contrast, the function $Y_{\gamma_i} = \chi_{\gamma_i}$ is analytically continued outside the fundamental strip and replaced by
\be
X_{-\gamma_i}|_{\mathscr{D}_{-}} = 1/X_{\gamma_i}|_{\mathscr{D}_{-}} = 1/\tilde{\chi}_{\gamma_{i}}\, ,
\ee
using~\eqref{eq:inversion-law} and~\eqref{eq:tilde-chi}. Since $1/\tilde{\chi}_{\gamma_i} = \bar{\chi}^{++}_{\gamma_i} = \bar{Y}^{++}_{\gamma_{i}}$ (see eq.~\eqref{eq:tilde-chi-chi++}), we conclude that the mutation amounts to analytically continuing $Y_{\gamma_i}$ under the shift $\theta\rightarrow \theta+i\pi$, up to a permutation of $Y$ and $\bar{Y}$. In this sense, the flip of the edge $i$ can be interpreted as a $++$ shift of the corresponding $Y$-function. This statement holds in a chamber where $\gamma_i$ is the highest element. In a generic chamber, one must first bring $\gamma_i$ to the top of the ordering via wall-crossing transformations before applying the shift.

\begin{figure}[h]
    \centering
    \includegraphics[width=0.5\linewidth]{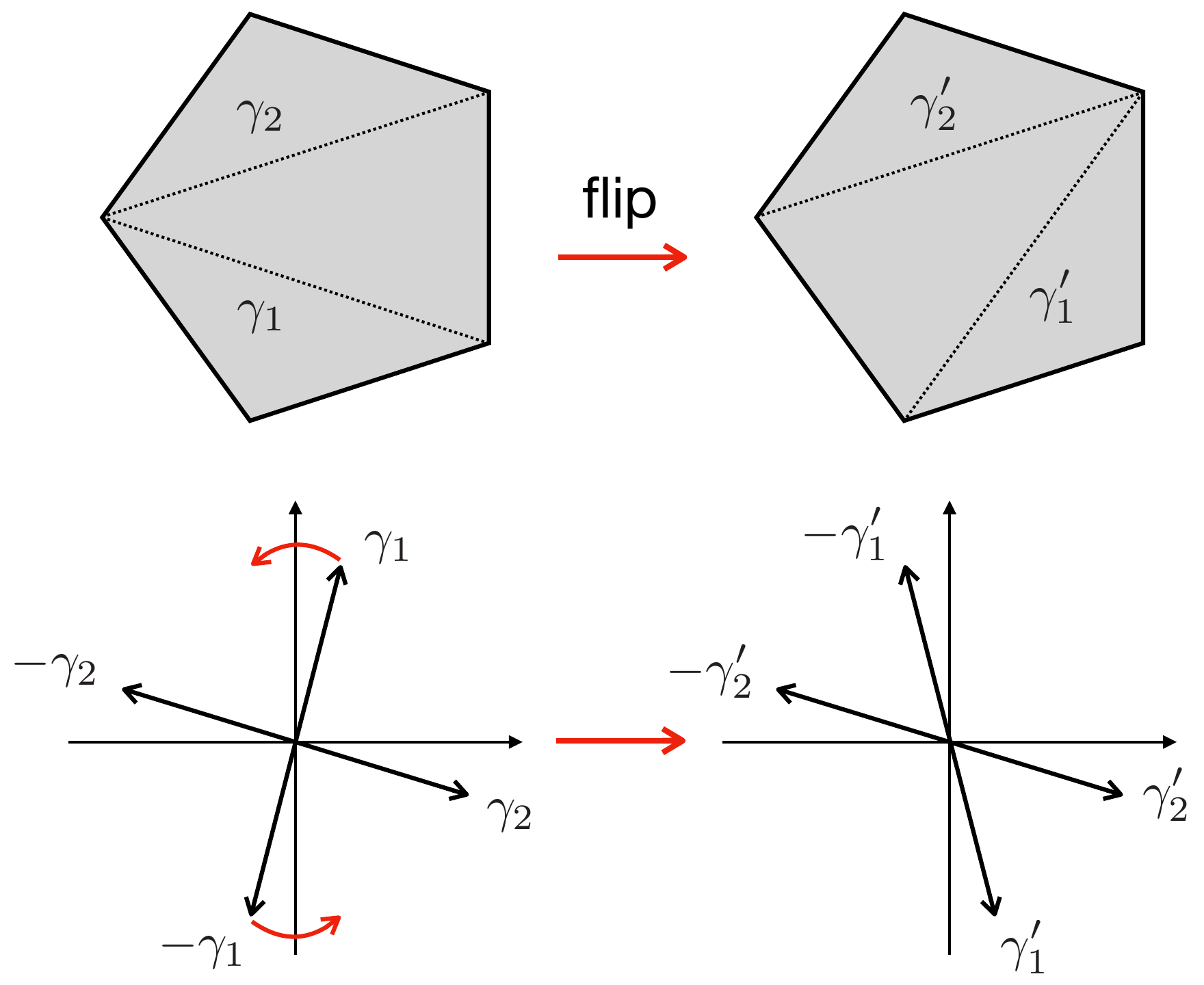}
    \caption{Mutation of a two-cut configuration. Here $\langle \gamma_{1}, \gamma_{2}\rangle = 1$. After relabeling, the relative orientation is reversed, $\langle \gamma'_{1}, \gamma'_{2}\rangle = -1$, with $\gamma'_{1} = -\gamma_{1}, \gamma'_{2} = \gamma_{2}$. In the bottom panel, we use $\gamma$ as shorthand for $\cC_{\gamma}$ to simplify notation.}
    \label{fig:two-cut-Fock-Goncharov}
\end{figure}

\begin{figure}[h]
    \centering
    \includegraphics[width=0.8\linewidth]{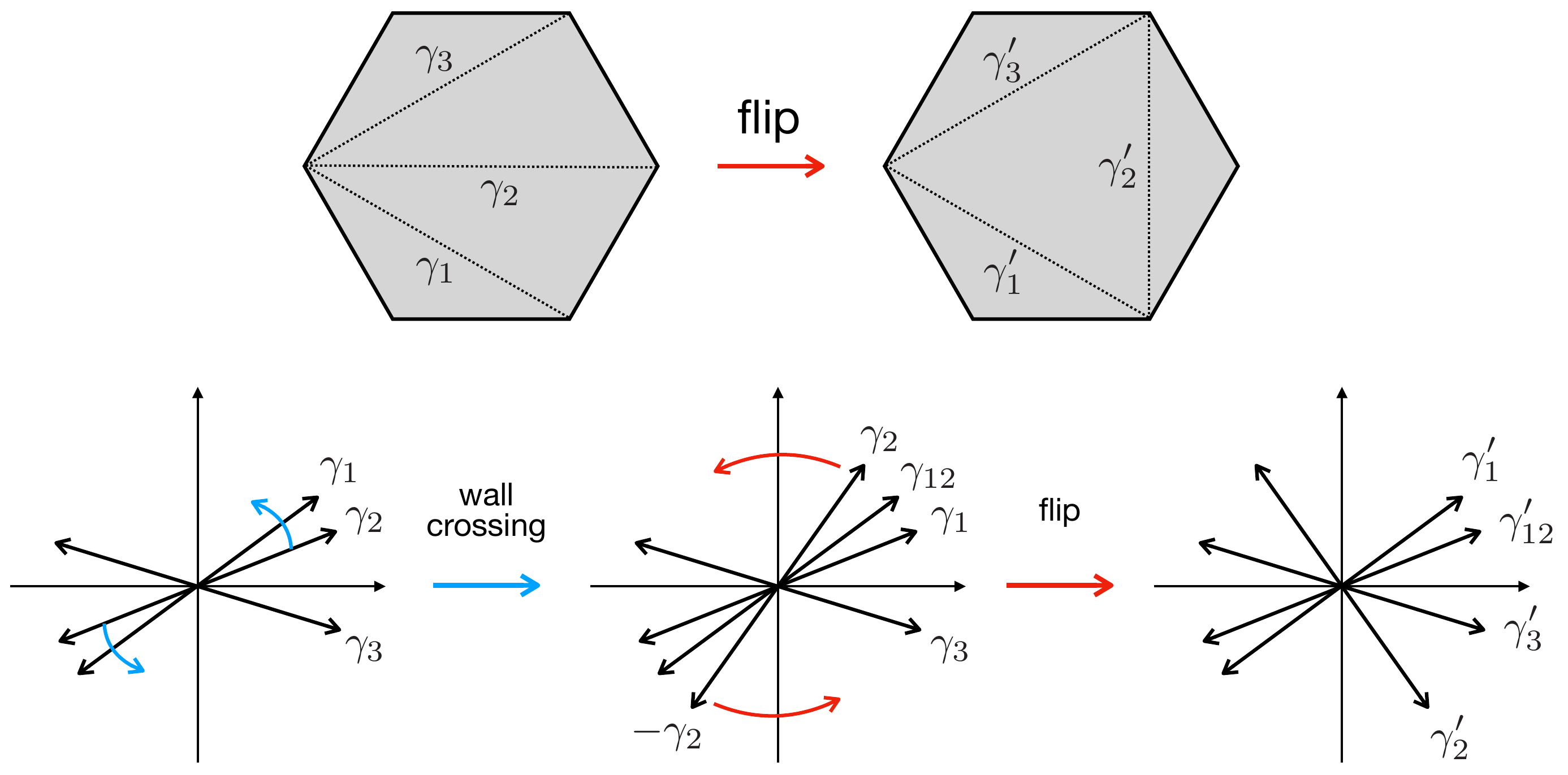}
    \caption{Mutation from the fan triangulation to the cyclic quiver. Here $\langle \gamma_{1}, \gamma_{2}\rangle = \langle \gamma_{2}, \gamma_{3}\rangle = 1$ and $\langle \gamma_{1}, \gamma_{3}\rangle =0$. A wall-crossing is required to bring $\gamma_{2}$ to the top position, generating the composite state $\gamma_{12} = \gamma_{1}+\gamma_{2}$. After flipping $\gamma_{2}$, one obtains the minimal chamber of the cyclic $A_{3}$ quiver, with $\langle \gamma'_{1}, \gamma'_{3}\rangle = \langle \gamma'_{3}, \gamma'_{2}\rangle = \langle \gamma'_{2}, \gamma'_{1}\rangle = 1$ and $\gamma'_{1} = \gamma_{1}+\gamma_{2}, \gamma'_{2} = -\gamma_{2}, \gamma'_{3} = \gamma_{3},\gamma'_{12} = \gamma'_{1}+\gamma'_{2} = \gamma_{1}$. Note that, after relabeling, the composite state $\gamma_{12}$ becomes the elementary state $\gamma'_1$ in the flipped triangulation.}
    \label{fig:three-cut-fan-to-cyclic-mutation}
\end{figure}

To conclude, we illustrate this construction in the simplest two-cut configuration in Figure~\ref{fig:two-cut-Fock-Goncharov} with mutation at $\gamma_1$. A more instructive example is the mutation from the fan triangulation in the three-cut configuration to the cyclic quiver, shown in Figure~\ref{fig:three-cut-fan-to-cyclic-mutation}. In this case, the mutation is associated with the state $\gamma_2$, which is not initially the highest charge. One must first perform a wall-crossing between $\gamma_1$ and $\gamma_2$, as depicted in the left panel of that figure, bringing $\gamma_2$ to the top of the ordering. The above procedure then applies. In this example, the additional state $\gamma_{12}$ of the cyclic quiver emerges naturally: although absent in the initial fan triangulation, it is generated by the combined action of wall-crossing and mutation~\cite{Gaiotto:2010okc}.

\subsection{Flip invariance}

Finally, let us discuss the flip invariance of the free energy. As recalled in Section~\ref{sec:review}, when the bridge length $l_{\gamma_i} =0$, one may switch to a different triangulation by flipping edge $i$ without affecting the result. As discussed above, this operation corresponds to an analytic continuation of the associated $Y$-functions under the $++$ shift. The previous analysis ensures that the TBA equations transform consistently under this operation, namely through a mutation of the geometric data entering the driving terms.

To make this precise, it is convenient to view the driving terms as functions of the cross ratios,
\be
\cY_{\gamma}(\theta) = \cY(z_{\gamma}, \bar{z}_{\gamma}, \alpha_{\gamma}, \bar{\alpha}, l_{\gamma}; \theta)\, .
\ee
Then the claim is that upon analytically continuing all functions of $Y_{\gamma_i}$ outside of the fundamental strip, using the $++$ shift, one finds that the TBA equations retain their form, up to the replacement
\be\label{eq:transf-calY}
\cY_{\gamma}(\theta) \rightarrow \cY_{\gamma}'(\theta) = \cY(z'_{\gamma}, \bar{z}'_{\gamma}, \alpha'_{\gamma}, \bar{\alpha}'_{\gamma}, l'_{\gamma}; \theta)\, , \qquad \gamma \neq \gamma_i\, ,
\ee
with
\be
l'_{\gamma} = l_{\gamma}\, ,
\ee
and where the primed cross ratios follow from the KS transformation,
\be\label{eq:z-prime}
z'_{\gamma} = z_{\gamma} (1+z_{\gamma_{i}})^{\langle \gamma_i, \gamma\rangle}\, .
\ee
A straightforward calculation on the driving terms for $\gamma\neq \gamma_i$ confirms that this is indeed the case. Instead, for $\gamma_i$, the transformation is induced by the shift itself,
\be
\cY_{\gamma_i}(\theta)\rightarrow \cY_{-\gamma_i}'(\theta) = \cY^{++}_{\gamma_i}(\theta)\, .
\ee
We stress that we work in the same setup as before, choosing the shifted state to lie at the top of the chamber. In particular, when shifting $\gamma_i$-quantities by $++$, we do not cross any contours and so no wall-crossing is involved here. At the end, the resulting contour for the new state $-\gamma_i$ lie \textit{below} all the others in the fundamental strip, as in the Figure~\ref{fig:two-cut-Fock-Goncharov} for instance.

The TBA equations with these new source terms and contour configuration can be identified with those of the flipped triangulation. To see that this is the result of a mutation requires changing the basis of charges, as recalled earlier. This follows from the mutation~\eqref{eq:mutationlatticebasis}, which maps the charge $\gamma_j$ for the edge $j$ in the initial triangulation to the charge $\gamma'_j = \mu_{\gamma_i}(\gamma_j)$ of the corresponding edge in the flipped triangulation. By linearity, we also extend this transformation to all other charges $\gamma$ in the chamber, using
\be
\gamma \rightarrow  \gamma' = \mu_{\gamma_{i}}(\gamma) = \gamma + \max\bigl(0, \langle \gamma_i, \gamma\rangle\bigr) \gamma_i\, .
\ee
There is an important point worth mentioning here: since no transformation really occurs on the charges $\gamma \neq \gamma_i$, as we transport $\gamma_i$ outside the fundamental strip, the set of charges $\cB\backslash \{\gamma_i\}$ in the chamber must be same before and after we flip. This means that the latter set must be invariant under the action of $\mu_{\gamma_i}$. This observation, and similar ones, are genuine constraints on the set of allowed chambers for a given quiver, and can be turned into efficient algorithms for building them, such as the mutation method. Here, we simply admit them.

That being said, we may formulate the flip invariance as the statement that
\be
Y_{\gamma'}' = Y_{\mu_{\gamma_i}(\gamma)}\, , \qquad \gamma' \neq \gamma_i\, ,
\ee
and
\be
Y'_{\gamma'_{i}} = \bar{Y}_{\gamma_i}^{++}\, , \qquad \bar{Y}'_{\gamma'_{i}} = Y_{\gamma_i}^{++}\, ,
\ee
solve the TBA equations associated with the flipped triangulation, with the driving terms
\be
\cY'_{\gamma'}(\theta) = \cY(z'_{\gamma'}, \ldots , l'_{\gamma'}; \theta)\, ,
\ee
where $z'_{\gamma'} = \mu_{\gamma_i}(z_{\gamma}), \ldots,$ are the mutated cross ratios, and $l'_{\gamma'} = l_{\gamma}$. We stress that this holds for $l_{\gamma_i} = 0$ here; when $l_{\gamma_i}\neq 0$, additional terms show up in the transformation of the driving terms, inducing shifts of the bridge length $l_{\gamma}$ notably.

To finish establishing flip symmetry, it remains to check whether the free energy transforms in the same way. As this calculation is slightly more subtle, we discuss it in more detail. Starting form the free energy,
\be\label{eq:free-energy-flip}
\cA = \cA_{\gamma_i} + \sum_{\gamma \, \neq\, \gamma_i} \cA_{\gamma}\, ,
\ee
we deform the integration contour in
\be
\cA_{\gamma_i} = \str \int \frac{d\theta}{2\pi \cosh^2{\theta}} R(Y_{\gamma_i})\, ,
\ee
across the line $\textrm{Im}\, \theta = i\pi/2$, while keeping all other contributions in~\eqref{eq:free-energy-flip} unchanged. Due to the pole in the measure, this deformation produces an additional contribution from the residue at $\theta = i\pi/2$,
\be
\cA_{\gamma_i} = \str \int \frac{d\theta}{2\pi \cosh^2{\theta}} R(Y^{++}_{\gamma_i}) + \textrm{Res}\, ,
\ee
with the residue given by
\be\label{eq:Res}
\textrm{Res} = \frac{1}{2i}\str \left[\log{(Y_{\gamma_i}/\cY_{\gamma_i})} \partial_{\theta}\log{(1+Y_{\gamma_i})} - \partial_{\theta}\log{(Y_{\gamma_i} \cY_{\gamma_i})} \log{(1+Y_{\gamma_i})}\right]_{\theta \, =\,  i\pi/2}\, .
\ee
An important simplification comes form the fact that the first term in brackets vanishes at $\theta = i\pi/2$. To see that, recall that since $\gamma_i$ is the highest element in the poset, one has $Y_{\gamma_i} = \chi_{\gamma_i}$, and hence $Y_{\gamma_i}/\cY_{\gamma_i} \rightarrow 1$ at this point. This holds exactly in the AdS sector. In the sphere sector, there is a slight mismatch due to dressing effects. Fortunately, the mismatch is such that whenever $\log{(Y_{\gamma_i}/\cY_{\gamma_i})}$ is non zero, the factor multiplying it in~\eqref{eq:Res} vanishes at $\theta = i\pi/2$ if $l_{\gamma_i} = 0$. Thus, when $l_{\gamma_i} = 0$, we can discard the first term in~\eqref{eq:Res}. A similar simplification applies to part of the second term, since  
\be
\partial_{\theta}\log{\cY_{\gamma_i} (i\pi/2)} = -il_{\gamma_i} = 0\, .
\ee
As a result, the residue reduces to
\be\label{eq:residue}
\textrm{Res} = \frac{i}{2}\str \left[\log{(1+Y_{\gamma_i})} \partial_{\theta}\log{Y_{\gamma_i}}\right]_{\theta \, =\, i\pi/2}\, .
\ee
The remaining derivative can be evaluated using the TBA equations~\eqref{eq:TBA-GMN}. The algebra is straightforward and shows that the net effect of~\eqref{eq:residue} is to shift the contributions proportional to $\log{\cY_{\gamma}}$ in $\sum_{\gamma\, \neq\, \gamma_i} \cA_{\gamma}$ to their final, mutated form given in~\eqref{eq:transf-calY}. In other words, both the free energy and the TBA equations transform identically under this shift, preserving their general structure up to the KS action on the kinematic data~\eqref{eq:z-prime}. This completes the proof of flip invariance.

Finally, let us comment on convergence. To justify the above manipulations, one must ensure that the $Y$-functions remain well behaved at large rapidity when analytically continued across the line $\textrm{Im}\, \theta = \pi/2$. This can be achieved by taking $\sigma_{\gamma_i}$ to be sufficiently large and positive, so that convergence is ensured by the factor $e^{i\sigma_{\gamma_{i}} \sinh{\theta}}$. In the opposite regime, where $\sigma_{\gamma_i}$ is negative, one should instead perform the analytic continuation in the opposite direction, that is $\theta\rightarrow \theta-i\pi$.

Throughout the analysis, we have also implicitly assumed that $1+Y_{\gamma_i}$ does not vanish in the relevant domain. Otherwise, extra residues---analogous to excited-state contributions in conventional TBA analyses---would arise from the contour deformation.

\section{Mutation method}\label{app:mutation-method}

In order to construct chambers in a more systematic manner, one may employ the powerful methods developed in the study of BPS quivers, in particular the mutation method proposed in~\cite{Alim:2011kw,Alim:2011ae}. This approach provides a systematic algorithm for determining chambers directly from the quiver data. It relies on general structural constraints that the chambers must satisfy, together with the mutation properties of the underlying quiver. We refer the reader to the original references for a detailed discussion of these aspects and give in this appendix only a brief outline of the method.

The mutation method involves performing successive mutations on the charges $\gamma$ associated with the nodes of the quiver. After each mutation, these charges transform linearly according to the general rules described in \eqref{eq:mutationlatticebasis}. To construct a chamber, one performs mutations on nodes associated with positive linear combinations of the elementary $\gamma$'s, continuing the process until the quiver returns to its original form, up to an overall reversal of the signs of the initial $\gamma$'s.

In the case of a linear quiver, this procedure readily reproduces the minimal chamber associated with the canonical ordering, starting from the sinks (nodes with all arrows directed toward them) and proceeding recursively until the sources (nodes with all arrows directed outward) are reached. Other chambers, corresponding to configurations with additional states, can be obtained by following alternative sequences of mutations. 

\begin{figure}[h]
    \centering
    \includegraphics[width=0.7\linewidth]{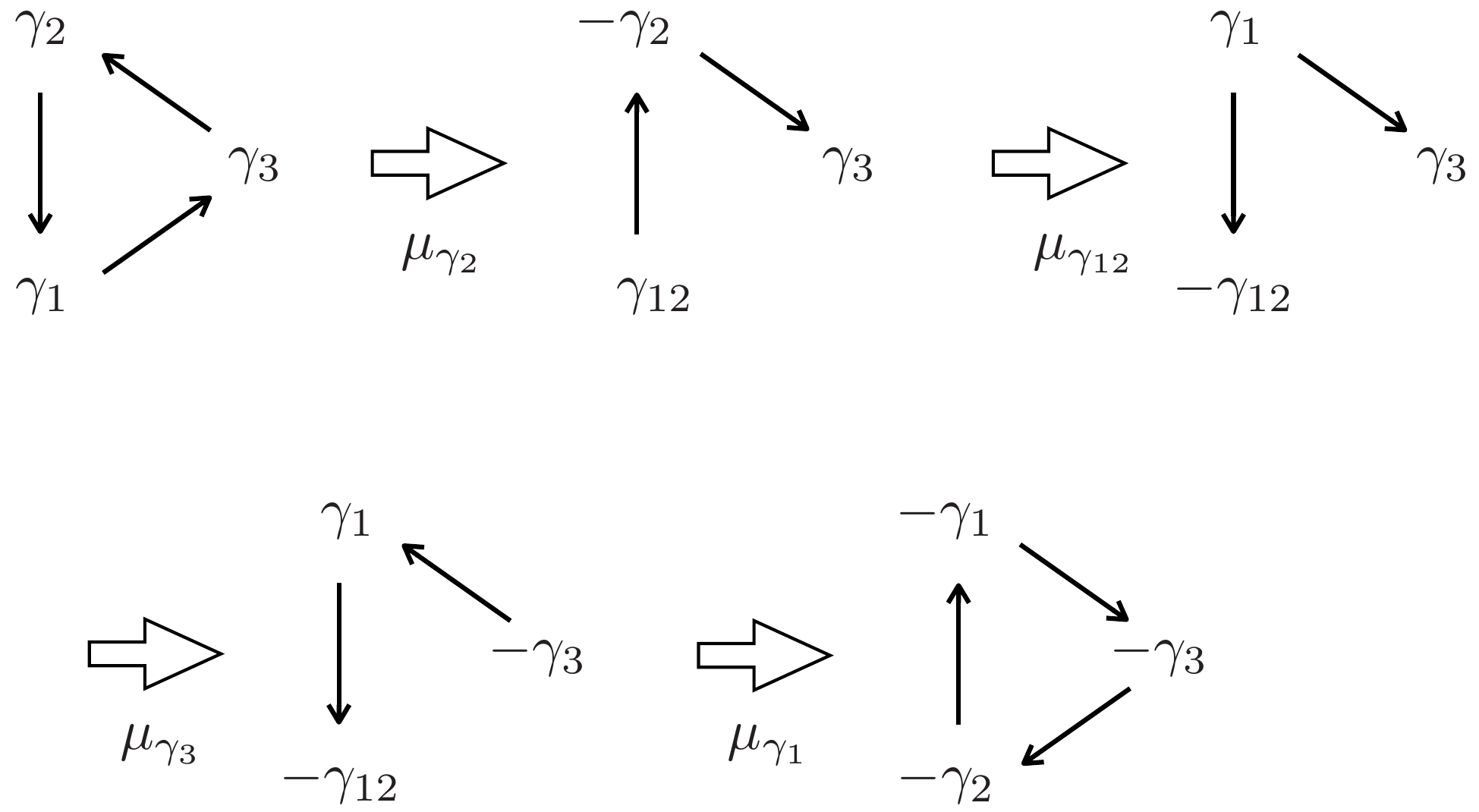}
    \caption{Illustration of the mutation method for a minimal chamber of cyclic $A_{3}$. Mutations are performed successively on the nodes according to the rules given in~\eqref{eq:mutationlatticebasis}. The process terminates when no positive combinations of $\gamma$'s remain to be mutated, at which point the quiver returns to its original form, up to an overall reversal of the signs of the $\gamma$'s.}\label{fig:mutationmethod}
\end{figure}

For instance, in the maximal chamber of the $A_2$ quiver $\gamma_1 \leftarrow \gamma_2$, one starts by mutating on $\gamma_2$, and then continues mutating as follows
\be
\gamma_1 \leftarrow \gamma_2 \,\, \underset{\mu_{\gamma_2}}{\Rightarrow} \,\, \gamma_1+\gamma_2 \rightarrow -\gamma_2 \,\, \underset{\mu_{\gamma_1+\gamma_2}}{\Rightarrow} \,\, -\gamma_1-\gamma_2 \leftarrow \gamma_1 \,\, \underset{\mu_{\gamma_1}}{\Rightarrow}\,\,  -\gamma_2 \rightarrow -\gamma_1\, .
\ee
This sequence corresponds to the ordering $\gamma_2 \succ \gamma_1+\gamma_2 \succ \gamma_1$, encountered earlier. 
Had we started to mutate on $\gamma_1$, we would have obtained 
the minimal chamber $\gamma_1\succ \gamma_2$,
\be
\gamma_1 \leftarrow \gamma_2 \,\, \underset{\mu_{\gamma_1}}{\Rightarrow} \,\, -\gamma_1 \rightarrow \gamma_2 \,\, \underset{\mu_{\gamma_2}}{\Rightarrow} \,\,   -\gamma_1 \leftarrow -\gamma_2\, .
\ee
Figure~\ref{fig:mutationmethod} illustrates another example: the construction of the minimal chamber for cyclic $A_3$. In this case, the mutations are performed in the order $\mu_{\gamma_{2}}, \mu_{\gamma_{1}+\gamma_{2}}, \mu_{\gamma_{3}}$ and $\mu_{\gamma_{1}}$. The ordering between the two middle elements can, however, be relaxed, since the corresponding mutations commute,
\be
\langle \gamma_{3}, \gamma_{1}+\gamma_{2}\rangle = 0 \qquad \Rightarrow \qquad \mu_{\gamma_{1}+\gamma_{2}}\,\mu_{\gamma_{3}} = \mu_{\gamma_{3}} \,\mu_{\gamma_{1}+\gamma_{2}}\, .
\ee 
At the level of the TBA equations, this corresponds to the earlier observation that there is no interaction between the states $\gamma_{3}$ and $\gamma_{12} = \gamma_{1}+\gamma_{2}$.

We may then generalize this computation to a tree of 3-cycles such as the one discussed in Section~\ref{sec:generalsolution}. In particular, by replacing $\gamma_3$ by $Q'$ in the above example, one readily reproduces the recurrence relation given in~\eqref{eq:recursion-S} for the spectrum generator, showing agreement with our former algorithm. The mutation method is, however, more flexible and systematic. It can be applied to classify all chambers for polygons and extends naturally to closed geometries.


\providecommand{\href}[2]{#2}\begingroup\raggedright\endgroup

\bibliographystyle{JHEP}

\end{document}